\RequirePackage{snapshot}
\documentclass[aps,prb,twocolumn,amsfonts,showpacs,longbibliography,superscriptaddress,floatfix]{revtex4-2}
\usepackage{verbatim,epsfig,amsmath,amssymb,bm,epsf,graphicx,psfrag,bbold,amsthm,amsfonts}
\usepackage[bottom]{footmisc}
\usepackage{hyperref,array}
\usepackage[capitalize]{cleveref} 
\usepackage{enumitem}
\usepackage{multirow}
\usepackage{grffile}
\usepackage{framed}
\usepackage{esint}
\setlist{nosep}
\usepackage{placeins}
\usepackage[all]{xy}
\usepackage{color}
\usepackage[utf8]{inputenc}
\usepackage{float}
\usepackage{natbib}
\usepackage{tikz-cd}
\usepackage{verbatim}
\usepackage{leftidx}
\usepackage[normalem]{ulem}
\usepackage{notes2bib}

\newcommand{\moy}[1]{\langle #1 \rangle}
\newcommand{\ket}[1]{\mbox{$ | #1 \rangle $}}

\newcommand{\half}{\frac{1}{2}}

\newcommand{\longsquiggly}{\xymatrix{{}\ar@{~>}[r]&{}}}

\newcommand\A {{\cal A}}
\newcommand\B {{\cal B}}
\newcommand\C {{\cal C}}
\newcommand\D {{\cal D}}

\newcommand\bR {{\mathbb R}}

\newcommand\bZ {{\mathbb Z}}

\newcommand{\cV}{\mathcal{V}}
\newcommand{\cW}{\mathcal{W}}
\newcommand{\cX}{\mathcal{X}}

\newcommand\beq {\begin{equation}}
	\newcommand\eeq {\end{equation}}
\newcommand\beqa {\begin{equatiobn}\begin{array}}
		\newcommand\eeqa {\end{array}\end{equation}}
\newcommand\bal {\begin{align}}
	\newcommand\eal {\end{align}}
\newcommand{\bea}{\begin{eqnarray}}
	\newcommand{\eea}{\end{eqnarray}}

\newcommand{\innerproduct}[2]{\langle #1| #2\rangle}
\newcommand{\outerproduct}[2]{|#1\rangle\langle #2|}

\newcommand{\ztwo}{\mathbb{Z}_2}

\theoremstyle{plain}

\theoremstyle{definition}

\theoremstyle{remark}

\makeatletter
\def\l@subsubsection#1#2{}
\makeatother
\begin{document}

	\title{Symmetry-Enriched Criticality in a Coupled Spin-Ladder}
	
	\author{Suman Mondal}
	\email{suman.mondal@theorie.physik.uni-goettingen.de}
	\affiliation{Institut f\"{u}r Theoretische Physik, Georg-August-Universit\"{a}t G\"{o}ttingen, D-37077 G\"{o}ttingen, Germany}
	\author{Adhip Agarwala}
	\email{adhip@iitk.ac.in}
	\affiliation{Department of Physics, Indian Institute of Technology, Kanpur 208016, India}
	\author{Tapan Mishra }
	\email{mishratapan@niser.ac.in}
	\affiliation{ School of Physical Sciences, National Institute of Science Education and Research, Jatni 752050, India}
	
	\affiliation{Homi Bhabha National Institute, Training School Complex, Anushaktinagar, Mumbai 400094, India}
	
	\author{Abhishodh Prakash}
	\email{abhishodh.prakash@physics.ox.ac.uk (he/him/his)}
	\affiliation{Rudolf Peierls Centre for Theoretical Physics, University of Oxford, Oxford OX1 3PU, United Kingdom}

	
	\begin{abstract}
		We study a one-dimensional ladder of two coupled XXZ spin chains and identify several distinct gapless symmetry-enriched critical phases. These have the same unbroken symmetries and long-wavelength description, but cannot be connected without encountering either a phase transition or other intermediate phases. Using bosonizaion, we analyze the nature of their distinction by determining how microscopic symmetries are manifested in the long-wavelength fields, the behavior of charged local and nonlocal operators, and identify the universality class of all direct continuous phase transitions between them. One of these phases is a gapless topological phase with protected edge modes. We characterize its precise nature and place it within the broader classification. We also find the occurrence of `multiversality' in the phase diagram, wherein two fixed phases are separated by continuous transitions with different universality classes in different parameter regimes. We determine the phase diagram and all its aspects, as well as verify our predictions numerically using density matrix renormalization group and a mapping onto an effective spin-1 model. 
	\end{abstract}
	
	\maketitle
	
	\tableofcontents
	
	\section{Introduction}
	One of the most remarkable characteristics of quantum and classical many-body physical systems is the emergence of distinct, stable \emph{phases} that are divided by sharp \emph{phase transitions}. There is tremendous theoretical and experimental interest in enumerating all possible phases and transitions and characterizing their properties. Symmetries have provided a guiding principle to facilitate this. It was realized that distinct phases of matter occur when microscopic symmetries are spontaneously broken at long distances~\cite{LanduLifshitzVol5}. The knowledge of microscopic symmetries allows us to enumerate the different ways it can be spontaneously broken, the properties of the resulting long-range order, and sometimes even the nature of the phase transition. The concept of `topological' ordering that falls outside the symmetry-breaking framework~\cite{Wen1990topological} following the discovery of the quantum Hall effect~\cite{PrangeGirvinQHEbook} has expanded the mechanisms by which distinct phases can arise. This has spurred a flurry of intense research activity over the past decades in classifying and characterizing gapped phases of matter \cite{Haldane_Nobel_RevModPhys.89.040502}. These new phases represent multiple ways in which symmetries can be unbroken and yet result in different phases. The distinguishing features are detectable in subtle signatures present in entanglement patterns and boundary/ topology effects.
	
	Gapless phases, on the other hand, have been left by the wayside in these recent developments. Despite being ubiquitous in nature and making frequent appearances in the phase diagrams of many known physical systems, the mechanisms by which they arise and are stabilized are relatively unclear although various descriptive frameworks have been successfully devised to understand them. For example, when noninteracting bands of fermions are partially filled they lead to the formation of Fermi liquids~\cite{LanduLifshitzVol9}, Dirac~\cite{DiracMaterials_review} / Weyl~\cite{WeylMaterialsReview} semimetals. Using partons and emergent gauge fields to describe systems has also been useful in accessing non-Fermi-liquid phases~\cite{Sachdev_TopologicalOrder_2018, Schulz1995fermi}. The most systematic known mechanism is arguably the spontaneous breaking of continuous symmetries, e.g., which results in the formation of superfluids. The program of classifying gapless states of matter with unbroken symmetries is still in its early stages.

	Examples of gapless states hosting edge modes have been reported in various works~\cite{DasSharmaGaplessSPTPhysRevB.83.174409,Fidkowski_GaplessSPT_PhysRevB.84.195436,MengCheng_GaplesSPT_PhysRevB.84.094503,AltmanGaplessSPT_PhysRevB.85.125121,KeselmanBerg_TLL_PhysRevB.91.235309,ParkerScaffidiVasseur_GaplessSPT_PhysRevX.7.041048,ParkerScaffidiVasseur_TLL_DDW_PhysRevB.97.165114,DHLee_GaplessSPT_jiang2017symmetry,VerresenThorngrenJonesPollmann_SymmetryEnrichedCriticality_PhysRevX.11.041059,ThorngrenVishwanathVerresen_IntrinsicallyGapless_PhysRevB.104.075132} and was developed into the notion of gapless symmetry protected topological (SPT) phases in refs.~\cite{ParkerScaffidiVasseur_GaplessSPT_PhysRevX.7.041048,ParkerScaffidiVasseur_TLL_DDW_PhysRevB.97.165114}. This was generalized in ref.~\cite{VerresenThorngrenJonesPollmann_SymmetryEnrichedCriticality_PhysRevX.11.041059} to the concept of `symmetry-enriched criticality' where the authors ask the following question--- given a critical state corresponding to a fixed universality class,  how many ways can an unbroken symmetry \emph{enrich} it? In other words, can microscopic symmetries manifest themselves in inequivalent ways at long distances when the physics is described by conformal field theory (CFT)? The authors demonstrate that the answer is yes and that distinct symmetry-enriched critical states exist that cannot be connected without encountering an abrupt change in universality class or intermediate phases. These critical states may be topological and host edge modes, or may not.

	It is desirable to study models and phase diagrams which demonstrate the existence of symmetry-enriched critical phases and transitions between them. The most common critical phases are the so-called `Luttinger liquids'~\cite{Haldane_LuttingerLiquid_1981} which is described by the compact-boson CFT~\cite{Ginsparg1988applied} and arise as the long-wavelength description for many one-dimensional interacting systems of bosons or fermions. Coupled Luttinger liquids, which naturally arise in spin-ladder models, provide a much richer playground and will be used in this work to investigate subtle symmetry and topological properties of gapless phases. In this paper, we study the phase diagram of a microscopic one-dimensional spin ladder that stabilizes multiple symmetry-enriched Luttinger liquid phases protected by the symmetries of the model. One of these, dubbed XY$_2^*$, is topological, i.e. it has stable symmetry-protected edge modes. Using Abelian bosonization, we give a comprehensive treatment of their symmetry distinction and features, as well as describe local and nonlocal observables that can differentiate between them.  We also study this rich variety of phases and phase transitions numerically using density matrix renormalization group (DMRG)  as well as an effective low-energy mapping to spin-1 Hamiltonians. We also discuss additional interesting features of the phase diagram such as the presence of `multiversality'~\cite{BiSenthil_Multiversality_PhysRevX.9.021034,AP2022multiversality} wherein the same two phases (Haldane and trivial) are separated by different stable universality classes in different parameter regimes. 
	
	The paper is organized as follows --- in \cref{sec:Hamiltonian}, we introduce our model, list its symmetries, summarize the phase diagram and its important elements. We use Abelian bosonization in \cref{sec:Bosonization1} to establish the symmetry distinction between various gapless phases and in \cref{sec:Bosonization2} to analyze the topological Luttinger liquid phase XY$_2^*$. We numerically analyze our model in \cref{sec:Numerics} and reproduce aspects of our phase diagram using an effective spin -1 model in \cref{sec:Spin1}. Various additional details are relegated to \cref{app:bosonization,app:phasedig,app:String}.

	\section{Model Hamiltonian and phase diagram}
	\label{sec:Hamiltonian}
	\subsection{Two presentations of the model}
	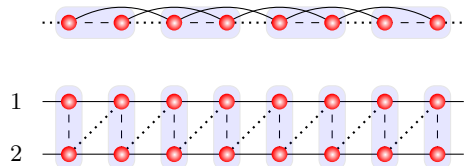
\begin{figure}[!ht]
		\begin{tikzpicture}[scale=.7]
			\foreach \x in {0,2,4,6}
			\filldraw[rounded corners,blue,opacity=0.1,draw=black,thin] (\x-0.25,2.5-0.3) rectangle (\x+1.25,2.8);
			\foreach \x in {0,2,4,6} 
			\draw[dashed] (\x,2.5) -- (\x+1,2.5);
			\foreach \x in {1,3,5} 
			\draw[dotted,thick] (\x,2.5) -- (\x+1,2.5);
			\draw[dotted,thick] (-0.5,2.5) -- (0,2.5);
			\draw[dotted,thick] (7,2.5) -- (7.5,2.5);
			\foreach \x in {0,...,5} 
			\draw[] (\x,2.5) to[bend left] (\x+2,2.5);
			\foreach \x in {0,...,7} 
			\shade[inner color=white, outer color = red] (\x,2.5) circle (0.15);
			
			\draw[] (-0.5,0) -- (7.5,0);
			\draw[] (-0.5,1) -- (7.5,1);
			\foreach \x in {0,...,7} 
			\draw[dashed] (\x,0) -- (\x,1);
			\foreach \x in {0,...,6} 
			\draw[dotted,thick] (\x,0) -- (\x+1,1);
			\foreach \x in {0,...,7} 
			\filldraw[rounded corners,blue,opacity=0.1,draw=black,thin] (\x-0.25,-0.3) rectangle (\x+0.25,1.3);
			\foreach \x in {0,...,7} 
			\shade[inner color=white, outer color = red] (\x,0) circle (0.15);
			\foreach \x in {0,...,7} 
			\shade[inner color=white, outer color = red] (\x,1) circle (0.15);
			\node[align=left,above] at (-1.,0.7)  {1};
			\node[align=left,above] at (-1.,-0.3)  {2};			
		\end{tikzpicture}
		\caption{Schematic representation of the Hamiltonian in the small-$J$ limit shown in  \cref{eq:Hamiltonian_smallJ} (top) and the large-$J$ limit shown in \cref{eq:Hamiltonian_largeJ} (bottom). The solid and broken lines represent the various two-spin interaction terms.}
		\label{fig:model}
	\end{figure}
	
\begin{figure}[!ht]
		\centering
		\begin{tabular}{cc}
			\includegraphics[width=0.72\linewidth]{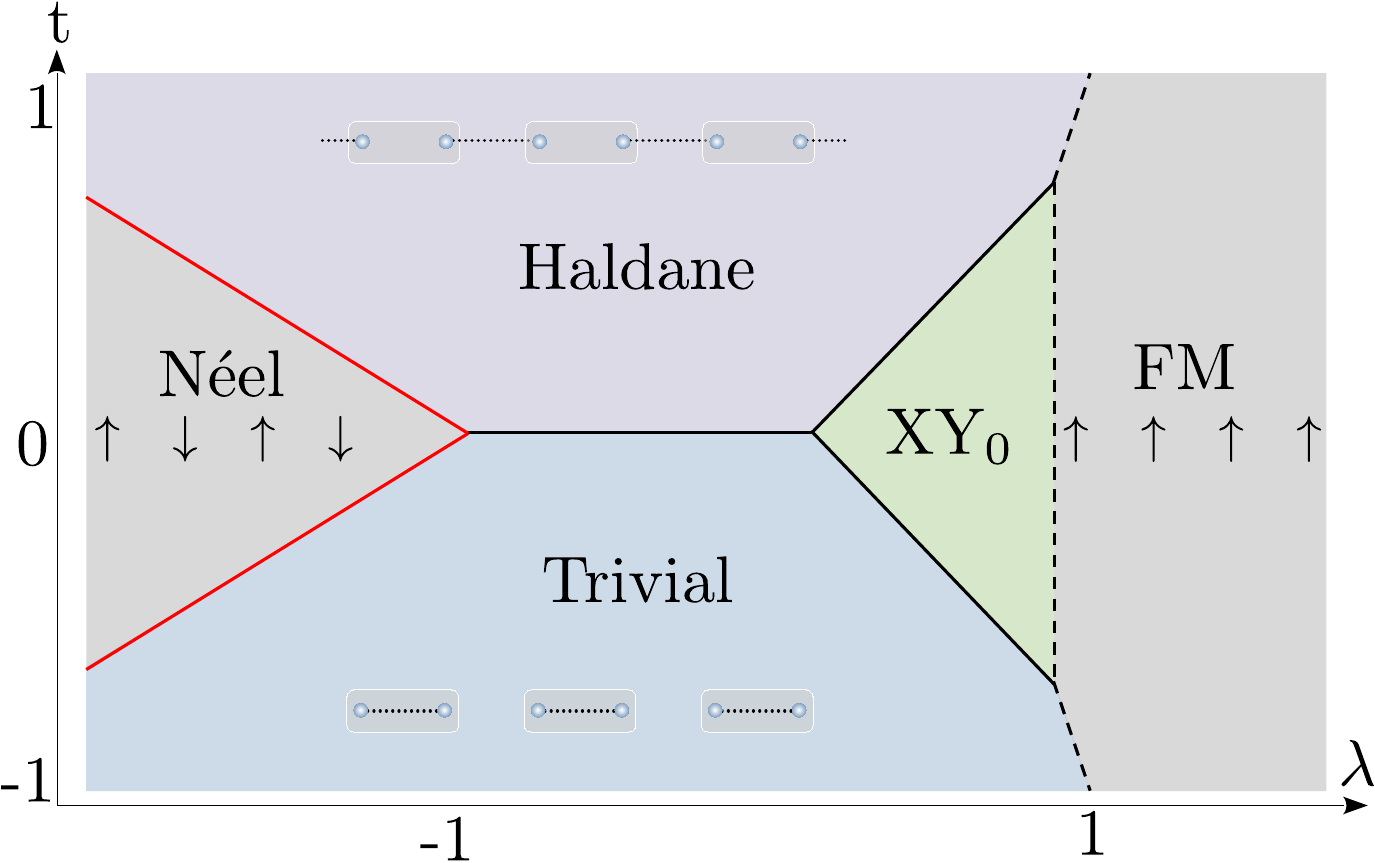} \\
			\includegraphics[width=0.72\linewidth]{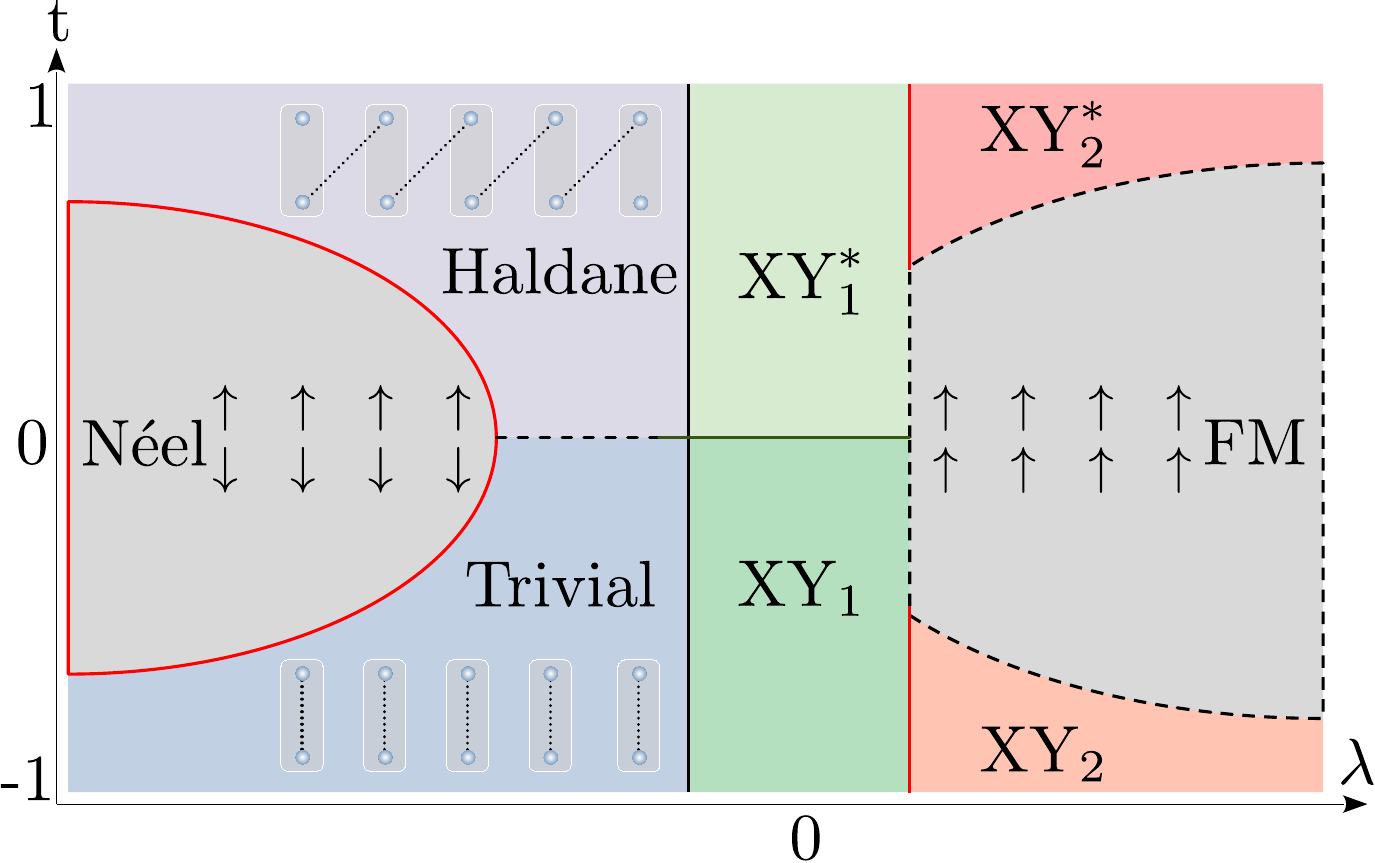} 
		\end{tabular}
		\caption{The schematic phase diagram for the small-$J$ Hamiltonian shown in \cref{eq:Hamiltonian_smallJ} (top) and the  large-$J$  Hamiltonian shown in \cref{eq:Hamiltonian_largeJ} (bottom). Continuous lines indicate second-order phase transitions and broken lines indicate first-order transitions. Cartoon ground states are shown for the gapped phases.}
		\label{fig:phasedig}
	\end{figure}

	We study a one-dimensional chain of qubits (spin halves). There are two ways to view the system. The first, shown in the top panel of \cref{fig:model}  is to regard the system as a single chain where the Hamiltonian can be written as an XXZ chain with alternating bond strength and next-nearest-neighbor coupling as follows (the  $S^zS^z$ coupling constants $\lambda$ and $\Delta$ are reversed in sign compared to the usual convention for convenience)
	\begin{multline}
		H = \sum_j \left(1+ (-1)^j t\right) \left( S^x_{ j} S^x_{ j+1} +  S^y_{j} S^y_{j+1} - \lambda  S^z_{j} S^z_{ j+1} \right)\\ + J \sum_j \left( S^x_{ j} S^x_{ j+2} +  S^y_{j} S^y_{j+2} - \Delta  S^z_{j} S^z_{ j+2} \right), \label{eq:Hamiltonian_smallJ}
	\end{multline}
	$\vec{S}_{j}$ are spin $\frac{1}{2}$ operators, defined as usual in terms of Pauli matrices: $\vec{S}_{ j} = \frac{1}{2} \vec{\sigma}_{ j}$. The model has four parameters: $\{J, \Delta, \lambda\} \in \bR$ and $t \in [-1,1]$. We will be interested in two-dimensional phase diagrams varying $\lambda$ and $t$ with $J$ and $\Delta$ fixed. The representation in \cref{eq:Hamiltonian_smallJ}  appropriate in the limit of small $J$ when the next-nearest neighbor (nnn) term can be regarded as a perturbation of the bond-dimerized XXZ spin chain. The phase diagram in this limit is well known~\cite{KohmotoNijsKadanoff_XXZDimerization_PhysRevB.24.5229,AP2022multiversality}, and is schematically shown in \cref{fig:phasedig}. We are interested in the gapless Luttinger liquid phase labeled XY$_0$ which can be adiabatically connected to the one found in the phase diagram of the XXZ model (i.e. $1/\sqrt{2}<\lambda<1$ for $t=J=0$). 

	For large $J$, the Hamiltonian is appropriately visualized as a two-rung spin ladder as shown in the bottom panel of \cref{fig:model}, with the following presentation: 
	\begin{align}
		H &= H_1 + H_2 + H_\perp + H'_\perp,  \label{eq:Hamiltonian_largeJ} \text{ where, } \\
		H_{\alpha} &=  J \sum_j \left( S^x_{\alpha j} S^x_{\alpha j+1} +  S^y_{\alpha j} S^y_{\alpha j+1} - \Delta  S^z_{\alpha j} S^z_{\alpha j+1} \right), \nonumber\\ 
		H_\perp &=  (1-t) \sum_j \left(S^x_{1 j} S^x_{2 j}+S^y_{1 j} S^y_{2 j}-\lambda S^z_{1 j} S^z_{2 j} \right), \nonumber \\
		H'_\perp &=  (1+t) \sum_j \left( S^x_{2 j} S^x_{1 j+1}+ S^y_{2 j} S^y_{1 j+1}-\lambda  S^z_{2 j} S^z_{1 j+1} \right).  \nonumber
	\end{align}
	
	$\alpha = 1,2$ labels the rungs of the ladder and contains, respectively, the even and odd lattice spins of \cref{eq:Hamiltonian_smallJ}. $H_\alpha$ represents the intra-rung   and $H_\perp,~H'_\perp$ represent the inter-rung XXZ couplings. In this limit, it is appropriate to treat $H_\perp,~H'_\perp$ as perturbations to $H_\alpha$. The schematic phase diagram in this limit which we find is shown in \cref{fig:phasedig}. Our prime interest in this phase diagram are the four Luttinger liquid phases labelled XY$_1$, XY$_1^*$, XY$_2$ and XY$_2^*$.  We will show that all five gapless phases found in the large and small $J$ phase diagrams are distinct from each other meaning they cannot be connected without encountering a phase transition. Furthermore, we will also show that one of these, XY$_2^*$ is a topological Luttinger liquid containing stable edge modes~\cite{KeselmanBerg_TLL_PhysRevB.91.235309,ThorngrenVishwanathVerresen_IntrinsicallyGapless_PhysRevB.104.075132,VerresenThorngrenJonesPollmann_SymmetryEnrichedCriticality_PhysRevX.11.041059}. A positive finite $\Delta$, introduces intra-chain ferromagnetic correlations, which is crucial to open up various gapless phases as will be discussed in detail.

	Parts of the large-$J$ phase diagram have appeared in previous studies~\cite{Haldane_XXZNNN_PhysRevB.25.4925,StrongMillis_Spinladder_PhysRevLett.69.2419,OrignacGiamarchi_SpinLadder_PhysRevB.57.5812,SommaAligia_XXZNNN_PhysRevB.64.024410,HirataNomura_XXZNNN_PhysRevB.61.9453,Jiang_XXZNNN,Furukawa_XXZNNN_PhysRevB.86.094417,SATO_XXZNNN_2011}. However, the complete set of gapless phases, their symmetry distinction and topological  properties have not been identified to the best of our knowledge. This will be the focus of our work. We will understand these (a) using bosonization  in \cref{sec:Bosonization1,sec:Bosonization2}, (b)numerically, using density matrix renormalization group (DMRG) in \cref{sec:Numerics} and (c)  by mapping \cref{eq:Hamiltonian_largeJ} to  effective spin-1 models in \cref{sec:Spin1}.

	\subsection{Symmetries}
	\label{sec:symmetries}
	
	\begin{table}[!ht]
		\begin{center}
			\begin{tabular}{|l|l|l|}
				\hline
				\textbf{Symmetry} & \textbf{Small $J$} & \textbf{Large $J$} \\				
				\hline
				\hline
				$U(1)$ spin rotations& $\begin{array}{l}
					S^{\pm}_{j} \mapsto e^{\pm i \chi}S^{\pm}_{j}  \\
					S^z_j \mapsto S^z_j
				\end{array}$  & $\begin{array}{l}
					S^{\pm}_{\alpha j} \mapsto e^{\pm i \chi}S^{\pm}_{\alpha j}\\
					S^z_{\alpha j} \mapsto  S^z_{\alpha j}
				\end{array}$ \\			
				\hline
				$\ztwo^R$ spin reflection & $\begin{array}{l}
					S^{\pm}_{ j} \mapsto S^{\mp}_{ j}  \\
					S^{z}_{j} \mapsto - S^{z}_{ j}
				\end{array} $ & $\begin{array}{l}
					S^{\pm}_{ j} \mapsto S^{\mp}_{\alpha j}  \\
					S^{z}_{j} \mapsto - S^{z}_{ \alpha j}
				\end{array} $ \\
				\hline
				$\ztwo^P$ lattice parity & $\vec{S}_{j} \mapsto \vec{S}_{-j+1}$ & $\vec{S}_{1, j} \leftrightarrow \vec{S}_{2, -j}$ \\
				\hline 	 			
				$\bZ$ lattice translation& $\vec{S}_{j} \mapsto \vec{S}_{j+2}$ & $\vec{S}_{\alpha, j} \mapsto \vec{S}_{\alpha,j+1}$\\
				\hline								    
			\end{tabular}
		\end{center}
		\caption{Symmetries of the model in both the small $J$ and large $J$ representations of local operators of the Hamiltonian shown in \cref{eq:Hamiltonian_smallJ,eq:Hamiltonian_largeJ}. \label{tab:symmetries}}
	\end{table}
	Global symmetries of the system will play an important role. Four symmetries are sufficient to characterize all phases and transitions: (i) on-site $U(1)$ symmetry that corresponds to spin rotations, (ii) on-site $\ztwo^R$ spin reflections (iii) $\ztwo^P$ lattice symmetry that corresponds to a bond-centred reflection in the small-$J$ version and site-centred reflection followed by layer-exchange in the large-$J$ version, (iv) $\bZ$ lattice translations. The symmetry action on spin operators is shown in \cref{tab:symmetries}. Altogether, the full symmetry group is~\footnote{The $U(1)$ and $\ztwo^R$ symmetries do not commute and together form the non-abelian group $O(2) 
		\cong U(1) \rtimes \ztwo^R$} $G \cong O(2) \times \ztwo^P \times \bZ$.  Additional symmetries are present in the model (eg: time-reversal) but are not needed for our purposes. In other words, they can be explicitly broken without changing the nature of the phase diagram.

	\subsection{Phases and  transitions}
	The main focus of our work are the five symmetry enriched Luttinger liquid phases XY$_{0}$, XY$_{1,2}$ and XY$_{1,2}^*$ shown in \cref{fig:phasedig}. At long distances, all five of these are described by a compact boson conformal field theory with central charge $c=1$. However, the presence of global symmetries results in distinctions between them. The microscopic symmetries shown in \cref{sec:symmetries} are imprinted on the long-wavelength degrees of freedom in different ways in each of the five phases, and as a consequence, they cannot be connected without encountering a phase transition or an intermediate phase. Conversely, the distinction can be eliminated between the phases, and they can be connected by explicitly breaking appropriate symmetries.  This will be explained in detail using bosonization analysis in \cref{sec:Bosonization1}.
	
	More operationally, we will show that the distinction between these phases can be demonstrated using appropriate local and string operators. While XY$_0$, XY$_1$ and XY$_1^*$ can be distinguished by local operators only, XY$_2$ is distinguished from XY$_2^*$ using string operators. This is comparable to the situation with gapped phases, where symmetry protected topological (SPT) phases~\cite{ChenGuWen_MPS_PhysRevB.84.235128} are distinguished by string operators. The phase diagrams shown in \cref{fig:phasedig} contain a non-trivial SPT phase, the Haldane phase, which is distinguished from the trivial paramagnet using an appropriate string operator. We will see that the same string operator can be used to distinguish between XY$_2$ and XY$_2^*$. Furthermore, like the Haldane phase, the XY$_2^*$ phase will also contain protected edge modes but with reduced degeneracy. This will also be explained in \cref{sec:Bosonization2} using bosonization and confirmed numerically in \cref{sec:Numerics}. 
	
	We are also interested in the phase transitions between the gapless phases shown in \cref{fig:phasedig}. These are summarized below along with the universality class.
	
	\begin{itemize}
		\itemsep1em
		\item XY$_1$ to XY$_1^*$: c = 2 theory of two compact bosons.
		\item XY$_1$ to XY$_{2}$ and XY$^*_1$ to XY$^*_{2}$: c= $\frac{3}{2}$ theory of a $c=1$ compact boson CFT combined with a $c=\frac{1}{2}$ Ising CFT.
	\end{itemize} 
	
	The second order transitions out of the gapless phases to either the Haldane or trivial phase in \cref{fig:phasedig} is of the BKT type, when the value of the Luttinger parameter is such that the perturbation that drives the gapped phase becomes relevant. We will also understand these using bosonization in \cref{app:phasedig} and confirm them numerically in \cref{sec:Numerics}. 
	
	Finally, the gapped phases present in \cref{fig:phasedig} (Haldane SPT, trivial paramagnet and symmetry breaking N\'{e}el and ferromagnet) as well as transitions between them are well understood. We mention them for completeness-- The Haldane and trivial phases are separated by a compact boson CFT for small $J$ and by a first-order transition for large $J$.  The  N\'{e}el phase is separated from the trivial and Haldane phases by an Ising CFT and its symmetry-enriched variant, respectively~\cite{VerresenThorngrenJonesPollmann_SymmetryEnrichedCriticality_PhysRevX.11.041059} for both small and large $J$. Finally, the FM is separated from the Haldane and Trivial phases through a first-order transition for small $J$.

	\section{Bosonization analysis I: characterizing the gapless phases}
	\label{sec:Bosonization1}
	In this section, we will study the properties of various gapless phases and transitions between them using abelian bosonization. We begin by reviewing the framework applicable to the parameter regimes for small and large $J$ and then proceed to understand the various gapless phases in two ways: (i) by using the effective action of microscopic symmetries on the CFT and (ii) the behavior of local and non-local operators carrying appropriate symmetry charges. We delay a thorough analysis of the topological aspects of the XY$_2^*$ phase to \cref{sec:Bosonization2}.
	
	\subsection{Bosonization formulas for small and large $J$ and conventional description of phases} 	
	
	For small $J$, the Hamiltonian shown in \cref{eq:Hamiltonian_smallJ}, can be treated as a single XXZ spin chain with perturbations. In the regime of our interest, it can be bosonized using standard arguments~\cite{Haldane_LuttingerLiquid_1981,Giamarchi} as follows (see  \cref{app:bosonization} for more details)
	\begin{multline}
		H  \approx  \frac{v}{2 \pi} \int dx \left[\frac{1}{4K} \left(\partial_x \phi\right)^2 + K \left(\partial_x \theta\right)^2\right] \\+ 2 \mathcal{A} \mathcal{C} t  \int dx~    \cos \phi  - \frac{\mathcal{B}^2 \lambda}{2}   \int dx~    \cos 2\phi  + \ldots \label{eq:H_bosonize_smallJ}
	\end{multline}
	$\phi \cong \phi + 2 \pi $ and $\theta \cong \theta + 2 \pi $ are canonically conjugate compact boson fields with unit radii satisfying the algebra~\cite{Ginsparg1988applied}
	\begin{equation}
		[\partial_x \phi (x), \theta (x') ] = 2 \pi i  \delta(x-x'), \label{eq:KacMoody_singlecomponent}
	\end{equation}
	and  $\mathcal{A}, \mathcal{B}, \mathcal{C}$ etc are bosonization prefactors whose precise values are not important. The Luttinger parameter $K$ and the velocity $v$ are related to the Hamiltonian parameters~\cite{Haldane_BetheLL_1981153} (see \cref{app:bosonization}).	The bosonized forms of the spin operators are
	\begin{align}
		S^{\pm}_{ j} &\approx \exp{\left(\pm i \theta(x) \right)}\left(  (-1)^j \A~  +\C \cos \phi(x) + \ldots \right),  \nonumber \\		
		S^z_{ j} &\approx \frac{1}{2 \pi} \partial_x \phi(x) + (-1)^j \B \sin \phi(x) + \ldots \label{eq:Bosonization_single}
	\end{align}

	\Cref{eq:H_bosonize_smallJ} is a compact boson conformal field theory (CFT) with central charge $c=1$ perturbed by vertex operators $\mathcal{U}_p \equiv \cos p\phi$ with scaling dimensions~\cite{Francesco2012conformal,Ginsparg1988applied} 
	\begin{equation}
		\left[ \mathcal{U}_p \right] = \left[ \cos p \phi \right] =  p^2 K.
	\end{equation}
	Note that we have only shown the most relevant operators, with the smallest scaling dimensions in \cref{eq:H_bosonize_smallJ}. The ellipses $\ldots$ represent other operators that are not important for our purposes. The the small-$J$ phase diagram shown in \cref{fig:phasedig} can be qualitatively reproduced from \cref{eq:Bosonization_single} by tracking the relevance~\footnote{Recall that a scaling operator $\mathcal{O}$ is relevant in d+1 dimensions if its scaling dimensions satisfies $[\mathcal{O}] <d+1$. }  of $\mathcal{U}_p$ (see \cref{app:phasedig} for a detailed discussion). Bond dimerization $t$ introduces the vertex operator $\mathcal{U}_1$ and the interaction $S^zS^z$ while $\lambda$ introduces $\mathcal{U}_2$.    
	For now, we note that in the regime when $K >2$, \emph{all} perturbations are irrelevant and correspond to the XY$_0$ gapless phase.

	A different starting point is useful in the large $J$ limit. We now interpret the Hamiltonian in \cref{eq:Hamiltonian_largeJ} as two XXZ spin chains with intra- and inter-rung perturbations. Each leg can be bosonized appropriately to obtain the following two-component compact-boson theory~\cite{Giamarchi}  
	\begin{multline}
		H \approx \frac{v}{2 \pi} \sum_{\alpha = 1,2} \int dx \left(\frac{1}{4K}( \partial_x \phi_\alpha)^2 + K ( \partial_x \theta_\alpha)^2\right) \\- \frac{\lambda}{2 \pi^2} \int dx~   \partial_x \phi_1 \partial_x \phi_2 -4\mathcal{A}^2t \int dx~   ~\cos \left(\theta_1-\theta_2\right) \\
		- \mathcal{B}^2  t \int dx~  \lambda~ \left(\cos \left(\phi_1+\phi_2\right) - \cos \left(\phi_1-\phi_2\right)\right) \\ +2 \mathcal{C}^2 \int dx \cos\left(\theta_1 - \theta_2\right) \cos\left(\phi_1 + \phi_2\right)+ \ldots \label{eq:H_bosonize_largeJ}
	\end{multline}	
	where $\phi_\alpha \cong \phi_\alpha + 2 \pi $ and $\theta_\alpha \cong \theta_\alpha + 2 \pi $ are  compact boson fields satisfying
	\begin{equation}
		[\partial_x \phi_\alpha (x), \theta_\beta (x') ] = 2 \pi i \delta_{\alpha\beta} \delta(x-x'),
		\label{eq:KacMoody_2component}
	\end{equation}
	$\mathcal{A}$ and $\mathcal{B}$ are again unimportant bosonization prefactors and we have only shown the most important operators. The bosonized forms of the spin operators are
	\begin{align}
		S^{\pm}_{\alpha j} &\approx  \exp{\left(\pm i \theta_\alpha(x) \right)}\left(  (-1)^j \A~  + \C \cos \phi_\alpha(x) + \ldots \right) \nonumber\\
		S^z_{\alpha j} &\approx \frac{1}{2 \pi} \partial_x \phi_\alpha + (-1)^j \mathcal{B} \sin \phi_\alpha + \ldots \label{eq:Bosonization_ladder}
	\end{align}
	The above theory represents a $c=2$ CFT with perturbations. We have only retained primary~\cite{Ginsparg1988applied,Francesco2012conformal} scaling operators in \cref{eq:Bosonization_ladder}. This is sufficient to determine the structure of the phases and transitions, which is our focus. However, it is known~\cite{FabianIncommensurate_PhysRevLett.81.910} that descendant operators must be considered to understand certain \emph{incommensurability} aspects of correlations. The large-$J$ phase diagram can be qualitatively reproduced using \cref{eq:Bosonization_ladder} by carefully tracking the relevance of the operators $\mathcal{V}_\pm \equiv \cos \left( \phi_1 \pm \phi_2\right)$, $\mathcal{W}_- \equiv \cos \left(\theta_1 - \theta_2\right)$ and $ \cW_- \cV_+\equiv \cos\left(\theta_1 - \theta_2\right) \left(\phi_1 + \phi_2\right)$ (details of this can be found in \cref{app:phasedig}). Here, we again focus only on how the four gapless phases can emerge. An important fact is that the scaling dimensions of the various operators listed above are not all independent. In particular, we have $[\cV_-]  = ([\cW_-])^{-1}$. Therefore, it is impossible for both $\cV_-$ and $\cW_-$ to be irrelevant at the same time, and for any $t \neq 0$, the $c=2$ theory is unstable and flows to a Luttinger liquid phase with $c=1$ or a gapped phase ~\cite{StrongMillis_Spinladder_PhysRevLett.69.2419,OrignacGiamarchi_SpinLadder_PhysRevB.57.5812,Giamarchi} as seen in \cref{fig:phasedig}. The first, which is of our main interest, occurs when all other operators, especially $\cV_+$ are irrelevant. The nature of the resulting gapless phase depends on: (i) which among $[\cV_-]$ and $[\cW_-]$ has the smaller scaling dimensions at $t=0$. This dominates long-distance physics for $t \neq 0$ resulting in the pinning of either $\moy{\phi_1 - \phi_2}$ or $\moy{\theta_1 - \theta_2}$ and (ii) the value to which $\moy{\phi_1 - \phi_2} = 0/ \pi$ or $\moy{\theta_1 - \theta_2} = 0/\pi$ is pinned, depending on the sign of $t$ and $\lambda$. The four possibilities result in the four gapless phases shown in the large-$J$ phase diagram of \cref{fig:phasedig} as follows.
	\begin{enumerate}
		\item XY$_1$:   $[\cV_-] > [\cW_-]$, $\moy{\theta_1 - \theta_2} = \pi$
		\item XY$_1^*$:   $[\cV_-] > [\cW_-]$, $\moy{\theta_1 - \theta_2} = 0$
		\item XY$_2$:   $[\cV_-] < [\cW_-]$, $\moy{\phi_1 - \phi_2} = 0$
		\item XY$_2^*$:   $[\cV_-] < [\cW_-]$, $\moy{\phi_1 - \phi_2} = \pi$ .   
	\end{enumerate}
	There are two critical scenarios which we now discuss: when  $[\cV_-]=[\cW_-]$, the theory flows to a $c= \frac{3}{2}$ theory corresponding to a compact boson with Ising CFT ~\cite{Schulz_Higherspinbosinization_PhysRevB.34.6372}. For $[\cV_-] \neq [\cW_-]$  $t=0$ corresponds to a phase transition described by the parent $c=2$ two-component compact boson theory when the pinned value of the appropriate fields changes. At this stage, let us point out elements of the discussion above that already exist in the literature. The competition between $[\cV_-]$ and $[\cW_-]$ leading to different phases was discussed in refs.~\cite{Schulz_Higherspinbosinization_PhysRevB.34.6372,OrignacGiamarchi_SpinLadder_PhysRevB.57.5812,Giamarchi}. The importance of the precise values to which the fields are pinned was appreciated relatively recently~~\cite{KeselmanBerg_TLL_PhysRevB.91.235309,ThorngrenVishwanathVerresen_IntrinsicallyGapless_PhysRevB.104.075132,VerresenThorngrenJonesPollmann_SymmetryEnrichedCriticality_PhysRevX.11.041059} where it was shown that $\moy{\phi_1 - \phi_2} =\pi$ produces a gapless phase with edge modes. 
	
	However, we must be careful in using these pieces of information to conclude that we have distinct phases of matter. It was recently pointed out that this kind of distinction can disappear suddenly ~\cite{BiSenthil_Multiversality_PhysRevX.9.021034,AP2022multiversality,Yuchi2023terminable}. A more robust characterization arises out of symmetry considerations which we now turn to. We do this in two complementary ways. First, we establish the fate of the microscopic symmetries shown in \cref{tab:symmetries,tab:symmetries_bosonization} in the deep IR for each of the gapless phases. The effective theory for all of them is that of a single compact boson. We show that in each of the five phases, the microscopic symmetries act in inequivalent ways that cannot be deformed into each other. Second, we study how appropriately charged local and nonlocal operators behave in the different phases and show a complete list of operators with distinct charges that can serve as order parameters to distinguish the different gapless phases. Our work therefore characterizes a rather subtle interplay of symmetries and topology leading to the emergence of novel gapless phases.
	
	\subsection{Multiversality along the $t=0$ surface}
	An interesting feature of the phase diagrams shown in \cref{fig:phasedig} is the nature of transition separating the Haldane and Trivial phases along parts of the $t=0$ surface. In the small-$J$ limit, we see from \cref{eq:H_bosonize_smallJ} that the critical theory corresponds to a compact boson CFT with central charge $c=1$. In the large-$J$ diagram, the situation is different. Consider the effective theory in \cref{eq:H_bosonize_largeJ} and set $t=0$. This is a $c=2$ CFT with perturbations and describes various transitions and phases along the $t=0$ surface. In particular, the transition between XY$_1$ and XY$_1^*$ corresponds to the $c=2$ theory when all perturbations are irrelevant or tuned away. As we move along this surface, the operator $\cW_- \cV_+$ becomes relevant and gives us a gapped theory with two ground states which precisely correspond to those of the Haldane and trivial phases and therefore represent a first-order transition between them (see \cref{app:phasedig} for a detailed discussion). Now consider the transition between XY$_1$ and the trivial phase. This is driven by the operator $\cV_+$ becoming relevant. Since $\cV_+$ has a smaller scaling dimension than $\cW_- \cV_+$, the XX$_1$ -to- trivial $c=1$ critical line strikes the $t=0$ line well before the first-order transition sets in. The same is true for the XY$_1^*$ -to- Haldane transition. Consequently, we expect that a segment of the transition (close to the gapless phases) between the Haldane and the trivial phase will also be described by the $c=2$ CFT before becoming first-order as shown in \cref{fig:phasedig}. This situation is unusual because it is a different universality class (with a different central charge) compared to the small-$J$ transition between the same phases. Furthermore, in both cases, the transitions are reached by tuning only a single parameter, without additional fine-tuning. 
	
	The presence of multiple stable universality classes that separate the same two phases has been termed `multiversality' ~\cite{BiSenthil_Multiversality_PhysRevX.9.021034,AP2022multiversality}. Although there are no physical reasons forbidding multiversality, models that exhibit it are surprisingly rare. We see that the spin ladder model considered in this work exhibits the phenomenon under relatively generic conditions and symmetries (compare this to the example in Ref.\cite{AP2022multiversality} where multiversality was observed under more restrictive symmetries and destroyed when symmetries were reduced).

	\subsection{Distinguishing gapless phases through effective symmetries}
	\label{sec:symmetry_distinction}
	\begin{table}[!ht]
		\begin{center}
			\begin{tabular}{|c|c|c|}
				\hline
				& \textbf{Small $J$} & \textbf{Large $J$} \\				
				\hline
				\hline
				$U(1)$ &  $\begin{array}{l}
					\theta(x) \mapsto \theta(x) + \chi\\
					\phi(x) \mapsto \phi(x)
				\end{array} $ & $\begin{array}{l}
					\theta_\alpha(x) \mapsto \theta_\alpha(x) + \chi  \\
					\phi_\alpha(x) \mapsto \phi_\alpha(x)
				\end{array} $ \\			
				\hline
				$\ztwo^R$ & $\begin{array}{l}
					\theta(x) \mapsto -\theta(x) \\
					\phi(x) \mapsto -\phi(x)
				\end{array}$ &  $\begin{array}{l}
					\theta_\alpha(x) \mapsto -\theta_\alpha(x)\\
					\phi_\alpha(x) \mapsto -\phi_\alpha(x)
				\end{array}$   \\
				\hline
				$\ztwo^P$  & $\begin{array}{l}
					\theta(x) \mapsto \theta(-x) + \pi\\
					\phi(x) \mapsto -\phi(-x)
				\end{array}$   & $\begin{array}{l}
					\theta_\alpha(x) \mapsto \tau^x_{\alpha \beta} \theta_\beta(-x)  \\
					\phi_\alpha(x) \mapsto \pi - \tau^x_{\alpha \beta} \phi_\beta(-x)  
				\end{array}$ \\
				\hline
				$\bZ$ & $\begin{array}{l}
					\theta(x) \mapsto \theta(x)\\
					\phi(x) \mapsto \phi(x)
				\end{array}$ & $\begin{array}{l}
					\theta_\alpha(x) \mapsto  \theta_\alpha(x) + \pi  \\
					\phi_\alpha(x) \mapsto   \phi_\alpha(x) + \pi 
				\end{array}$ \\			
				\hline												    
			\end{tabular}
		\end{center}
		\caption{Representation of symmetries (see \cref{tab:symmetries}) on the boson fields applicable in the small $J$ and large $J$ limits of the Hamiltonian shown in \cref{eq:Hamiltonian_smallJ,eq:Hamiltonian_largeJ}. $\tau^x$ is the Pauli X matrix. \label{tab:symmetries_bosonization} }
	\end{table}
	We begin this subsection by listing the action of symmetries listed in  \cref{tab:symmetries} on the compact boson fields in both the small and large $J$ versions~\cite{Yuchi_Parity_PhysRevB.104.195126,Giamarchi}. This is shown in \cref{tab:symmetries_bosonization} and is obtained by comparing the action on the lattice operators shown in \cref{tab:symmetries} with the dictionary shown in \cref{eq:Bosonization_single,eq:Bosonization_ladder} (see \cref{app:bosonization} for more details). We want to understand the fate of these symmetries in various gapless phases. The long wavelength physics of each of these gapless phases is identical and corresponds to that of a single compact boson  with a Hamiltonian of the form
	\begin{equation}
		H  =  \frac{v_{\text{eff}}}{2 \pi} \int dx \left[\frac{1}{4K_{\text{eff}}} \left(\partial_x \phi\right)^2 + K_{\text{eff}} \left(\partial_x \theta\right)^2\right]. \label{eq:Effective_compactboson}
	\end{equation}
	How do the microscopic symmetries act on the long wavelength effective fields? Observe that the compact boson theory itself has various symmetries such as 
	\begin{align}
		U(1)_\theta &: \theta \mapsto \theta+ \chi,~\ztwo^\theta: \theta \mapsto -\theta, \nonumber\\
		U(1)_\phi &: \phi \mapsto \phi+ \xi,~\ztwo^\phi: \phi \mapsto -\phi, \label{eq:Symmetries_CompactBoson}
	\end{align}
	which form the group $G_{IR} \cong O(2)_\theta \times O(2)_\phi$~\footnote{The two $U(1)$ symmetries also contains a well known mixed anomaly~\cite{Fradkin2021quantum}. The full collection of symmetries of the compact boson CFT is larger and contains the group of conformal transformations as well as other generators that may not even form a group structure~\cite{NonInvertible_Thorngren2021fusion,NonInvertible_PhysRevD.105.125016}. For our purposes, the ones listed in \cref{eq:Symmetries_CompactBoson} that form $G_{IR}$ are sufficient.}.   
	
	The action of symmetries can also be studied in the spectrum of local scaling operators $\cX_{m,n} \equiv \exp\left( i \left(m \theta + n \phi \right)\right)$ with scaling dimensions $\left[\cX_{m,n} \right] = m^2K_{\text{eff}} + \frac{n^2}{4K_{\text{eff}}}$ where $m$ and $n$ are integers. These read as follows
	\begin{align}
		U(1)_\theta &: \cX_{m,n}  \mapsto e^{im\chi} \cX_{m,n},~~\ztwo^\theta:\cX_{m,n}  \mapsto \cX_{-m,n}, \nonumber\\
		U(1)_\phi &: \cX_{m,n}  \mapsto e^{i n\xi} \cX_{m,n},~~~\ztwo^\phi:\cX_{m,n}  \mapsto \cX_{m,-n}. \label{eq:Symmetries_CompactBoson_operators}
	\end{align}
	
	The question we are interested in is how the microscopic symmetries of spins $G_{UV}$ listed in \cref{tab:symmetries} attach themselves to those of  compact boson degrees of freedom, $G_{IR}$. In other words, we we are interested in the \emph{homomorphisms} $G_{UV} \rightarrow G_{IR}$. Distinct homomorphisms will lead to inequivalent symmetry enriched Luttinger liquids that cannot be adiabatically connected. We will determine this for each phase one-by-one to confirm this. 
	
	\subsubsection{Effective symmetries of XY$_0$}
	
	\begin{table}[!ht]
		\begin{center}
			\begin{tabular}{|c|c|c|c|}
				\hline
				$U(1)$  & $\ztwo^R$ & $\bZ$ & $\ztwo^P$ \\				
				\hline
				\hline				
				$  \begin{pmatrix}
					\theta(x) + \chi \\
					\phi(x)
				\end{pmatrix}$
				& $  \begin{pmatrix}
					-\theta(x) \\
					-\phi(x)
				\end{pmatrix}$ & 
				$  \begin{pmatrix}
					\theta(x)  \\
					\phi(x)
				\end{pmatrix}$ &  $  \begin{pmatrix}
					\pi +     \theta(-x)  \\
					-\phi(-x) 
				\end{pmatrix}$  \\
				\hline
				$e^{im \chi}~ \cX_{m,n}(x)$
				& $\cX_{-m,-n}(x)$ & $\cX_{m,n}(x)$
				& $e^{i \pi m} ~\cX_{m,-n}(-x)$  \\
				\hline							
			\end{tabular}
		\end{center}
		\caption{Effective action of symmetries in the XY$_0$ phase. \label{tab:symmetries_bosonization_XY0}}
	\end{table}
	
	Let us begin with the gapless phase seen in the small-$J$ limit, XY$_0$. The effective action of symmetries were already obtained using the bosonization formulas as listed in \cref{tab:symmetries_bosonization}. This can also be used to determine the action on various scaling operators as shown in \cref{tab:symmetries_bosonization_XY0}. We see that the microscopic $U(1)$ attaches itself to $U(1)_\theta$, $\ztwo^R$ to a simultaneous action of $\ztwo^\theta$ and $\ztwo^\phi$, and $\ztwo^P$ to a composite action of simultaneous $\pi$ rotation of $U(1)_\theta$ and  $\ztwo^\phi$ action while UV lattice translations $\bZ$ have no effect in the IR.

	\subsubsection{Effective symmetries of XY$_1$ and XY$_1^*$}	
	\begin{table}[!ht]
		\begin{center}
			\begin{tabular}{|c|c|c|c|}
				\hline
				$U(1)$  & $\ztwo^R$ & $\bZ$ & $\ztwo^P$ \\				
				\hline
				\multicolumn{4}{|c|}{XY$_1$}\\
				\hline				
				$  \begin{pmatrix}
					\theta(x) + \chi \\
					\phi(x)
				\end{pmatrix}$
				& $  \begin{pmatrix}
					-\theta(x) \\
					-\phi(x)
				\end{pmatrix}$ & 
				$  \begin{pmatrix}
					\theta(x) + \pi \\
					\phi(x)
				\end{pmatrix}$ &  $  \begin{pmatrix}
					\theta(-x) + \pi \\
					-\phi(-x)
				\end{pmatrix}$ \\	
				\hline
				$e^{im \chi}~ \cX_{m,n}(x)$
				& $\cX_{-m,-n}(x)$ & $e^{i \pi m} ~\cX_{m,n}(x)$
				& $e^{i \pi m} ~\cX_{m,-n}(-x)$ \\
				\hline
				\multicolumn{4}{|c|}{XY$_1^*$}\\
				\hline				
				$  \begin{pmatrix}
					\theta(x) + \chi \\
					\phi(x)
				\end{pmatrix}$
				& $  \begin{pmatrix}
					-\theta(x) \\
					-\phi(x)
				\end{pmatrix}$ & 
				$  \begin{pmatrix}
					\theta(x) + \pi \\
					\phi(x)
				\end{pmatrix}$ &  $  \begin{pmatrix}
					\theta(-x)  \\
					-\phi(-x)
				\end{pmatrix}$ \\	
				\hline
				$e^{im \chi}~ \cX_{m,n}(x)$
				& $\cX_{-m,-n}(x)$ & $e^{i \pi m} ~\cX_{m,n}(x)$
				& $ \cX_{m,-n}(-x)$ \\
				\hline
			\end{tabular}
		\end{center}
		\caption{Effective action of symmetries in the XY$_1$ $\left(\moy{\vartheta} = \pi \right),$ and XY$_1^*$ $\left(\moy{\vartheta} = 0 \right)~$ phases. \label{tab:symmetries_bosonization_XY1}}
	\end{table}
	We now consider the gapless phases in the large-$J$ limit obtained when $\cW_- \equiv \cos \left(\theta_1 - \theta_2\right)$	dominates at long distances pinning $\vartheta \equiv \theta_1 - \theta_2$.  To determine the nature of the resulting compact boson CFT the system flows to, we perform the following $SL(2,\bZ)$ transformation which preserves the unit compactification radius of the fields as well as the canonical commutation relation~\cref{eq:KacMoody_2component}
	\begin{equation}
		\begin{pmatrix}
			\vartheta \\
			\theta
		\end{pmatrix} \equiv \begin{pmatrix}
			\theta_1 - \theta_2 \\
			\theta_2
		\end{pmatrix},~~ \begin{pmatrix}
			\varphi \\
			\phi
		\end{pmatrix} \equiv \begin{pmatrix}
			\phi_1  \\
			\phi_1 + \phi_2
		\end{pmatrix} \label{eq:SL2Z XY1}.
	\end{equation}
	When $\vartheta \equiv \theta_1 - \theta_2$ is pinned, its conjugate $\varphi$ is disordered and we obtain physics at  long distances  by setting
	\begin{equation}
		e^{i m \vartheta} \approx \moy{e^{i m \vartheta}} \approx 	e^{i m \moy{\vartheta}} \text{ and }	e^{i n \varphi} \approx \moy{e^{i n \varphi}} \approx 0. \label{eq:XY1_pinning_effecive}
	\end{equation}
	The effective theory is simply that of the unpinned canonically conjugate pair of fields, $\theta$ = $\theta_2$ and $\phi$ $=\phi_1+\phi_2$  with a Hamiltonian of the form shown in \cref{eq:Effective_compactboson}. 
	
	Using \cref{eq:XY1_pinning_effecive} and the action of the symmetries on the compact bosons obtained from bosonization at large $J$ shown in \cref{tab:symmetries_bosonization}, we can read off the effective symmetry action on the $\theta$ and $\phi$ as well as on the spectrum of scaling operators as shown in \cref{tab:symmetries_bosonization_XY1}. First, compare these with \cref{tab:symmetries_bosonization_XY0}. We see that the actions of $U(1)$ and $\bZ_2^R$ are identical in all three phases. However, the action of $\bZ$ distinguishes XY$_0$ from the other two. Finally, the symmetry action of $\ztwo^P$ depends on the value of $\moy{\vartheta}$ and distinguishes between XY$_1$ ($\moy{\vartheta} = \pi$) and XY$_1^*$ ($\moy{\vartheta} = 0$). Observe that both \emph{electric} scaling operators (i.e., those carrying $U(1)$ charge) with the smallest scaling dimensions, $\cos \theta$ and $\sin \theta$ are pseudo-scalars for XY$_1$ and scalars in XY$_1^*$ respectively. Thus, we have succeeded in establishing that XY$_0$, XY$_1$ and XY$_1^*$ are distinct from each other.

	\subsubsection{Effective symmetries of XY$_2$ and XY$_2^*$}
	\begin{table}[!ht]
		\begin{center}
			\begin{tabular}{|c|c|c|c|}
				\hline
				$U(1)$  & $\ztwo^R$ & $\bZ$ & $\ztwo^P$ \\				
				\hline
				\multicolumn{4}{|c|}{XY$_2$}\\
				\hline			
				$  \begin{pmatrix}
					\theta(x) + 2\chi \\
					\phi(x)
				\end{pmatrix}$
				& $  \begin{pmatrix}
					-\theta(x) \\
					-\phi(x)
				\end{pmatrix}$ & 
				$  \begin{pmatrix}
					\theta(x)  \\
					\phi(x) + \pi
				\end{pmatrix}$ &  $  \begin{pmatrix}
					\theta(-x)  \\
					\pi-\phi(-x)
				\end{pmatrix}$  \\
				\hline
				$e^{2 i m \chi}~ \cX_{m,n}(x)$
				& $\cX_{-m,-n}(x)$ & $e^{i \pi n} ~\cX_{m,n}(x)$
				& $e^{i \pi n}~\cX_{m,-n}(-x)$  \\
				\hline
				\multicolumn{4}{|c|}{XY$_2^*$}\\
				\hline			
				$  \begin{pmatrix}
					\theta(x) + 2\chi \\
					\phi(x)
				\end{pmatrix}$
				& $  \begin{pmatrix}
					-\theta(x) \\
					-\phi(x)
				\end{pmatrix}$ & 
				$  \begin{pmatrix}
					\theta(x)  \\
					\phi(x) + \pi
				\end{pmatrix}$ &  $  \begin{pmatrix}
					\theta(-x)  \\
					-\phi(-x)
				\end{pmatrix}$  \\
				\hline
				$e^{2 i m \chi}~ \cX_{m,n}(x)$
				& $\cX_{-m,-n}(x)$ & $e^{i \pi n} ~\cX_{m,n}(x)$
				& $ \cX_{m,-n}(-x)$  \\
				\hline
			\end{tabular}
		\end{center}
		\caption{Effective action of symmetries in the XY$_2$ $\left(\moy{\vartheta} = 0 \right),$ and XY$_2^*$ $\left(\moy{\vartheta} = \pi \right)~$ phases. \label{tab:symmetries_bosonization_XY2}}
	\end{table}
	Finally, we turn to the large-$J$ gapless phases obtained when $\cV_-$ dominates at long distances and pins $\phi_1 - \phi_2$. To get the effective symmetries of the resulting compact boson CFT the system flows to, we perform a different $SL(2,\bZ)$ transformation from~\cref{eq:SL2Z XY1}
	
	\begin{equation}
		\begin{pmatrix}
			\vartheta \\
			\theta
		\end{pmatrix} \equiv \begin{pmatrix}
			\theta_1  \\
			\theta_1+ \theta_2
		\end{pmatrix},~~ \begin{pmatrix}
			\varphi \\
			\phi
		\end{pmatrix} \equiv \begin{pmatrix}
			\phi_1 - \phi_2  \\
			\phi_2 
		\end{pmatrix} \label{eq:SL2Z XY2}.
	\end{equation}
	When $\varphi \equiv \phi_1 - \phi_2$ is pinned, its conjugate $\vartheta$ is disordered and we obtain the long distance physics by setting
	\begin{equation}
		e^{i m \vartheta} \approx \moy{e^{i m \vartheta}} \approx 0, \text{ and }	e^{i n \varphi} \approx \moy{e^{i n \varphi}} \approx e^{i n \moy{\varphi}} \label{eq:XY2_pinning_effecive}.
	\end{equation}
	The effective theory is simply that of the unpinned fields $\theta$ and $\phi$ with an effective Hamiltonian of the form shown in \cref{eq:Effective_compactboson}. 
	
	Using \cref{eq:XY2_pinning_effecive} the symmetry action on the effective low-energy fields and on the spectrum of operators can be read off from \cref{tab:symmetries_bosonization} and is summarized in \cref{tab:symmetries_bosonization_XY2}. The most striking feature is that the $\theta$ field is a charge 2 operator for the $U(1)$ symmetry. Consequently, the smallest $U(1)$ charge carried by the spectrum of scaling operators is $2$. This immediately shows that XY$_2$ and XY$_2^*$ are distinct from XY$_0$, XY$_1$ and XY$_1^*$. Let us now focus on the effective action of $\ztwo^P$ which seemingly depends on the value of $\moy{\varphi}$ and distinguishes XY$_2$ from XY$_2^*$. This is \emph{not} true-- the symmetry actions are merely related by a change of basis. However, keeping track of other symmetry charges exposes the distinction. Consider magnetic scaling operators (those without any $U(1)$ charge) with the smallest scaling dimensions, $\cos \phi$ and $\sin \phi$. We see that in XY$_2$, the operator with $\ztwo^R$ charge ($\sin \phi$) transforms as a scalar under $\ztwo^P$ whereas the operator without $\ztwo^R$ charge ($\cos \phi$) transforms as a pseudoscalar. This situation is precisely reversed for XY$_2^*$ where the $\ztwo^R$ charged operator is a $\ztwo^P$ pseudoscalar, whereas the $\ztwo^R$ neutral operator is a $\ztwo^P$ scalar. This completes the proof that the five gapless phases are distinct. 
	
	\subsubsection{Explicit symmetry breaking}
	Observe that all four microscopic symmetries were important in establishing these distinctions. Explicitly breaking certain symmetries eliminates the distinction between certain phases and opens a potential path to connect them without phase transitions or intermediate phases. Let us look at a few instances. 
	\begin{enumerate}
		\item If we break $\ztwo^R$, the distinction between XY$_2$ and XY$_2^*$ is eliminated and reduces five phases to four: XY$_0$,  XY$_1$, XY$_1^*$ and (XY$_2=$ XY$_2^*$). 
		\item If we break $\ztwo^P$, the distinction between $XY_1$ and XY$_1^*$ is eliminated, as well as between $XY_2$ and XY$_2^*$ and reduces the five phases to three: XY$_0$,  ( XY$_1=$ XY$_1^*$) and (XY$_2=$ XY$_2^*$). 
		\item If we only preserve $U(1)$ and break all other symmetries, the five phases reduce to two: (XY$_0=$ XY$_1=$ XY$_1^*$) and (XY$_2=$ XY$_2^*$).  
	\end{enumerate}

	\subsection{Local and non-local observables}
	\begin{table}[!ht]
		\begin{center}
			\begin{tabular}{|c||c|c|c||c|c|}
				\hline
				& XY$_0$ & XY$_1$ & XY$_1^*$ & XY$_2$ & XY$_2^*$ \\
				\hline
				\hline
				$\moy{O^{+}_s(x)~O^{-}_s(y)}$ & alg & exp & alg & exp & exp \\				
				\hline
				$\moy{O^{+}_a(x)~O^{-}_a(y)}$ & alg & alg  & exp & exp& exp \\
				\hline
				\hline
				$\moy{C(x,y)}$ & 0 &  0& 0 & 0 & $\neq$ 0\\
				\hline						
			\end{tabular}
		\end{center}
		\caption{Local and nonlocal order observables for large $|x-y|$. We denote algebraic and exponential decay by `alg' and `exp'.  \label{tab:orderparameters}}
	\end{table}	
	
	We now turn to how we can physically characterize various gapless phases using local and non-local observables. We will use the previously determined effective symmetry action listed in \cref{tab:symmetries_bosonization_XY0,tab:symmetries_bosonization_XY1,tab:symmetries_bosonization_XY2} to guide us in this. We will focus on two local operators $O^{s \pm}_x,~O^{a \pm}_x$ and a non-local string operator $C_{x,y}$ defined as follows (in both the small $J$ ($J_<$) and large $J$ ($J_>$) representations)
	
	\begin{align}
		O^{ \pm}_s(j) &\equiv \begin{cases}
			S^\pm_{2j-1} + S^\pm_{2j} &J_<\\
			~~S^\pm_{1,j}~ + S^\pm_{2,j} &J_>
		\end{cases},\label{eq:Os}\\ O^{ \pm}_a(j) &\equiv \begin{cases}
			S^\pm_{2j-1} - S^\pm_{2j} &J_<\\
			~~S^\pm_{1,j} ~- S^\pm_{2,j} &J_>
		\end{cases}, \label{eq:Oa}\\
		C{(j,k)} &\equiv \begin{cases}
			\sigma^z_{2j-1} ~\left(\prod_{l=2j}^{2k}\sigma^z_l\right)~ \sigma^z_{2k+1} &J_<\\
			\sigma^z_{2,j} ~\left(\prod_{l=j+1}^{k-1} \sigma^z_{1,j} \sigma^z_{2,j}\right)~ \sigma^z_{1,k} &J_>
		\end{cases}.\label{eq:C}
	\end{align}
	
	The nature of two-point correlation functions of the local operators and the expectation value of the string operator are summarized in \cref{tab:orderparameters} and completely characterize the phases.   We see in \cref{tab:orderparameters} that local operators uniquely identify the XY$_0$, XY$_1$ and XY$_2$ phases but cannot distinguish between the XY$_2$ and XY$_2^*$ phases, which the nonlocal operator can. In this section, we will see how this behavior can be determined using the bosonization formulas as well as using the effective symmetry action shown in \cref{tab:symmetries_bosonization_XY0,tab:symmetries_bosonization_XY1,tab:symmetries_bosonization_XY2}. These predictions will also be confirmed numerically in \cref{sec:Numerics}.  
	
	\subsubsection{Local operator behavior from bosonization}
	Let us begin with XY$_0$ where, using \cref{eq:Bosonization_single} the local operators can be bosonized as 
	\begin{equation}
		O^{ \pm}_s(x) \sim e^{\pm i \theta(x)} \cos \phi(x),~ O^{ \pm}_a(x) \sim e^{\pm i \theta(x)} .\label{eq:local_XY0}
	\end{equation}
	In \cref{eq:local_XY0}, we have suppressed the bosonization prefactors and retained only the most relevant scaling operators the lattice operators have an overlap with. Clearly, the two point functions of $O^\pm_s$ and $O^\pm_a$ are expected to have algebraic decay governed by the parameters of the effective compact-boson CFT that describe the phase at long distances. Recall that for a CFT, the correlation functions of the scaling operators $\cX(x)$ with scaling dimensions $\Delta_\cX$ scale as 
	\begin{equation}
		\moy{\cX(x) \cX^\dagger(y)} \sim |x-y|^{2 \Delta_\cX}.
	\end{equation}
	Thus, at long distances $|x-y|$, we expect
	\begin{align}
		|\moy{O^+_s(x) O^-_s(y)}| &\sim |x-y|^{- \left(2K  + \frac{1}{2K}\right)},\nonumber\\
		|\moy{O^+_a(x) O^-_a(y)}| &\sim |x-y|^{- \frac{1}{2K} }.\label{eq:local_correlations_XY0}
	\end{align}
	Let us now consider the large-$J$ phases where, using \cref{eq:Bosonization_ladder}, we get
	\begin{align}
		O^{ \pm}_s(x) &\sim\left( e^{\pm i \theta_1(x)} +  e^{\pm i \theta_2(x)}\right), \nonumber\\
		O^{ \pm}_a(x) &\sim\left( e^{\pm i \theta_1(x)} -  e^{\pm i \theta_2(x)}\right). \label{eq:local_largeJ}
	\end{align}
	We have again suppressed bosonization prefactors and retained only the most relevant scaling operators. When we have the full $c=2$ theory along the $t=0$ line shown in \cref{fig:phasedig_largeJ_bosonization} we see that both local operators have algebraic correlations. However, for $t \neq 0$, when $\cV_-$ or $\cW_-$ are relevant resulting in the different gapless phases, this changes. Consider the case where $\cW_-$ is the most relevant operator and pins $\vartheta \equiv \theta_1 - \theta_2$. We can use the $SL(2,\bZ)$ transformation shown in \cref{eq:SL2Z XY1} and \cref{eq:XY1_pinning_effecive} to obtain the following.
	
	\begin{align}
		O^{ \pm}_s(x) &\sim\left( e^{\pm i \theta_1} +  e^{\pm i \theta_2}\right) \approx e^{\pm i \theta} \left( 1+ e^{\pm i \moy{\vartheta}}\right) \nonumber \\ &\approx \begin{cases}
			0 &\text{ for }\moy{\vartheta} = \pi~ (\text{XY}_1) \\
			e^{\pm i \theta}&  \text{ for }\moy{\vartheta} = 0 ~(\text{XY}_1^*) \\
		\end{cases}, \nonumber\\
		O^{ \pm}_a(x) &\sim\left( e^{\pm i \theta_1} -  e^{\pm i \theta_2}\right) \approx e^{\pm i \theta} \left( 1-  e^{\pm i \moy{\vartheta}}\right) \nonumber \\ &\approx \begin{cases}
			e^{\pm i \theta}&  \text{ for }\moy{\vartheta} = \pi ~(\text{XY}_1) \\
			0 &\text{ for }\moy{\vartheta} = 0 ~(\text{XY}_1^*) \\
		\end{cases}.\label{eq:local_XY1}
	\end{align}
	We see that for each case $\moy{\vartheta} = \pi/ 0$, only one of the two operators $O^{ \pm}_s(x)$ / $O^{ \pm}_s(x)$ has vanishing overlap with scaling operators and has algebraic correlations whereas the other has exponential correlations:
	\begin{align}
		|\moy{O^+_s(x) O^-_s(y)}| &\sim \begin{cases}
			e^{-\frac{|x-y|}{\xi}} &(\text{XY}_1)\\
			|x-y|^{-\frac{1}{2K_{\text{eff}}}} &(\text{XY}_1^*)
		\end{cases} \nonumber, \\
		|\moy{O^+_a(x) O^-_a(y)}| &\sim \begin{cases}
			|x-y|^{-\frac{1}{2K_{\text{eff}}}}&(\text{XY}_1)\\
			e^{-\frac{|x-y|}{\xi}} &(\text{XY}_1^*)
		\end{cases}. \label{eq:local_correlations_XY1}
	\end{align}
	$K_{\text{eff}}$ is the effective Luttinger parameter shown in \cref{eq:Effective_compactboson} that characterizes the effective compact boson CFT at long distances. We may wonder if the calculations above are modified if we include the corrections to the bosonization formulas represented by ellipses in \cref{eq:Bosonization_ladder}. It turns out that the answer is no and can be verified by including all higher terms explicitly. A more powerful way is using symmetries, as will be discussed in the next subsection.
	
	We now turn to the phases obtained when $\cV_-$ is dominant and pins $\varphi \equiv \phi_1 - \phi_2$. Using the $SL(2,\bZ)$ transformation shown in \cref{eq:SL2Z XY2} as well as \cref{eq:XY2_pinning_effecive}, we get
	\begin{align}
		O^{ \pm}_s &\sim e^{\pm i \theta_1} +  e^{\pm i \theta_2} \approx  \left(  \moy{e^{\pm i \vartheta}}+ e^{\pm i \theta}~\moy{e^{\mp i \vartheta}}  \right) \approx 0 ,\nonumber\\
		O^{ \pm}_a &\sim e^{\pm i \theta_1} -  e^{\pm i \theta_2} \approx  \left(  \moy{e^{\pm i \vartheta}}- e^{\pm i \theta}~\moy{e^{\mp i \vartheta}}  \right) \approx 0.\label{eq:local_XY2}
	\end{align}
	We see that both $O^{ \pm}_s(x)$ and $O^{ \pm}_a(x)$ have no overlap with any scaling functions and therefore their correlation functions decay exponentially
	\begin{equation}
		|\moy{O^+_s(x) O^-_s(y)}| \sim |\moy{O^+_a(x) O^-_a(y)}| \sim e^{-\frac{|x-y|}{\xi}}.\label{eq:local_correlations_XY2}
	\end{equation}
	We can check that this behaviour does not change even when corrections represented by ellipses in \cref{eq:Bosonization_ladder} are included. This can also be justified using symmetry arguments as we will now see.

	\subsubsection{Local operator behaviour from effective symmetry action}
	
	\begin{table}[!ht]
		\begin{center}
			\begin{tabular}{|c||c|c|c|}
				\hline 
				& $O^{\pm}_s(x) \mapsto$ & $O^{\pm}_a(x) \mapsto$ & $C(x,y) \mapsto$ \\
				\hline
				\hline
				$U(1)$ & $ e^{ \pm i \chi} ~O^{\pm}_s(x)$ & $ e^{ \pm i\chi}~ O^{\pm}_a(x)$ & $ C(x,y)$ \\				
				\hline			
				$\ztwo^R$ & $ O^{\mp}_s(x)$ & $  O^{\mp}_a(x)$ & $C(x,y)$ \\
				\hline			
				$\ztwo^P$ & $ O^{\pm}_s(-x)$ & $ - O^{\pm}_a(-x)$ & $C(-y,-x)$ \\
				\hline			
			\end{tabular}
		\end{center}
		\caption{Symmetries transformations of local and non-local operators defined in \cref{eq:Oa,eq:Os,eq:C}. \label{tab:symmetries_orderparameters}}
	\end{table}	
	
	The correlations of local operators shown in \cref{tab:orderparameters} can also be understood directly by using symmetries. Let us begin by noting down the transformations of the local operators under the $U(1)$, $\ztwo^R$ and $\ztwo^P$ symmetries. This is shown in \cref{tab:symmetries_orderparameters}. At this point, let us remark that all local operators are charged under various internal symmetries. The non-local operator, on the other hand, although is neutral overall, has end points that carry  $\ztwo^R$ charge. This is important to establish the topological nature of phases and will be discussed in \cref{sec:Bosonization2}. Now, we can ask if the transformations shown in \cref{tab:symmetries_orderparameters} can be obtained in each of the five gapless phases using combinations of the scaling operators $\cX_{mn}(x)$ whose transformations are shown in \cref{tab:symmetries_bosonization_XY0,tab:symmetries_bosonization_XY1,tab:symmetries_bosonization_XY2}. If the answer is yes, it will mean that the local operator will have algebraic correlations at long distances with the exponent determined by the scaling dimensions of the said operators with smallest scaling dimensions. If not, then the operators will have exponentially decaying correlations
	
	\emph{XY$_0$}: Comparing the $U(1)$ transformations shown in \cref{tab:symmetries_orderparameters} and \cref{tab:symmetries_bosonization_XY0} tells us that $O^{\pm}_s$ and $O^\pm_a$ can have overlap with $\cX_{\pm 1,n}$. Comparing the $\ztwo^P$ action tells us that the smallest operators that transform correctly are
	\begin{align}
		O^\pm_s(x) &\sim \cX_{\pm1,1} + \cX_{\pm1,-1} \sim e^{\pm i \theta} \cos \phi,\\
		O^\pm_a(x) &\sim \cX_{\pm1,0} \sim e^{\pm i \theta},
	\end{align} 
	which is precisely what was obtained from the bosonization formulas in \cref{eq:local_XY0} and \cref{eq:local_correlations_XY0}. This combination also transforms correctly under $\ztwo^R$.
	
	\emph{XY$_1$ and XY$_1^*$}: Comparing the $U(1)$ transformations shown in \cref{tab:symmetries_orderparameters} and \cref{tab:symmetries_bosonization_XY1} again tells us that $O^{\pm}_s$ and $O^\pm_a$ can overlap with $\cX_{\pm 1,n}$. It is easy to check that no combination of scaling operators $\cX_{\pm 1,n}$ can simultaneously reproduce the $\ztwo^P$ and $\ztwo^R$ transformations of $O^\pm_s(x)$ (for $\moy{\vartheta} = \pi$ that is, XY$_1$) and $O^\pm_a(x)$ (for $\moy{\vartheta} = 0$ that is, XY$_1^*$) and therefore have correlations that decay exponentially. On the other hand, $\cX_{\pm 1,0} \sim e^{\pm i \theta}$ has the right transformation properties as $O^\pm_a(x)$ (for $\moy{\vartheta} = \pi$ i.e. XY$_1$) and $O^\pm_s(x)$ (for $\moy{\vartheta} = 0$ i.e. XY$_1^*$). This reproduces \cref{eq:local_XY1,eq:local_correlations_XY1}.
	
	\emph{XY$_2$ and XY$_2^*$}: The effective $U(1)$ transformations in \cref{tab:symmetries_bosonization_XY2} tell us that all scaling operators have a minimum $U(1)$ charge of $2$ and therefore there are no combinations of scaling operators that have the transformation properties of $O^{\pm}_s$ and $O^\pm_a$ and that have a unit $U(1)$ charge as seen in \cref{tab:symmetries_orderparameters}. Consequently, the correlations of $O^{\pm}_s$ and $O^\pm_a$ have exponential decay in both  XY$_2$ and XY$_2^*$ phases~\footnote{Note that lattice operators with $U(1)$ charge 2 such as $S^\pm_{1,j}S^\pm_{2,j}$ will have algebraic correlations in  XY$_2$ and XY$_2^*$ phases as shown in \cite{Schulz_Higherspinbosinization_PhysRevB.34.6372}.}. This reproduces \cref{eq:local_XY2,eq:local_correlations_XY2}.

	\subsubsection{Behaviour of the non-local operator}
	
	We now turn to the nonlocal string operator $C(x,y)$ defined in \cref{eq:C} which can be bosonized in both the small-$J$ ($J_<$) and large-$J$ ($J_>$) limits as follows (see \cref{app:String} for details)
	\begin{align}
		&C(x,y) \sim C_L(x)~ C_R(y) \label{eq:C_bosonized}\\
		&C_{L/R}  \approx \begin{cases}
			\gamma~ \sin \left(\frac{\phi}{2}\right)~ &(J_<)\\
			\alpha \sin\left(\frac{\phi_1+\phi_2}{2}\right)  +  \beta \sin\left(\frac{\phi_1-\phi_2}{2}\right) &(J_>)
		\end{cases}  \label{eq:CLR_bosonized} 
	\end{align}
	where we have only shown operators with the smallest scaling dimensions and $\alpha,\beta,\gamma$ are non-zero coefficients whose values we do not fix. It is now easy to verify how $\moy{C(x,y)}$ behaves at large $|x-y|$. From \cref{eq:C_bosonized}, we have
	\begin{equation}
		\moy{C(x,y)} \sim \moy{C_L(x)}~\moy{ C_R(y)}.
	\end{equation}
	Therefore, when $\moy{C_{L/R}} \neq 0$, we have $\moy{C(x,y)} \neq 0$. Among the phases without spontaneous symmetry breaking,  this happens when $\moy{\phi} = \pi$ for small $J$ and $\moy{\phi_1 \pm \phi_2} = \pi$ for large $J$. From \cref{fig:phasedig_smallJ_bosonization,fig:phasedig_largeJ_bosonization}, we see that $\moy{C(x,y)} \neq 0$ in the Haldane and XY$_2^*$ phases whereas in the trivial gapped phase and other gapless phases XY$_0$, XY$_1$ and XY$_1^*$, $C(x,y) \rightarrow 0$ for sufficiently large $|x-y|$. This confirms the remaining entries of \cref{tab:orderparameters}.

	\section{Bosonization analysis II: the topological nature of XY$_2^*$}
	\label{sec:Bosonization2}
	We now focus on the XY$_2^*$ gapless phase and study its topological nature. First, we show that it has protected edge modes and then discuss the nature of the topological phase. In particular, we show that the gapless topological phase is not `intrinsically gapless' and briefly discuss a related model where it is. 
	
	\subsection{Edge modes}
	A hallmark of gapped symmetry protected topological phases such as topological insulators and superconductors is the presence of protected edge modes degenerate with the ground state, which have exponentially small splitting at finite system sizes.  Gapless topological phases are defined as those that have edge modes protected by symmetries and can be sharply identified at finite volumes by exponential or algebraic splitting with coefficients different from bulk states~\cite{KeselmanBerg_TLL_PhysRevB.91.235309,ThorngrenVishwanathVerresen_IntrinsicallyGapless_PhysRevB.104.075132,DHLee_Gapless_jiang2017symmetry,ParkerScaffidiVasseur_GaplessSPT_PhysRevX.7.041048,ParkerScaffidiVasseur_TLL_DDW_PhysRevB.97.165114} . Recall that XY$_2^*$ was characterized by a nonzero expectation value of the string operator $C(x,y)$ whose endpoints were charged under $\ztwo^R$. The following argument presented in \cite{ThorngrenVishwanathVerresen_IntrinsicallyGapless_PhysRevB.104.075132,VerresenThorngrenJonesPollmann_SymmetryEnrichedCriticality_PhysRevX.11.041059} shows that this automatically implies the presence of edge modes. Let us first present the argument using lattice operators and then using bosonization. 
	\subsubsection{Argument using lattice operators}
	Let $\ket{\psi}$ be the ground state of the XY$_2^*$ Luttinger liquid which has string order
	\begin{equation}
		\innerproduct{\psi}{C(x,y)|\psi} \neq 0.
	\end{equation}
	We also know that $\ket{\psi}$ is invariant under the symmetries shown in ~\cref{tab:symmetries}. Let us consider $U(1)$ rotation by angle $\chi = \pi$ generated by the following operator (we only consider the large $J$ notation for convenience).
	\begin{equation}
		U(\pi)  \propto ~ \prod_{j=1}^L\left(  \sigma^z_{j,1} \sigma^z_{j,2} \right),~ U(\pi)\ket{\psi} \propto \ket{\psi}.
	\end{equation} 
	This operator acts on a finite chain of length $L$ with unit cells labeled $j=0,\ldots,L$. Let us now consider the action of the string operator defined on the full length of the chain, i.e.
	\begin{equation}
		C(0,L) \equiv \sigma^z_{0,2} \left(\prod_{j=2}^{L-1} \sigma^z_{j,1} \sigma^z_{j,2}\right) \sigma^z_{L,1}.
	\end{equation}
	We use this along with $U(\pi)$ to get the following result.
	\begin{multline}
		\innerproduct{\psi}{C(x,y)|\psi} \neq 0 \implies \innerproduct{\psi}{C(x,y) U(\pi)|\psi} \neq 0 \\ \implies \innerproduct{\psi}{\sigma^z_{1,1}\sigma^z_{L,2}|\psi} \neq 0.    
	\end{multline}
	By cluster decomposition, we have
	\begin{equation}
		\innerproduct{\psi}{\sigma^z_{1,1}|\psi} \neq 0~~ \text{ and }~~ \innerproduct{\psi}{\sigma^z_{L,2}|\psi}\neq 0.
	\end{equation}
	This proves that when we have string order, we also have edge magnetization. Since $\sigma^z_{1,1}$ and $\sigma^z_{L,2}$ are charged under symmetry $\ztwo^R$, we can interpret this result as spontaneously breaking the symmetry at the edges and resulting in degenerate edge modes. 
	
	\subsubsection{Argument using bosonization}
	It is nice to obtain the same result using bosonization. Let us first write down the bosonized version of $U(\pi)$ (see \cref{app:String})
	\begin{align}
		U(\pi) & \sim U_{L}(\pi) U_{R}(\pi) \\
		U_{L/R} &\approx \begin{cases}
			\cos\left( \frac{\phi}{2}\right) &(J_<)\\
			\gamma \cos\left(\frac{\phi_1+\phi_2}{2}\right)  +  \delta \cos\left(\frac{\phi_1-\phi_2}{2}\right) &(J_>)
		\end{cases} \label{eq:ULR_bosonized}
	\end{align}
	In the phases with $\moy{C(x,y)} \neq 0$, letting the string operator span the length of the system setting $x=0,~y=L$ we have 
	\begin{align}
		\moy{C(0,L) U(\pi)}  = \moy{C_LU_L)}_{x=0} \moy{C_RU_R}_{x=L} \neq 0.
	\end{align}
	By cluster-decomposition, we get
	\begin{equation}
		\moy{C_{L/R}~U_{L/R}(\pi))} \neq 0.
	\end{equation}
	Using \cref{eq:CLR_bosonized,eq:ULR_bosonized}, this reduces to
	\begin{multline*}
		C_{L/R}~U_{L/R}(\pi) \sim \sin \phi + \ldots \text{ for small $J$} \\
		C_{L/R}~U_{L/R}(\pi) \sim \tilde{\alpha} \sin \phi_1 +  \tilde{\beta} \sin \phi_2 + \tilde{\gamma} \sin\left(\phi_1 + \phi_2\right) \\ + \tilde{\delta} \sin\left(\phi_1 - \phi_2\right)+ \ldots \text{ for large $J$}.
	\end{multline*}
	$\tilde{\alpha},\ldots,\tilde{\delta}$ are some constants whose precise values are irrelevant. We see that $C_{L/R}U_{L/R}(\pi)$ are proper local operators (without fractional coefficients) carrying $\ztwo^R$ charge. Therefore, we have spontaneous symmetry breaking at the edges and associated boundary degeneracy whenever we have $\moy{C(x,y)}\neq 0$ and unbroken $U(\pi)$ symmetry, such as the Haldane and XY$_2^*$ phases.
	
	\subsection{Why XY$_2^*$ is \emph{not} an intrinsically gapless topological phase?}
	In the taxonomy of gapless topological phases~\cite{KeselmanBerg_TLL_PhysRevB.91.235309,ThorngrenVishwanathVerresen_IntrinsicallyGapless_PhysRevB.104.075132,DHLee_GaplessSPT_jiang2017symmetry,ParkerScaffidiVasseur_GaplessSPT_PhysRevX.7.041048,ParkerScaffidiVasseur_TLL_DDW_PhysRevB.97.165114}, a special role is played by so-called intrinsically gapless topological phases~\cite{ThorngrenVishwanathVerresen_IntrinsicallyGapless_PhysRevB.104.075132,Oshikawa_IntrisicallyGapless_2022symmetry,PotterWen_IntrinsicallyGapless_2023bulkboundary}. These are gapless phases with stable edge modes protected by symmetries that do not allow gapped topological phases. In this sense, the topological nature is intrinsically gapless. Phase diagrams in which intrinsically gapless topological phases can be found cannot, by definition, contain gapped topological phases. Therefore, the phase diagrams shown in \cref{fig:phasedig} that contain the Haldane phase, which is a gapped topological phase, make it clear that the XY$_2^*$ phase is not intrinsically gapless. This is because the symmetries of the model $G \cong O(2) \times \ztwo^P \times \bZ$ protect both gapless and gapped topological phases. We can ask whether we can break certain symmetries to preserve only the gapless topological phase but eliminate the gapped one. We now show using bosonization that this too is not possible.  
	
	Let us focus on the large $J$ limit where XY$_2^*$ is present. From \cref{fig:phasedig_largeJ_bosonization}, we see that the gapped Haldane phase is obtained when $\moy{\phi_1 + \phi_2} = \pi$  whereas the XY$_2^*$ obtains when $\moy{\phi_1 - \phi_2} = \pi$. Let us consider the possibility of eliminating the Haldane phase that has a gap while preserving XY$_2^*$ by adding an operator that ensures that $\moy{\phi_1 + \phi_2}$ can be tuned smoothly to zero, while $\moy{\phi_1 - \phi_2}$ can only be pinned to $0$ or $\pi$. The operator that achieves this is
	\begin{equation}
		\delta H \sim \int dx \sin \left(\phi_1 + \phi_2\right). \label{eq:Sine_perturbation}
	\end{equation}
	However, note that the addition of \cref{eq:Sine_perturbation} simultaneously breaks both $\ztwo^R$ and $\ztwo^P$ symmetries. Therefore, any lattice operator that produces \cref{eq:Sine_perturbation} also generically produces an operator of the form 
	\begin{equation}
		\delta H' \sim \int dx \sin \left(\phi_1 - \phi_2\right). \label{eq:Sine_perturbation'}
	\end{equation}
	which smoothly tunes the pinned value of $\moy{\phi_1 + \phi_2}$ to zero and therefore eliminates XY$_2^*$~\footnote{Note that \cref{eq:Sine_perturbation'} that breaks $\ztwo^R$ but preserves $\ztwo^P$ can be introduced to eliminate XY$_2^*$ but preserve the Haldane phase. But this is not interesting.}.

	\subsection{A related model where XY$_2^*$ \emph{is} an intrinisically gapless topological phase}
	\begin{figure}[!ht]
		\begin{tikzpicture}[scale=.7]
			\draw[] (-0.5,0) -- (7.5,0);
			\draw[] (-0.5,1) -- (7.5,1);
			\foreach \x in {0,...,7} 
			\draw[dashed] (\x,0) -- (\x,1);
			\foreach \x in {0,...,6} 
			\draw[dotted,thick] (\x,0) -- (\x+1,1);
			\foreach \x in {0,...,6} 
			\draw[dotted,thick] (\x,1) -- (\x+1,0);
			\foreach \x in {0,...,7} 
			\filldraw[rounded corners,blue,opacity=0.1,draw=black,thin] (\x-0.25,-0.3) rectangle (\x+0.25,1.3);
			\foreach \x in {0,...,7} 
			\shade[inner color=white, outer color = red] (\x,0) circle (0.15);
			\foreach \x in {0,...,7} 
			\shade[inner color=white, outer color = red] (\x,1) circle (0.15);
			\node[align=left,above] at (-1.,0.7)  {1};
			\node[align=left,above] at (-1.,-0.3)  {2};			
		\end{tikzpicture}
		\caption{Schematic representation of the Hamiltonian in \cref{eq:Hamiltonian_intrinsic_gapless_SPT} which can host an intrinsically gapless SPT.}
		\label{fig:model_2}
	\end{figure}
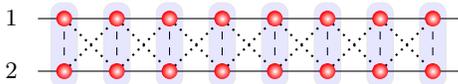
	
	We now present a model where XY$_2^*$ is an intrinsically gapless topological phase. We work in the large-$J$ limit and modify the Hamiltonian in \cref{eq:Hamiltonian_largeJ} as follows
	\begin{align}
		H &= H_1 + H_2 + H_\perp + H'_\perp+ H''_\perp,  \label{eq:Hamiltonian_intrinsic_gapless_SPT} \text{ where, } \\
		H_{\alpha} &=  J \sum_j \left( S^x_{\alpha j} S^x_{\alpha j+1} +  S^y_{\alpha j} S^y_{\alpha j+1} - \Delta  S^z_{\alpha j} S^z_{\alpha j+1} \right), \nonumber\\ 
		H_\perp &=  (1-t) \sum_j \left(S^x_{1 j} S^x_{2 j}+S^y_{1 j} S^y_{2 j}-\lambda S^z_{1 j} S^z_{2 j} \right), \nonumber \\
		H'_\perp &=  \frac{(1+t)}{2} \sum_j \left( S^x_{2 j} S^x_{1 j+1}+ S^y_{2 j} S^y_{1 j+1}-\lambda  S^z_{2 j} S^z_{1 j+1} \right),  \nonumber\\
		H''_\perp &=  \frac{(1+t)}{2} \sum_j \left( S^x_{1 j} S^x_{2 j+1}+ S^y_{1 j} S^y_{2 j+1}-\lambda  S^z_{1 j} S^z_{2 j+1} \right). \nonumber
	\end{align}
	The presence of the new term, $H''_\perp$ preserves all original symmetries shown in  \cref{tab:symmetries} but importantly introduces a new on-site symmetry which exchanges the two legs. The action on spin operators and large-$J$ bosonized variables is as follows.
	\begin{align}
		\ztwo^L:~ \vec{S}_{1,j} \leftrightarrow \vec{S}_{2,j},~\phi_1\leftrightarrow \phi_2,~\theta_1 \leftrightarrow\theta_2. \label{eq:layer_exchange}
	\end{align}
	Remarkably, the bosonized version of \cref{eq:Hamiltonian_intrinsic_gapless_SPT} is identical to \cref{eq:Bosonization_ladder} and therefore should contain the same phases although in different parameter regimes. Let us now consider including lattice operators that explicitly break the $\ztwo^R$ and $\ztwo^P$ symmetries but preserve the new $\ztwo^L$ symmetry shown in \cref{eq:layer_exchange}. In the continuum limit, this introduces only the perturbation shown in \cref{eq:Sine_perturbation} but not \cref{eq:Sine_perturbation'} since the latter breaks $\ztwo^L$. As explained above, this eliminates the Haldane phase. The equivalent of  XY$_2^*$  phase in this model is an intrinsically gapless topological phase. Indeed, the residual on-site unitary symmetry $U(1) \times \ztwo^L$ is known to not host any gapped symmetry protected topological phases in one dimension~\cite{ChenGuLiuWen_GroupCohomology_PhysRevB.87.155114}. We leave the numerical study of the model in \cref{eq:Hamiltonian_intrinsic_gapless_SPT} to future work.

	\section{Numerical Analysis}
	\label{sec:Numerics}
	
	In this section, we numerically analyze the system at hand and validate the analytical results predicted above. We map the spin system to hard-core bosons, where the on-site occupancy is restricted to $n=0/1$. The Hamiltonian in terms of hard core bosons (see \cref{eq:Hamiltonian_largeJ}) becomes
	\begin{align}
		H_\alpha =& J \Big[ \sum_{j} \frac{1}{2} \left(b_{\alpha, j}^{\dagger}b_{\alpha, j+1} + \text{H.c.} \right) \notag \\
		&-\Delta \left(\tilde{n}_{\alpha, j} \tilde{n}_{\alpha, j+1}\right) \Big] ,~~ \alpha = 1,2 \nonumber\\
		H_\perp =& (1-t) \Big[ \sum_{j} \frac{1}{2} \left(b_{1, j}^{\dagger}b_{2, j} + \text{H.c.} \right) -\lambda \left(\tilde{n}_{1, j} \tilde{n}_{2, j}\right) \Big] \nonumber\\
		H^{'}_{\perp} =& (1+t) \Big[ \sum_{j} \frac{1}{2} \left(b_{2, j}^{\dagger}b_{1, j+1} + \text{H.c.} \right) -\lambda \left(\tilde{n}_{2, j} \tilde{n}_{1, j+1}\right) \Big] 
		\label{eq:bosonHam}
	\end{align}
	where $b_j$ ($b_j^\dagger$) annihilation (creation) operators and $\tilde{n}_j = \left(n_j-\frac{1}{2}\right)$ with $n_j$ being the number operator for site $j$. The ground state of the model Hamiltonian is computed using the Density Matrix Renormalization Group (DMRG) method~\cite{whitedmrg,SCHOLLWOCK201196,ulrich2004}. The bond dimension is taken to be $\sim 500$, which is sufficient for convergence for typical system sizes $L = 200$ where $L$ is the total number of sites in the system. Unless otherwise stated, sites are labeled using a single-site label convention of \cref{eq:Hamiltonian_smallJ}.

	\begin{figure}[!ht]
		\centering
		\includegraphics[width=0.85\columnwidth]{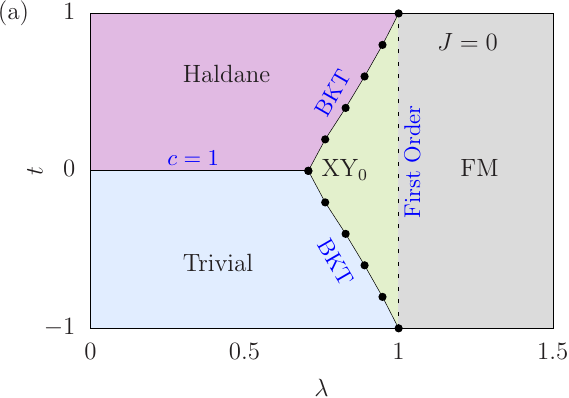}
		\includegraphics[width=0.85\columnwidth]{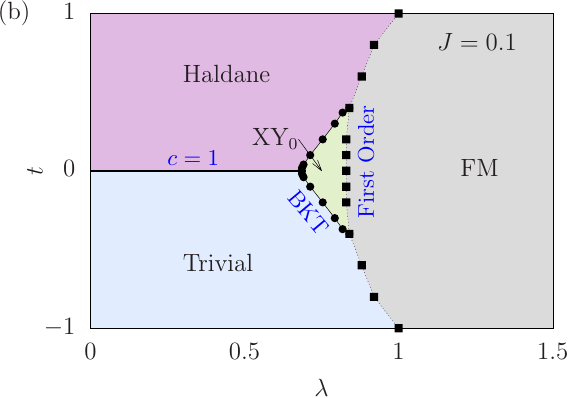}
		\includegraphics[width=0.85\columnwidth]{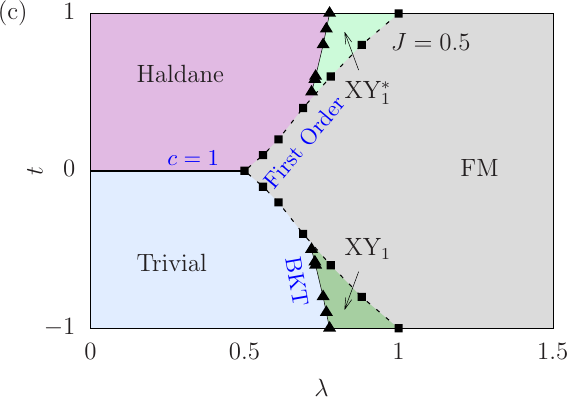}
		\caption{
			Phase diagram for small $J$ evaluated using DMRG corresponding to Hamiltonian (see \cref{eq:bosonHam}) for  (a) $J=0.0$, (b) $J=0.1$ and (c) $J=0.5$ $\Delta=\lambda$ in (a-c). Phase boundaries represent the second- and first-order transitions calculated using various diagnostics as mentioned in the text. The transition between the gapped, i.e. Trivial, Haldane and gapless, i.e. XY$_0$, XY$_1$ and XY$_1^*$, when they exist are BKT transitions and correspond to a single component compact boson theory with central charge $c=1$ and Luttinger parameter $K=2$ whereas the transition between the Haldane and Trivial phase is a single component compact boson theory with central charge $c=1$ and varying $K$. All transitions to the FM are first order. Symbols are the only calculated points, and lines connect the points for clarity.
		}
		\label{fig:pd1}
	\end{figure}

	\begin{figure}[!ht]
		\centering
		\includegraphics[width=0.95\columnwidth]{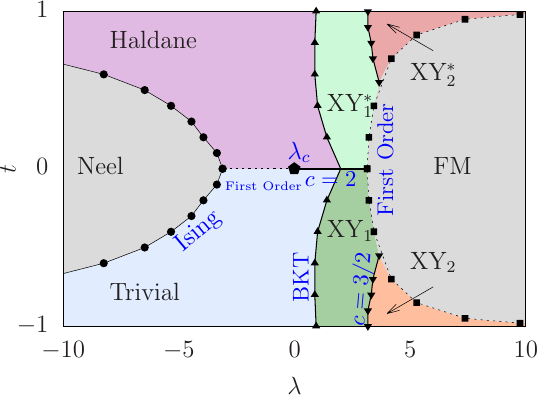}
		\caption{Phase diagram for large $J$ in $\lambda-t$ plane with fixed $J=2.5$ and $\Delta = 0.1$. Bold circles mark the Ising transition with central charge $c= \frac{1}{2}$ between N\'{e}el to trivial and N\'{e}el to Haldane phase. The transition between Trivial to XY$_1$ and Haldane to XY$_1^*$ (up triangles) are BKT transitions,  described by a single-component compact boson theory with $c=1$ and Luttinger parameter $K=2$. The transition from XY$_1$ to XY$_1^*$ is a two-component compact boson theory with $c=2$ with varying Luttinger parameters. The Trivial-to-Haldane phase transition through the MG points is first order, which changes to $c=2$ for larger $\lambda$ at $\lambda_c$ (pentagonal point). The transition from XY$_1$ to XY$_2$ and from XY$_1^*$ to XY$_2^*$ (down triangles) belong to the Ising universality class stacked on top of a compact boson with $c=\frac{3}{2}$. Finally, the transition from any phase to the FM phase is a first-order transition (squares). Note that symbols are the only calculated points, and lines connect the points for clarity. }
		\label{fig:pd2}
	\end{figure}
	
	\begin{figure*}[!ht]
		\centering
		\includegraphics[width=1.4\columnwidth]{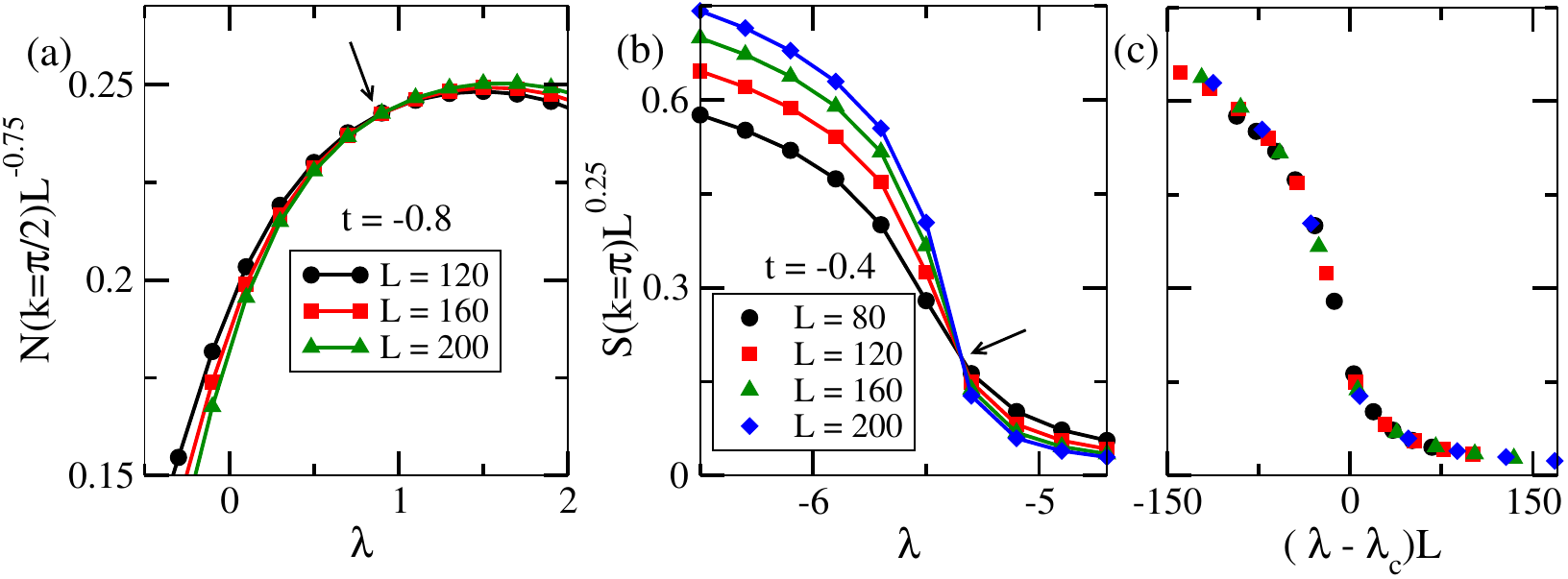}
		\includegraphics[width=0.71\columnwidth]{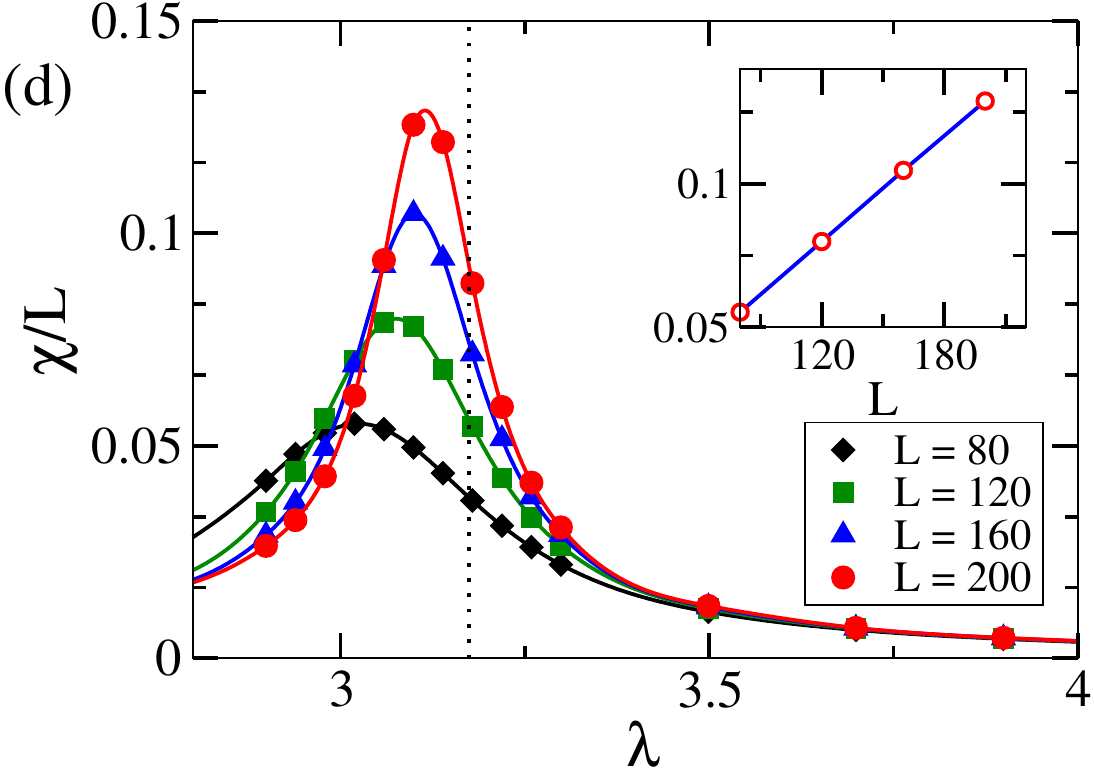}
		\includegraphics[width=0.68\columnwidth]{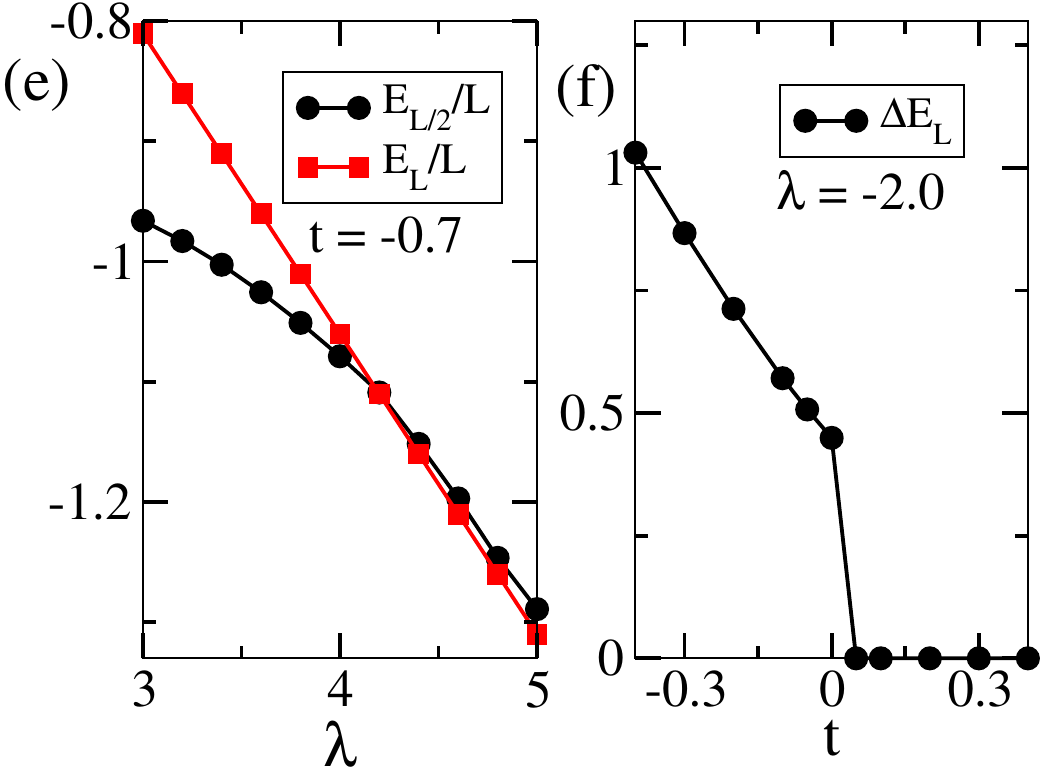}
		\caption{The first row of the figure demonstrates the finite size calling to determine (a) the BKT transition between the Trivial and XY$_1$ phase and (b-c) the Ising transition between the N\'{e}el and Trivial phase. The perfect crossing of different $N(\pi/2) L^{\frac{1}{2K} -1}$ curves in (a) with $t=-0.8$ for different $L$ implies the transition point with the Luttinger parameter $K=2$ for the BKT transition. The crossing of different $S(\pi)L^{\frac{2\beta}{\nu}}$ curves with $t=-0.4$ for different $L$ in (b) reveals the transition point with exponents $\nu=1$ and $\beta = 1/8$. The collapse of all the data points for different $L$ in $S(\pi)L^{\frac{2\beta}{\nu}}$ vs. $(\lambda-\lambda_c)L^\nu$ shown in (c) further confirms the Ising transition point at $\lambda_c \sim -5.34$.
			The Ising transition between two gapless phases, from XY$_1$ to XY$_2$ phase, at $t = -1$ using the finite size scaling of fidelity-susceptibility ($\chi$) shown in (d). The $\chi$ peaks at the transition point, and for the Ising transition, the peak height diverges linearly with $L$ (inset). The transition point at the thermodynamic limit (dotted line) is calculated by extrapolating the peak positions for different $L$. The eigenvalues are plotted to determine the first-order transitions. (e) The level crossing in the ground state energies $E_N$ at $t=-0.7$ with $N=L/2$ and $N=L$ implies the first-order transition between the XY$_2$ and FM phase. {{(f) The sharp jump in single particle excitation gap ($\Delta E_L$ with $\Delta n =1$) at $t =0$ for $\lambda = -2.0$ signifies the first order transition between the trivial and Haldane phases. See the phase diagram in \cref{fig:pd2}. In (e) and (f), we consider $L=200$.}}
		}
		\label{fig:transition}
	\end{figure*}

	\subsection{Diagnostics, Phases and Phase transitions}
	We explore the parameter space in the $\lambda-t$ plane with fixed $J$ and identify the phases and their transitions. The most illustrative limit is to first investigate when $J=0$ \cite{Yoshida_JPSJ_1989} where the system, in the absence of any dimerization ($t=0$), undergoes a first order phase transition at $\lambda=1$ (see \cref{fig:pd1}(a)). $t$ engineers gapped phases between $0<\lambda<
	\frac{1}{\sqrt{2}}$, however $t<0$ is trivial and $t>0$ is topological (Haldane phase) in nature. A gapless phase (XY$_0$) opens between $\frac{1}{\sqrt{2}}<\lambda<1$ where both perturbations $\lambda$ and $t$ are irrelevant. Introducing a small finite $J$, not unexpectedly, only renormalizes the phase boundaries (see $J=0.1$ $\lambda-t$ phase diagram in \cref{fig:pd1}(b)) reducing the size of the gapless XY$_0$ phase. A further increase in $J_2$ leads to the emergence of two new gapless phases (XY$_1$) and XY$^*_1$ as XY$_0$ disappears (see \cref{fig:pd1}(c)) and we get the large $J$ picture. 
	
	To explore the large $-J$ phase diagram schematically shown in \cref{fig:phasedig}, a particularly illustrative parameter choice is to explore the phase diagram for fixed $J = 2.5$ as shown in \cref{fig:pd2} ($\Delta = 0.1$). Four {\it distinct} symmetry enriched critical phases are clearly obtained. To conclusively characterize the phase boundaries and the nature of their transitions, we use a host of diagnostics which we now discuss.
	
	\subsubsection{BKT Transitions}
	Transitions from the trivial gapped phase to XY$_1$ and the Haldane phase to XY$_1^*$ belong to the BKT universality class. To characterize these transitions, it is useful to note that in the (hard-core) bosonic language, the XY$_1$ and XY$_1^*$ phases are $\pi/2$-superfluids (SF($\pi/2$)) phases~\cite{Mishra2015, Mishra2019}. In such systems, the momentum distribution is given by
	\begin{equation}
		N(k)=\frac{1}{L}\sum_{i,j} e^{ik|i-j|}\Gamma_{i,j}
		\label{eq:mom}
	\end{equation}
	(where $\Gamma_{i,j}= \langle b_i^\dagger b_j \rangle$) and is expected to show a sharp peak at $k=\pi/2$. At the transition itself, the finite-size scaling of $N(\pi/2)$ carries the signature of an underlying BKT transition. For example, at the critical point, the BKT ansatz predicts $N(k)\propto L^{1-\frac{1}{2K}}$ which can be used to extract the value of $K$ \cite{Dhar_PRA_2012,Mondal_PRA_2019}. 
	The perfect crossing of the $N(k=\pi/2)$ data for different lengths at $t = -0.8$ as shown in \cref{fig:transition}(a) indicates a BKT transition which has the Luttinger parameter of value $K=2$ as expected from \cref{sec:Bosonization1} (see also \cref{app:bosonization}). This is found to be true for all values of ${t}$ using which the phase boundaries to the SF($\pi/2$) (XY$_1$) phase of \cref{fig:pd2} have been obtained.

	\subsubsection{Ising transitions} The transitions from the N\'{e}el to trivial and Haldane (circles), from XY$_1$ to XY$_2$, and from XY$_1^*$ to XY$_2^*$ phase are found to be of Ising type (see \cref{fig:pd2}). Such Ising transitions can be characterized by analysing the finite size scaling of the structure factor $S(k)$ defined by
	\begin{equation}
		S(k) = \frac{1}{L^2}\sum_{l,m} e^{ik(l-m)} \left( \langle n_ln_m\rangle - \langle n_l \rangle\langle n_m\rangle  \right)
	\end{equation}
	In the N\'{e}el state, $S(k)$ shows a peak at $k=\pi$ signalling antiferromagnetic correlations. At the Ising transitions, it is known that $S(k=\pi)$ follows a scaling ansatz $ \propto L^{-\frac{2\beta}{\nu}}$ such that at the critical point $S(k) L^{\frac{2\beta}{\nu}}$ is invariant for different $L$ with exponents $\nu=1$ and $\beta = 1/8$ \cite{mishratvvp, sumanv1v2}. The perfect crossing of $S(k) L^{\frac{1}{4}}$ as shown in \cref{fig:transition}(b), and eventual collapse of all the data points, shown in \cref{fig:transition}(c),  for different $L$ in $S(\pi)L^{\frac{2\beta}{\nu}}$ vs. $(\lambda-\lambda_c)L^\nu$ plane near the transition point implies an Ising phase transition at $t=-0.4$ with a critical point $\lambda_c \sim -5.41$. We use the same approach to calculate the Ising phase boundaries in the phase diagram (\cref{fig:pd2}).
	
	\begin{figure}[!ht]
		\centering
		\begin{tabular}{c c}
			\includegraphics[height=0.43\columnwidth]{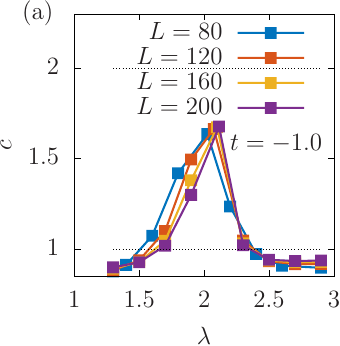} &
			\includegraphics[height=0.43\columnwidth]{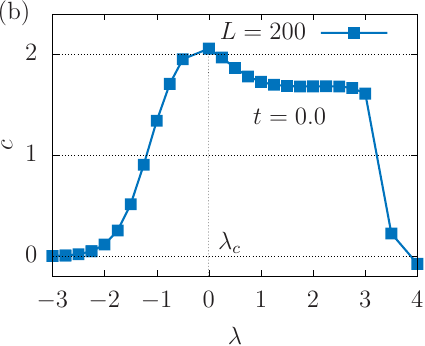}
		\end{tabular}
		
		\caption{The central charge ($c$) is plotted for cuts along (a) $t=-1.0$  that goes through the XY$_1$ and XY$_2$ phases for different $L$ corresponding to \cref{fig:pd2} where $J=2.5$ and $\Delta = 0.1$. In (b), we plot $c$ for a system of $L=200$ along a cut that goes through the interface between the phases XY$_1$ and XY$_1^*$ ($t=0$) (see \cref{fig:pd2}).}
		\label{fig:ccharge}
	\end{figure}

	\subsubsection{$c=\frac{3}{2}$ transition between gapless phases}
	
	Unlike the previous Ising transitions where one transits from a gapless to a gapped phase, the Ising transitions that appear between XY$_1$ to XY$_2$ and XY$_1^*$ to XY$_2^*$ are gapless-to-gapless transitions. Since this CFT appears in addition to the existing compact boson, the total central charge of the transition is expected to be $c = \frac{3}{2}$. Phase transition points are quantified by analyzing fidelity susceptibility ($\chi$) where
	\begin{equation}
		\label{eq:fid}
		\chi =  \lim_{(\lambda - \lambda^\prime) \to 0} \frac{-2 \mathrm{ln} |\langle\psi(\lambda)|\psi(\lambda^\prime)\rangle}{(\lambda - \lambda^\prime)^2}
	\end{equation}
	where $|\psi(\lambda)\rangle$ is the ground state at $\lambda$. At the phase transition point, $\chi/L$ develops a peak, and the height of the peak diverges linearly with $L$ for the Ising transition~\cite{Gu2010,Singh2018,Lahiri2020}. In \cref{fig:transition}(d), we plot $\chi/L$ for different system sizes, which shows an increase in the peak height with $L$. The inset of \cref{fig:transition}(d) shows the linear divergence of the peak height, implying the Ising transition. The critical point of the transition is determined by extrapolating the position of the peak to the thermodynamic limit, which is marked by the dashed line in \cref{fig:transition}(d).
	
	\subsubsection{Multiversality along the $t=0$ line}
\begin{figure}[!ht]
		\centering
		\includegraphics[width=0.9\linewidth]{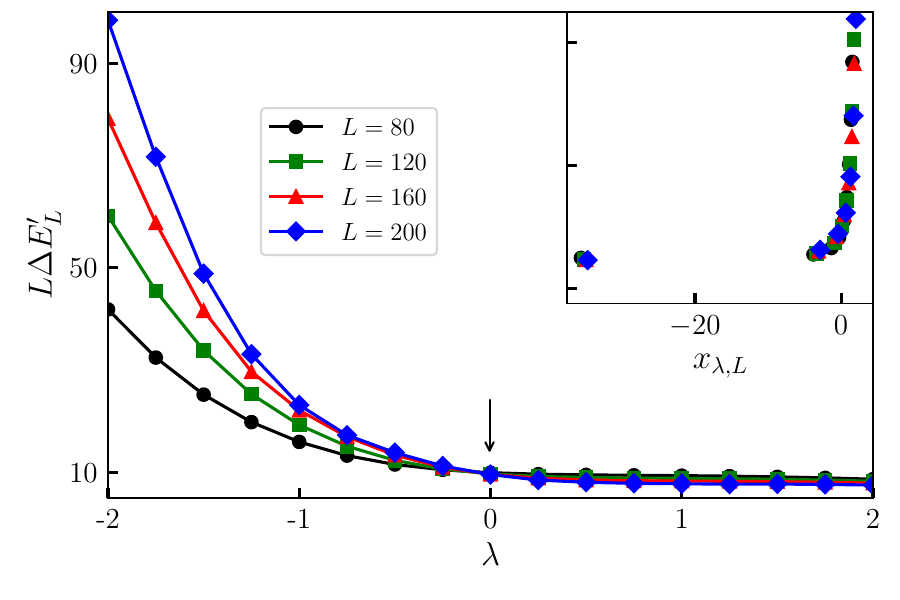}
		\caption{Finite size scaling of $\Delta E_L$ is shown to find the BKT transition point along $t=0$ line for large $J$ where the universality class changes from first order to second order with $c=2$. The crossing of all curves captures the critical point, marked by the arrow. (inset) The collapse of all the data for different $L$ complements the BKT transition with $\lambda_c = -0.01$.}
		\label{fig:gap_bkt}
	\end{figure}

	On $t=0$ line, the gapless phase with $c=2$ starts at $\lambda_c\sim -0.01$ which is a BKT transition point that can be calculated using finite size scaling of single particle excitation gap~\cite{mishratvvp}. 
	{{The excitation gap at half-filling can be defined as, 
			\begin{equation}
				\Delta E_L = (E_{N-\Delta n} + E_{N+\Delta n} -2E_{N})/\Delta n,
				\label{eq:gap}
			\end{equation}
			where $N=L/2$ and $\Delta n$ is the number of particles in an excitation.}}
	The invariance of $L\Delta E_L^\prime$ with $\Delta n = 1$ at the critical point and the collapse of all the data in $L\Delta E_L^\prime$ vs. $x_{\lambda,L}$ plane, where 
	\begin{align}
		\Delta E_L^\prime &= \Delta E_L \left[ 1+1/(2{\rm{ln}}L + C) \right] \nonumber\\
		x_{\lambda, L} &= {\rm{ln}} L -a/\sqrt{\lambda - \lambda_c},
	\end{align}
	at and near the critical point with a suitable choice of constants $C$ and $a$ predicts the BKT transition point $\lambda_c\sim -0.01$ (see Fig.~\ref{fig:gap_bkt}).
	
	From \cref{fig:pd2}, we see that the $c=2$ line separates the gapless phases XY$_1$ and XY$_1^*$ as well as the trivial and Haldane gapped phases. The latter phases are separated by a different universality class with $c=1$ for small $J$. This is a numerical confirmation of the `multiversality'~\cite{AP2022multiversality,BiSenthil_Multiversality_PhysRevX.9.021034} phenomenon discussed in \cref{sec:Bosonization1,app:bosonization}.
	
	\subsubsection{ First order transitions}
	Finally, the transition between the trivial and the Haldane gapped phase for negative values of $\lambda$ at large $J$, or that between any of the phases to FM is first-order in nature. These can be characterized by analyzing the level crossings between eigenstate energies. For instance, in the case of transitions to the FM phase, we plot the ground state energy at boson half-filling ($E_{L/2}$), which corresponds to zero magnetization sector, and completely filled ($E_{L}$), which is equivalent to fully magnetized case (FM phase) across the boundary in \cref{fig:transition}(e). The crossing between $E_{L/2}$ and $E_{L}$ determines the first-order transition points, which are marked by squares in the phase diagram (\cref{fig:pd2}). {{On the other hand, the sharp jump in single particle excitation gap $\Delta E_L$ (Eq.~\ref{eq:gap} with $\Delta n = 1$) at $t=0$ for the transition between the trivial gapped to Haldane gapped phase, as shown in \cref{fig:transition}(f), signifies a first-order transition (also see \cref{fig:pd2}).}}

	\subsubsection{Central charge}
	Now we want to give numerical evidence for the central charge predicted by the bosonization analysis. We find the central charge ($c$) by fitting the bipartite Von-Neumann entanglement entropy ($S_{vN}$) to its conformal expression \cite{calabrese2009entanglement}
	\begin{equation}
		S_{vN} = \frac{c}{6} {\rm{ln}} \left[ \frac{L}{\pi} {\rm{sin}} \frac{\pi l}{L} \right] + g.
	\end{equation}
	\Cref{fig:ccharge}(a) show how $c$ changes at the transition between XY$_1$ and XY$_2$ phases according to \cref{fig:pd2}. In \cref{fig:ccharge}(b), we represent $c$ along the $t=0$ line that cuts through the interface between the XY$_1$ and XY$_2$ phases. As discussed above in the analytical analysis, we find from the numerical analysis, although not exactly, that the $c$ is close to $2$ in the XY$_1$ $-$ XY$_1^*$ transition and at the Ising transition point between the phases XY$_1$ and XY$_2$, the $c$ is close to $1.5$ (upto finite size effects).

	\subsection{Characterising gapless phases}
	
	Since the gapped phases and particularly ordered phases are well understood and can be easily characterized by conventional order parameters, here we will focus our discussion on the gapless phases and their characterization.

	\begin{figure*}[!ht]
		\centering
		\includegraphics[width=0.65\columnwidth]{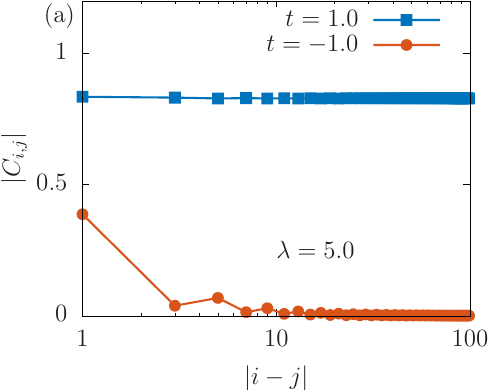}
		\includegraphics[width=0.65\columnwidth]{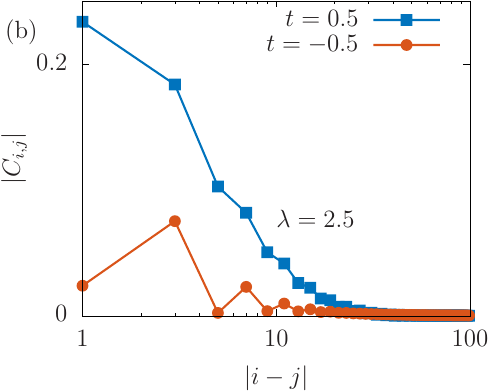}
		\includegraphics[width=0.65\columnwidth]{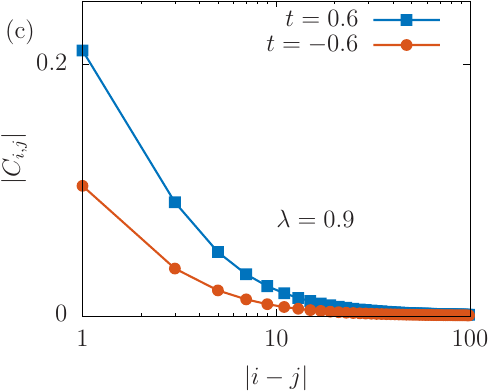}
		\includegraphics[width=0.48\columnwidth]{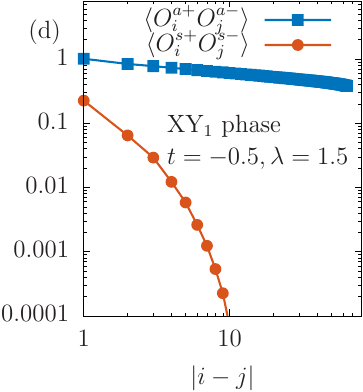}
		\includegraphics[width=0.48\columnwidth]{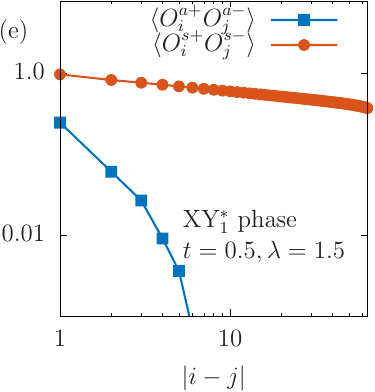}
		\includegraphics[width=0.48\columnwidth]{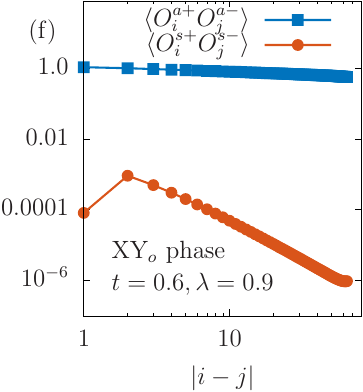}
		\includegraphics[width=0.48\columnwidth]{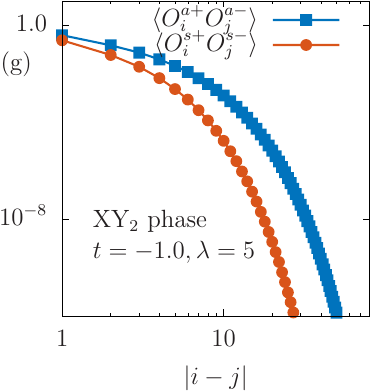}
		\caption{String order parameter $|C_{i,j}|$ is shown in (a-c) for different gapless phases to distinguish the trivial and non-trivial topological phases. (a) The parameter ($\lambda$, $t$) $= (5,-1.0)$ belongs to the XY$_2$ phase and $(5,1.0)$ belongs to the XY$_2^*$ phase (see \cref{fig:pd2}). The presence (absence) of long-range $|C_{i,j}|$ signifies the non-trivial (trivial) topological nature of the XY$_2^*$ (XY$_2$) phase. (b) The parameter $(2.5,-0.5)$ belongs to the XY$_1$ phase and $(2.5,0.5)$ belongs to the XY$_1^*$ phase (see \cref{fig:pd2}). The absence of long-range $|C_{i,j}|$ signifies the trivial nature of the phases XY$_1$ and XY$_1^*$. (c) $|C_{i,j}|$ for the XY$_0$ phase ($J=0$) (see \cref{fig:pd1} (a)).  For all cases, we calculate $C_{ij}$ with $i=L/4 +1$, and $j$ goes from $\frac{L}{4} +2$ to $L$ where $j-i \in \mathrm{odd}$ for a system of $L=200$. (d-g) shows the correlation functions $|\langle O_i^{s+} O_j^{s-}\rangle|$ (circles) and $|\langle O_i^{a+} O_j^{a-}\rangle|$ (squares) as a function of distance $|i-j|$ for four different gapless phases XY$_1$, XY$_1^*$, XY$_0$ and XY$_2$, respectively. (g) is also applicable to XY$_2^*$. We see different behaviors of these correlations in different gapless phases (see the main text). Note that in (d), (e), and (g), the parameters correspond to the phases in \cref{fig:pd2}, and the parameters in (f) correspond to the phase in \cref{fig:pd1}(a). Here, we also use a system of $L=200$ and calculate the correlations at the center of the system where $i=L/8$.}
		\label{fig:string}
	\end{figure*}

	\subsubsection{String Order Parameter}
	A particularly useful tool, which also helps in distillation of the topological features of the gapless phases, is the string order parameter $C_{i,j}$ (see equivalently~\cref{eq:C} upto a phase)
	where 
	\begin{equation}\label{eq:string}
		C_{i,j} = -\langle z_i  e^{i \frac{\pi}{2} \sum_{k=i+1}^{j-1} z_k}  z_{j}\rangle,
	\end{equation}
	and $z_i= 1- 2b_i^\dagger b_i$. The string order parameter $|C_{2, L-1}|$, not unexpectedly, shows a finite value in the Haldane (gapped) phases in the system (not shown)~\cite{hadaPRB1992,sumanv1v2}. Interestingly, the same order parameter also takes nontrivial values in XY$_2^*$, proving that it is a {\it gapless topological} phase. In \cref{fig:string}(a) the behavior of $|C_{i,j}|$ is shown for both XY$_2$ (circles) and XY$_2^*$ (squares) - one finds that, unlike XY$_2$, in XY$^*_2$, $|C_{i,j}|$ takes a finite value that does not decay with $|i-j|$.  Similarly in \cref{fig:string}(b) we plot $|C_{i,j}|$ within the phase XY$_1$ (circles) and XY$_1^*$ (squares), and in \cref{fig:string}(c), we plot it in the XY$_0$ phase. In both plots, the string-order parameter vanishes, showing that these phases are trivial in nature. 
	
	\subsubsection{Local Order Parameters}
	The nature of long-range correlations can also distinguish between the different phases, as shown in ~\cref{tab:orderparameters}. To this end, we calculate $|\langle O_i^{s+} O_j^{s-}\rangle|$ and $|\langle O_i^{a+} O_j^{a-}\rangle|$ where $ O_i^{s+} = b_{1,i}^\dagger + b_{2,i}^\dagger$ and $ O_i^{a+} = b_{1,i}^\dagger - b_{2,i}^\dagger$ to distinguish between the trivial, XY$_1$, XY$_1^*$ and XY$_0$ phases. The results are shown in \cref{fig:string}(d-g) for all the gapless phases. We see a contrast in the nature of these correlations in two phases. The $|\langle O_i^{s+} O_j^{s-}\rangle|$ ($|\langle O_i^{a+} O_j^{a-}\rangle|$) falls exponentially (algebraically) with distance $|i-j|$ in the XY$_1$ phase. Whereas, in the XY$_1^*$ phase, the behavior flips. However, in XY$_0$ and XY$_2$, both correlation functions are algebraic or exponential, as shown in Figs.~\ref{fig:string}(f) and (g), respectively.
	
	\begin{figure}[!ht]
		\centering
		\includegraphics[width=0.48\columnwidth]{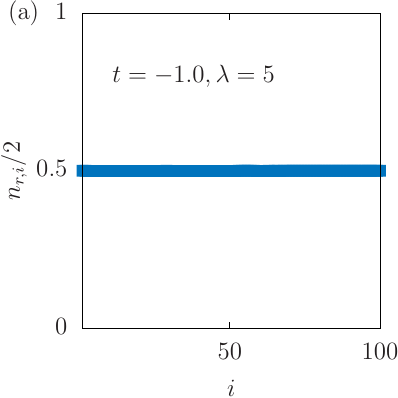}
		\includegraphics[width=0.48\columnwidth]{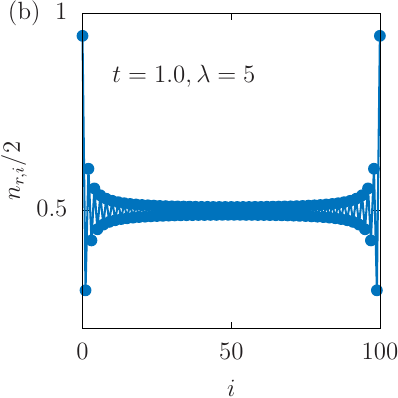}
		\caption{The presence of edge state in the topological gapless phase (XY$_2^*$), which is present in phase diagram \cref{fig:pd2}, is portrayed. Here, $J=2.5$ and $\Delta = 0.1$. (a) shows the density of particles $n_{r,i}/2$ on the rung $i$ with stronger coupling is plotted corresponding to XY$_2$ phase and (b) shows the same except the edge sites where $\langle n_i \rangle$ is plotted corresponding to XY$_2^*$ phase.}
		\label{fig:edge}
	\end{figure}
	
	\begin{figure}[!ht]
		\centering
		\includegraphics[width=0.7\columnwidth]{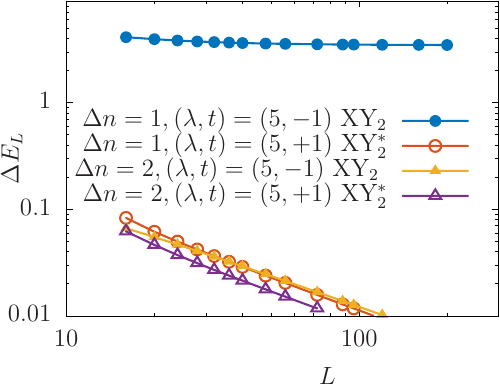}
		\caption{The energy gaps $\Delta E_L$ is plotted for XY$_2$ (solid symbols) and XY$_2^*$ (empty symbols) phases, emerged in \cref{fig:pd2}, as a function of $L$. Circles and triangles represent single ($\Delta n = 1$) and two-particle ($\Delta n = 2$) excitation gaps, respectively. Here, $J=2.5$ and $\Delta = 0.1$.} 
		\label{fig:gap}
	\end{figure}

	\subsubsection{Edge states}
	
	The topological XY$_2^*$ phase exhibits edge states, a hallmark property of such topological phases. In  \cref{fig:edge}(a) and \cref{fig:edge}(b), we plot the number of particles $n_{r,i} = \langle n_{1,i} +  n_{2,i} \rangle$ in strong rungs, where the hopping and interaction coupling are large (even or odd rungs) according to the construction of the system, for the phases XY$_2$ and XY$_2^*$, respectively. For XY$_2^*$, two edge sites do not belong to the strong bond where we plot $\langle n_i \rangle$. We can see that, only for the XY$_2^*$ phase (\cref{fig:edge}(b)), the system exhibits exponentially localized occupied edge states.
	
	The edge states manifest gapless excitations at the edges of the system. To confirm this property, we plot the energy gap for the excitation (Eq.~\ref{eq:gap}) at half-filling. In \cref{fig:gap} we plot $\Delta E_L$ for $\Delta n = 1$ (circles) and $\Delta n = 2$ (triangles) in the phases XY$_2$ (solid symbols) and XY$_2^*$ (empty symbols). In both phases, the elementary excitation in bulk is $\Delta n =2$ (a pair of particles on the strong rungs), which is gapless. This can be confirmed from the algebraic decay of $\Delta E_L$. In the XY$_2^*$ phase, due to the presence of edge states, which can be occupied by a single particle, we see algebraic decay of $\Delta E_L$ even with $\Delta n =1$. In the XY$_2$ phase, however, single particle excitation is gapped where $\Delta E_L$ saturates to a finite value.

	\section{Mapping to effective  spin--1 models}
	\label{sec:Spin1}
	\begin{figure}[!ht]
		\centering
		\includegraphics[width=0.3\textwidth]{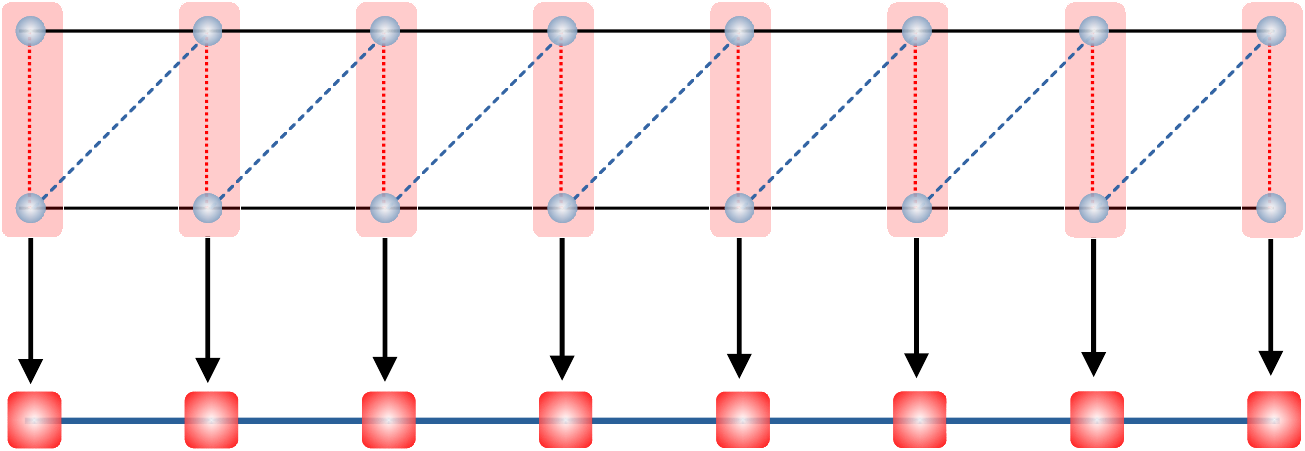}\\ 
		\medskip
		\includegraphics[width=0.3\textwidth]{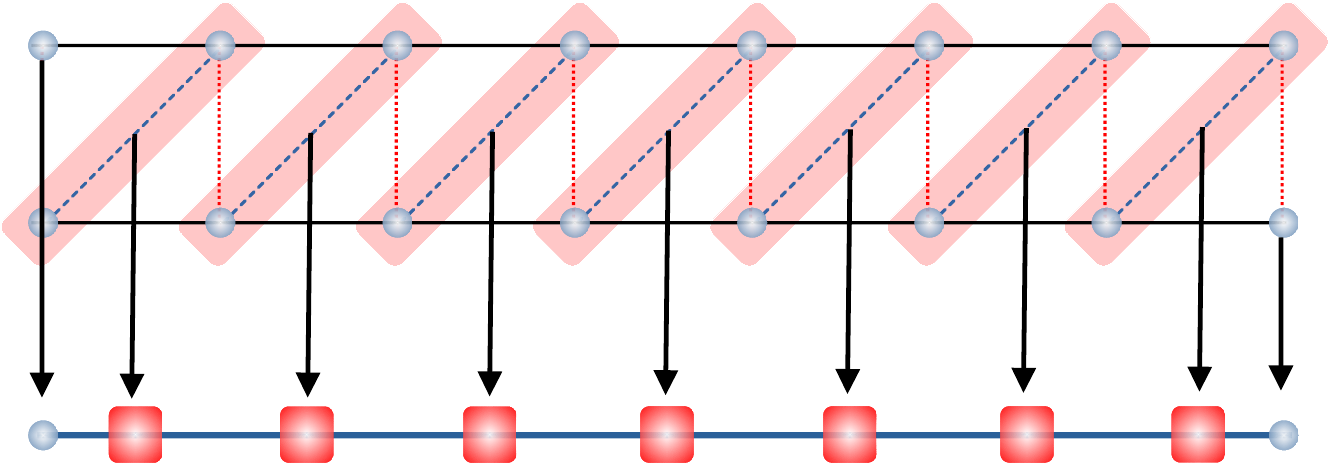} 
		\caption{Mapping to an effective spin-1 chain in the regime $t \approx -1$ (top) and $t \approx +1$ (bottom). Circles represent the qubits from the original Hilbert space and the boxes enclosing circles represent which pair of qubits are mapped to effective spin 1 entities (squares). Boundary effects are seen in the latter case where the mapping leaves behind a qubit on each end. }
		\label{fig:Spin1Map}
	\end{figure}	
	
	The connection between phase diagrams of spin $\half$ ladders and higher-spin chains is rather well known~\cite{Schulz_Higherspinbosinization_PhysRevB.34.6372}.
	Certain parts of the phase diagram shown in \cref{fig:phasedig} too can be determined using a mapping to an effective spin 1 model. This provides both consistency checks and physical insights into the phases. To do this, let us begin with the Hamiltonian in \cref{eq:Hamiltonian_largeJ} and perform a change of basis 
	\begin{equation}
		\begin{pmatrix}
			S^x_{1j}\\ 
			S^y_{1j}\\
			S^z_{1j}
		\end{pmatrix} \mapsto 		\begin{pmatrix}
			-	S^x_{1j}\\
			-	S^y_{1j}\\
			~~S^z_{1j}
		\end{pmatrix}, 		\begin{pmatrix}
			S^x_{2j}\\
			S^y_{2j}\\
			S^z_{2j}
		\end{pmatrix} \mapsto 		\begin{pmatrix}
			S^x_{2j}\\
			S^y_{2j}\\
			S^z_{2j}
		\end{pmatrix}.
	\end{equation} 
	which results in the following change in $H_\perp$ and $H'_\perp$
	\begin{align}
		H'_\perp &\mapsto  -(1+t) \sum_j \left( S^x_{2 j} S^x_{1 j+1}+ S^y_{2 j} S^y_{1 j+1}+\lambda  S^z_{2 j} S^z_{1 j+1} \right), \nonumber \\
		H_\perp &\mapsto  -(1-t) \sum_j \left(S^x_{1 j} S^x_{2 j}+S^y_{1 j} S^y_{2 j}+\lambda S^z_{1 j} S^z_{2 j} \right) \label{eq:Hperp_basischange}.  
	\end{align}
	Let us first consider the parameter regime when $H_\perp$ is dominant, i.e. $t \approx -1$. Since $H_\perp$ decouples into disjoint pieces each of which has support on two spins living on vertical bonds as shown in \cref{fig:Spin1Map} and takes the form
	\begin{equation}
		h_\perp = -(1-t)  \left(S^x_{1 j} S^x_{2 j}+S^y_{1 j} S^y_{2 j}+\lambda S^z_{1 j} S^z_{2 j} \right). \label{eq:Hperplocal} 
	\end{equation}
	it can be easily diagonalized as follows (suppressing site labels for clarity)
	\begin{multline}
		h_\perp = (1-t) \Big[(\lambda +2) \outerproduct{s}{s}+(\lambda -2) \outerproduct{0}{0} \\  -\lambda \left(\outerproduct{+1}{+1}+\outerproduct{-1}{-1}\right)\Big] \text{ where} \\
		\ket{+1} \equiv \ket{\uparrow_1 \uparrow_2},~ 		\ket{-1} \equiv \ket{\downarrow_1 \downarrow_2}, \\ ~ \ket{0} \equiv \frac{	\ket{\uparrow_1 \downarrow_2} +  \ket{\downarrow_1 \uparrow_2}}{\sqrt{2}}, ~ \ket{s} \equiv \frac{	\ket{\uparrow_1 \downarrow_2} -  \ket{\downarrow_1 \uparrow_2}}{\sqrt{2}}. \label{eq:local_spin1}
	\end{multline}
	and $\ket{\uparrow}, \ket{\downarrow}$ represent eigenstates of $S^z$ with eigenvalues $\pm \frac{1}{2}$ respectively. We see that for all values of $\lambda>-1$, $\ket{\pm 1},\ket{0}$ have the lowest energies. We can project the two-spin Hilbert space on the vertical bonds of every site onto this three-dimensional subspace using the following projection operator 
	\begin{equation}
		\mathbb{P} = \prod_j \left(\outerproduct{0}{0} + \outerproduct{+1}{+1} + \outerproduct{-1}{-1}\right)
	\end{equation}
	as schematically shown in the top figure of \cref{fig:Spin1Map} to get an effective spin-1 chain with Hamiltonian 
	\begin{multline}
		H_{eff} = \mathbb{P} H \mathbb{P}^\dagger = J_{xy} \sum_j \left(L^x_j L^x_{j+1} + L^y_j L^y_{j+1}\right) \\ + J_z \sum_j L^z_j L^z_{j+1} +  D \sum_j \left(L^z_j\right)^2 \label{eq:Heff_Spin1}  
	\end{multline}
	where
	\begin{multline}
		J_{xy} = \left(\frac{J}{2} - \frac{(1+t)}{4}\right) ,
		J_z =- \left( \frac{J \Delta}{2} +  \frac{ \lambda (1+t)}{4}\right), \\ \text{ and }
		D =  2(1-t) (1-\lambda). \label{eq:Spin1parameter_map}
	\end{multline}
	$L^x,L^y,L^z$ are the spin 1 representations of the angular momentum algebra with representations
	\begin{equation}
		\frac{1}{\sqrt{2}} \begin{pmatrix}
			0 & 1 & 0\\
			1 & 0 & 1\\
			0 & 1 & 0
		\end{pmatrix}, ~\frac{1}{\sqrt{2}} \begin{pmatrix}
			0 & -i & 0\\
			i & 0 & -i\\
			0 & i & 0
		\end{pmatrix},~  \begin{pmatrix}
			1 & 0 & 0\\
			0 & 0 & 0\\
			0 & 0 & -1
		\end{pmatrix}.
		\nonumber
	\end{equation}

\begin{figure}[!ht]
		\centering
		\includegraphics[width=0.6\columnwidth]{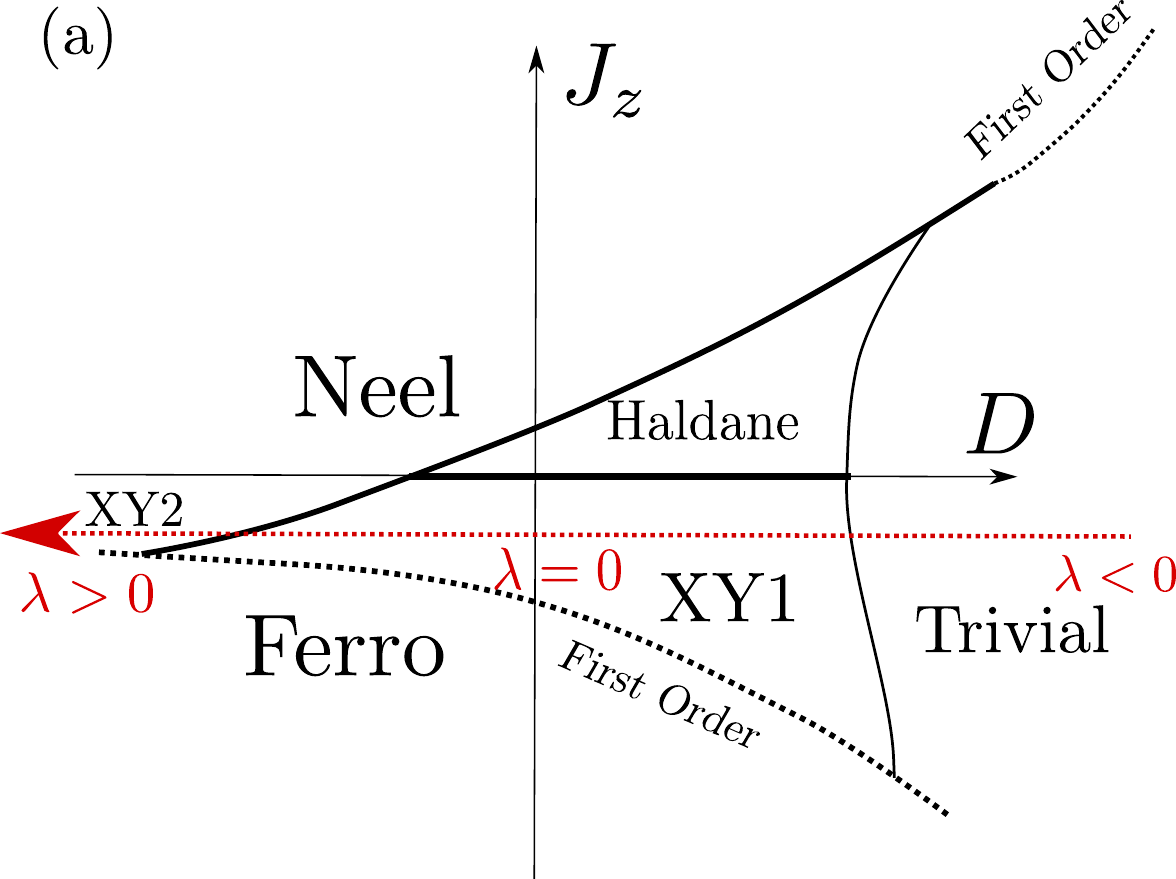}
		\includegraphics[width=0.6\columnwidth]{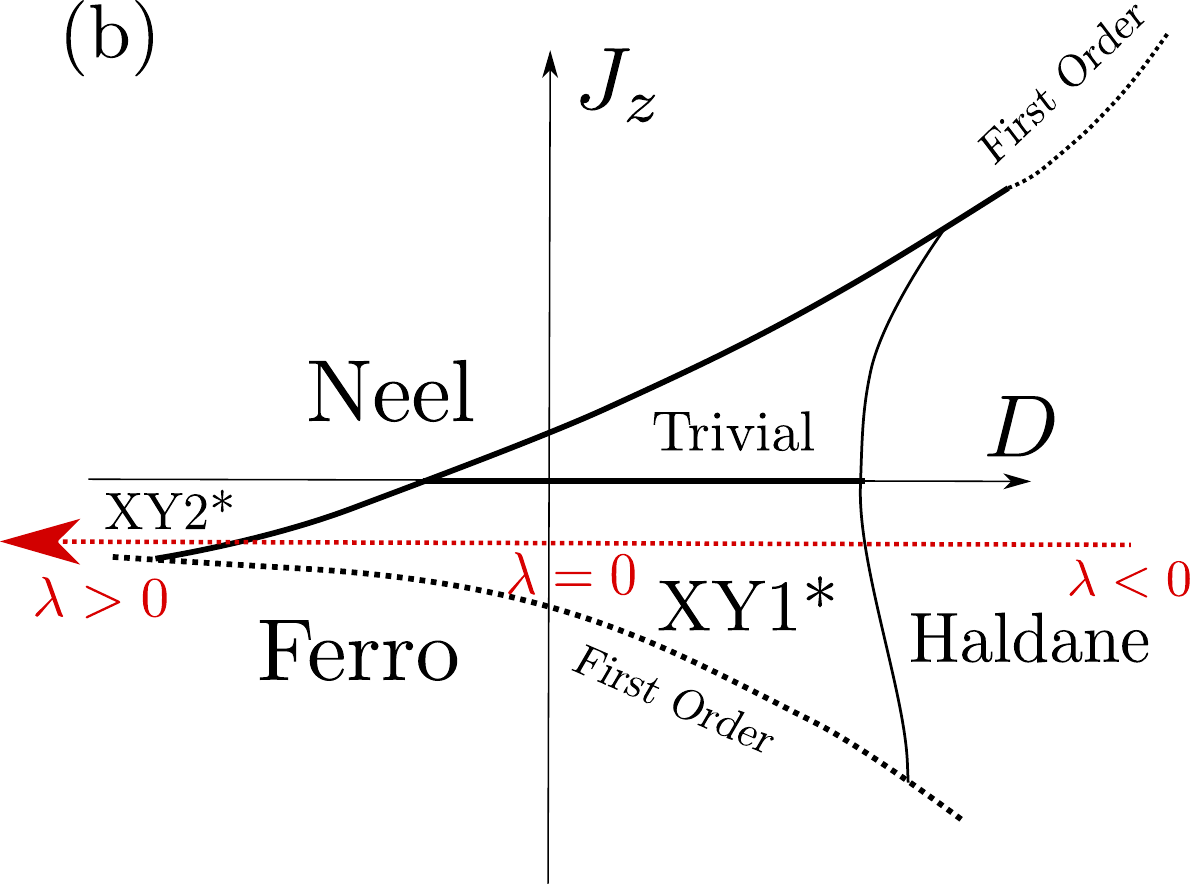}
		\caption{ Schematic phase diagrams of the spin 1 XXZ chain Hamiltonians shown in \cref{eq:Heff_Spin1,eq:Heffp_Spin1_OBC} applicable to the limits $t \sim -1$ (top) and $t \sim +1$ (bottom) of \cref{eq:Hamiltonian_largeJ} whose phase diagram is shown in \cref{fig:phasedig}.}
		\label{fig:spin1map}
	\end{figure}

	The Hamiltonian in \cref{eq:Heff_Spin1} is the familiar spin-1 XXZ model with uniaxial single-ion-type anisotropy whose phase diagram is known~\cite{ChenHidaSanctuary_Spin1XXZ_PhysRevB.67.104401} and is schematically reproduced in \cref{fig:spin1map}. For the parameter regime close to $t \approx -1$, the phases and transitions of the Hamiltonian in \cref{eq:Hamiltonian_largeJ} are qualitatively reproduced by that of \cref{eq:Heff_Spin1}. For example, consider the limit $t \rightarrow -1$ when \cref{eq:Heff_Spin1} reduces to 
	\begin{multline}
		H_{eff} \rightarrow \frac{J}{2} \sum_j \left(L^x_j L^x_{j+1} + L^y_j L^y_{j+1} -\Delta L^z_j L^z_{j+1}\right)  \\ +  4 (1-\lambda) \sum_j \left(L^z_j\right)^2. \label{eq:Heff_tminus1}
	\end{multline}
	If $\Delta$ is fixed to a small value, as $\lambda$ is tuned, we see from \cref{fig:spin1map} that \cref{eq:Heff_tminus1} passes through  the large-D (trivial), XY$_1$, XY$_2$ and the Ferromagnetic phases-- the same as what is seen in \cref{fig:phasedig}. It is worth emphasizing the crucial role of $\Delta$ which builds residual ferromagnetic correlations between effective spin-1s, thus leading to the realization of interesting gapless phases. Through the spin-1 mapping we are able to see that in order to access the XY$_2$ phase, we \emph{need} to fix $\Delta$ to be small as was done in our numerical investigations.

	Let us now consider the limit when the Hamiltonian \cref{eq:Hamiltonian_largeJ} is dominated by $H_\perp'$. First, let us observe that with periodic boundary conditions, $t \mapsto -t$ is induced by a unitary transformation generated by a single-site translation on one of the legs of the ladder $\vec{S}_{1j} \mapsto \vec{S}_{1j+1}$. As a result, the phase diagram for \cref{eq:Hamiltonian_largeJ} is perfectly symmetric under $t \mapsto -t$. The identity of the phases, however, can change under this map. In particular, the unitary transformation is ill defined with open boundary conditions and therefore it is conceivable that the distinction between the regions related by $t \mapsto -t$, is topological in nature. We will now map the $H_\perp'$ dominant Hamiltonian to a spin 1 chain. To do this, we repeat the steps above and observe that with periodic boundary conditions, $H_\perp'$ decouples into disjoint pieces, each of which has support on two spins, this time living on the diagonal bonds as schematically shown in the bottom figure of \cref{fig:Spin1Map}. We again perform a convenient change of basis similar to \cref{eq:Hperp_basischange} to get the following local term
	\begin{equation}
		h'_\perp = -(1+t)  \left(S^x_{2 j} S^x_{1 j+1}+S^y_{2 j} S^y_{1 j+1}+\lambda S^z_{2 j} S^z_{1 j+1} \right). \nonumber
	\end{equation}
	This is easily diagonalized as 
	\begin{multline}
		h'_\perp = (1+t) \Big[(\lambda +2) \outerproduct{s}{s}+(\lambda -2) \outerproduct{0}{0} \\ -  \lambda \left(\outerproduct{+1}{+1}+\outerproduct{-1}{-1}\right)\Big] 
	\end{multline}
	where $\ket{\pm 1}, \ket{0}$ and $\ket{s}$ are as defined as in \cref{eq:local_spin1}. Projecting onto the low-energy Hilbert space spanned by $\ket{\pm1},\ket{0}$ on each diagonal bond, we again get an effective spin-1 chain with the following Hamiltonian
	\begin{multline}
		H'_{eff} = J'_{xy} \sum_{\tilde{j}} \left(L^x_{\tilde{j}} L^x_{\tilde{j}+1} + L^y_{\tilde{j}} L^y_{\tilde{j}+1}\right) \\+ J'_z \sum_{\tilde{j}} L^z_{\tilde{j}} L^z_{\tilde{j}+1}  +   D' \sum_{\tilde{j}} \left(L^z_{\tilde{j}}\right)^2 \label{eq:Heffp_Spin1}
	\end{multline}
	with
	\begin{multline}
		J'_{xy} = \left(\frac{J}{2} - \frac{(1-t)}{4}\right), ~J'_z =- \Big( \frac{J \Delta}{2}  +  \frac{ \lambda (1-t)}{4}\Big),\\D' =  2(1+t) (1-\lambda). \label{eq:Spin1parameter_map_prime}
	\end{multline}
	We have denoted the bond between spins $(2,j)$ and $(1,j+1)$  by $\tilde{j}$. So far, \cref{eq:Heffp_Spin1} looks identical to \cref{eq:Heff_Spin1} with the replacement $t\mapsto -t$. However, a change occurs with open boundary conditions. There is no natural association of the boundary qubits with any diagonal bond. As a result, it survives the  the projection and remains as a qubit on the ends of the chain. The effective Hamiltonian with open boundary conditions is thus
	
	\begin{multline}
		H'_{eff} = J'_{xy} \sum_{\tilde{j}=1}^{L-1} \left(L^x_{\tilde{j}} L^x_{\tilde{j}+1} + L^y_{\tilde{j}} L^y_{\tilde{j}+1}\right)\\+ J'_z \sum_{\tilde{j}=1}^{L-1} L^z_{\tilde{j}} L^z_{\tilde{j}+1} +  D' \sum_{\tilde{j}=1}^{L} \left(L^z_{\tilde{j}}\right)^2 
		+H^\partial.
		\label{eq:Heffp_Spin1_OBC}  
	\end{multline}
	where $J'_{xy}, J'_z$ and $D'$ are the same as in \cref{eq:Spin1parameter_map_prime}. $H^\partial$ is the effective boundary Hamiltonian,
	\begin{multline}
		H^\partial = J^\partial_{xy}  \left(S^x_{11} L^x_{\tilde{1}} + S^y_{11} L^y_{\tilde{1}} +  L^x_{\tilde{L}} S^x_{2L+1} +  L^y_{\tilde{L}} S^y_{2L+1}\right) \\+ J^\partial_z \left( S^z_{11} L^z_{\tilde{1}} + L^z_{\tilde{L}}  S^z_{2L+1} \right)
	\end{multline}
	where the coupling constants to the boundary qubits  $\vec{S}_{11}$ and $\vec{S}_{2L+1}$ are
	\begin{equation}
		J^\partial_{xy} \equiv \left(\frac{J}{2} - \frac{(1-t)}{2}\right),~J^\partial_z =- \left( \frac{J \Delta}{2} +  \frac{ \lambda (1-t)}{2}\right). \nonumber
	\end{equation}
	The picture above suggests an interesting alternative method of analysis to the abelian bosonization of \cref{sec:Bosonization1} by treating the boundary spin 1/2 as a quantum impurity~\cite{Affleck_QuantumImpurity}, however,  we will not pursue this route in this work and leave it for future work. 
	
	Let us make a few comments on the limitations and utility of the mapping to a spin 1 chain before we proceed to a discussion of the phases in the effective Hamiltonian for the $t \sim 1$ limit. Recall that for the $t \sim 1$ limit, the phase diagram for the spin 1 XXZ chain accurately reproduces the phases of the spin ladder. To identify the phases of the spin 1 XXZ with that of \cref{eq:Hamiltonian_largeJ} in the $t \sim -1$ limit, we need additional tools, although plausible arguments can be made, especially for the gapped phases. For instance, it is clear that the identity of the Ferromagnet obtained for large $\lambda$ remains the same in \cref{eq:Heffp_Spin1_OBC,eq:Heff_Spin1} as can be easily seen by taking $\lambda$ to a large value in \cref{eq:Hamiltonian_largeJ}. The identities of the large-D and Haldane phase in \cref{eq:Heff_Spin1} are reversed in \cref{eq:Heffp_Spin1_OBC} and can be understood from the effect of additional end qubits appearing with open boundary conditions. On the one hand, the qubit hybridizes with the edge mode of the Haldane phase and gaps out the edge degeneracy, rendering it a trivial phase. On the other hand, the same qubits contribute to the edge degeneracy to the large D phase where the gapped bulk protects the hybridization between qubits on opposite ends of the chain, thus converting it to a topological phase. The effect of the qubits on the gapless phases is not straightforward to determine. One could extend the previous argument to justify the mapping of the XY$_2$ phase to the topological XY$_2^*$ phase, which has edge modes, but the absence of a bulk gap makes it heuristic at best. Indeed, the mapping of XY$_1$ to a different gapless phase XY$_1^*$ which does not have edge modes, is not easily explained within the spin 1 mapping. We need more sophisticated tools, such as bosonization and numerical analysis, to nail down the precise identity and nature of gapless phases, as has been achieved in the previous sections.
	
	In summary, spin 1 mapping presents an independent confirmation of distinct phases in the limits $t \sim 1$. It also guides us to fix the parameters to open up various gapless phases, especially XY$_2$. It also confirms that the topology of the $t \sim -1$ phase diagram is identical to that of $t \sim +1$. However, additional analysis, as has been shown in the previous sections, is needed to determine the identity of phases in the latter limit although heuristic arguments are consistent with detailed analysis.

	\section{Summary and Outlook}
	
	In this work we have studied a coupled spin model hosting several symmetry enriched gapless phases that exhibit an intricate interplay of symmetries, strong correlations, and topological features.  Our multipronged approach, which includes bosonization (\cref{sec:Bosonization1,sec:Bosonization2}), DMRG studies (\cref{sec:Numerics}) and effective low-energy modelling (\cref{sec:Spin1}) provides a comprehensive understanding of all aspects of the phase diagram. Our study points out that even the well-known Luttinger liquid state can appear in the form of distinct phases based on how the microscopic UV symmetries inherited from the underlying spin model get reflected in the low-energy IR (see \cref{sec:symmetry_distinction}).  Among these phases is an interesting {\it gapless} topological phase XY$^*_2$ that hosts symmetry-protected edge modes. Finally, our mapping to a Spin 1 XXZ chain (\cref{sec:Spin1}) provides an alternative view point to understand the nature of the gapless phases and their transitions. We also find the presence of multiple stable universality classes -- `multiversality' along the critical surface separating the gapped trivial and Haldane phases.
	
	There are many generalizations that can follow from our work. First, it would be useful to use more sophisticated tools of boundary CFT~\cite{ParkerScaffidiVasseur_TLL_DDW_PhysRevB.97.165114,YuConformalBCPhysRevLett.129.210601} to gain insight into the gapless phases seen in this work. Second, although in this work we have focused on a two-chain ladder, we believe that as the number of chains increases, a much wider variety of symmetry-enriched criticality may be realizable in such systems, leading to a host of unique gapless phases and  transitions~\cite{Cabra_manyN_PhysRevB.58.6241,Cabra_manyN_PhysRevLett.79.5126}. Another interesting direction is to couple such one-dimensional chains to realize possibly novel two-dimensional gapless states \cite{Kane_SLLPhysRevB.64.045120,Sondhi_SLLPhysRevB.63.054430,Kivelson_SLL_PhysRevLett.85.2160,Vishwanath_SLL_PhysRevLett.86.676} mimicking the success of gapped topological phases \cite{YakovenkoPRB1991, NeupertPRB2014, Meng_PRB_2015, liu2023assembling}. Finally, it would be interesting to see if the symmetry enriched gapless phenomena investigated in this work can be observed in Rydberg simulators~\cite{RydbergReview10.1116/5.0036562} where other gapless phenomena have been postulated to exist~\cite{TitasRydbergPhysRevB.105.155159,PaulLuisa_Rydberg_2023xxz,PaulLuisa_Rydberg_eck2023critical,Titas_RydbergPhysRevB.106.155411}. We leave these and other questions to future work.

	\section*{Acknowledgements}
	We thank  Masaki Oshikawa, Siddharth Parameswaran, Nick Bultinck, Sounak Biswas, Michele Fava, Nick Jones, Yuchi He and Diptarka Das for useful discussions. We are especially grateful to Fabian Essler for collaboration in the early stages of this work. During the initial stages of this work, A.P. was supported by a grant from the Simons Foundation (677895, R.G.) through the ICTS-Simons  postdoctoral fellowship. He is currently funded by the European Research Council under the European Union Horizon 2020 Research and Innovation Programme, Grant Agreement No. 804213-TMCS  and the Engineering and Physical Sciences Research Council, Grant number EP/S020527/1. SM acknowledges funding by the Deutsche Forschungsgemeinschaft (DFG, German Research Foundation) –  436382789, 493420525 via large-equipment grants (GOEGrid cluster). AA acknowledges support from IITK Initiation Grant (IITK/PHY/2022010). TM acknowledges support from DST-SERB, India through Project
	No. MTR/2022/000382. The authors acknowledge the hospitality of ICTS-TIFR, Bangalore, and ICTP, Trieste, where discussions and parts of this work were carried out. 
	
The data shown in the figures will be made available as ancilla files on the arxiv.

\begin{appendix}

		\section{Additional bosonization details}
		\label{app:bosonization}
		The subject of bosonization has been extensively discussed in several excellent books and reviews. In this appendix, we review a few details that are subtle and are easy to miss. The CFT term in \cref{eq:H_bosonize_smallJ,eq:H_bosonize_largeJ} is determined using standard techniques~\cite{Giamarchi} from the XXZ Hamiltonian. The various perturbations can be determined from the bosonized form of the spin operators shown in \cref{eq:Bosonization_single,eq:Bosonization_ladder} in a straightforward manner for the most part. Cases involving coincident field operators should be treated with care employing a `point-splitting' device to determine how coincident vertex operators are multiplied. Let us review this in the single component/ small $J$ limit:
		\begin{multline}
			e^{i  m \phi(x)} e^{i n \theta(x)} = \lim_{\epsilon \rightarrow 0}   e^{i  m \phi(x+\epsilon)} e^{i n \theta(x - \epsilon)}\\= \lim_{\epsilon \rightarrow 0} e^{i \left(m \phi(x + \epsilon) +n \theta(x-\epsilon) \right)}   e^{-\frac{mn}{2} [\phi(x+\epsilon),\theta(x - \epsilon)]} \\= \lim_{\epsilon \rightarrow 0} e^{i n \theta(x-\epsilon)}  e^{i  m \phi(x + \epsilon)}  e^{-mn [\phi(x+\epsilon),\theta(x - \epsilon)]}.
		\end{multline}
		
		This is determined using an integrated version of \cref{eq:KacMoody_singlecomponent,eq:KacMoody_2component}
		\begin{align}
			[\phi_\alpha (x), \theta_\beta (x') ] &= i \pi  \delta_{\alpha\beta} \text{sgn}(x-x'), \nonumber\\
			~[\phi (x), \theta (x') ] &= i \pi   \text{sgn}(x-x'). \label{eq:KacMoodyIntegrated}
		\end{align}
		using which we get
		\begin{equation}
			e^{i  m \phi(x)} e^{i n \theta(x)} = (-1)^{mn} e^{i n \theta(x)} e^{i  m \phi(x)}. \label{eq:VertexCommutation}
		\end{equation}
		\Cref{eq:VertexCommutation} is needed to obtain the correct bosonized form for operators involving products of $S^\pm$ such as the bond-dimerization term $ \propto \sum_j \left((-1)^j S^+_j S^-_{j+1} + h.c. \right)$ in \cref{eq:Hamiltonian_smallJ}. Another important place where point splitting is needed is in determining the correct symmetry action. The $U(1), \ztwo^R$ and $\bZ$ actions are easy to read off by directly comparing the action on the lattice operators shown in \cref{tab:symmetries} with \cref{eq:Bosonization_single,eq:Bosonization_ladder}. The action of lattice parity $\ztwo^R$ on the bosonized variables, on the other hand, needs some care. Let us review this again in the small $J$, single component version. Recall that the action of $\ztwo^P$ is bond inversion, which can be thought of as a composite of site inversion and single-site translation. Since translation is straightforward by direct comparison, let us focus on site inversion $\vec{S}_j \mapsto \vec{S}_{-j}$. On the continuum operators and simple vertex operators, this naively acts as 
		\begin{equation}
			\phi(x) \mapsto \phi(-x), \theta(x) \mapsto \theta(-x). \label{eq:Parity_naive}
		\end{equation}
		Let us look at how this naive action is reflected on products of non-commuting operators,
		\begin{multline}
			e^{im \theta(x) }  e^{i n\phi(x)}  =  \lim_{\epsilon \rightarrow 0} e^{i \frac{mn\pi}{2} \text{sgn}(\epsilon)} e^{i \left(m\theta(x-\epsilon)+ n\phi(x+\epsilon)  \right)}\\
			\mapsto \lim_{\epsilon \rightarrow 0} e^{i \frac{mn\pi}{2} \text{sgn}(\epsilon)} e^{i \left(m\theta(-x+\epsilon)+ n\phi(-x-\epsilon)  \right)}\\ = \lim_{\epsilon \rightarrow 0} e^{i mn\pi  \text{sgn}(\epsilon)} e^{i m\theta(-x+\epsilon)} e^{in\phi(-x-\epsilon)  } \\
			= (-1)^{mn} e^{im \theta(-x) }  e^{i n\phi(-x)}. \label{eq:Parity_vertexproduct}
		\end{multline}
		Using \cref{eq:Parity_naive,eq:Parity_vertexproduct} we get
		\begin{align}
			S^{\pm}_{ -j} &\approx \exp{\left(\pm i \theta(-x) \right)}\left(  (-1)^j \A~  -\C \cos \phi(x) + \ldots \right),  \nonumber \\		
			S^z_{-j} &\approx \frac{1}{2 \pi} \partial_x \phi(-x) + (-1)^j \B \sin \phi(-x) + \ldots \label{eq:Bosonization_single_parity}
		\end{align}	
		We can now read off the symmetry action corresponding to site reflection from \cref{eq:Bosonization_single_parity} as
		\begin{equation}
			\phi(x) \mapsto \pi - \phi(x),~\theta(x) \mapsto \theta(-x). \label{eq:site_reflection}
		\end{equation}
		Combining \cref{eq:site_reflection} with the action of translation shown in \cref{tab:symmetries_bosonization}, we get the final effective action of $\ztwo^R$ shown in \cref{tab:symmetries_bosonization}.

		\section{Phase diagrams from bosonization}
		\label{app:phasedig}
		In this appendix, we use bosonization to obtain the qualitative details of the phase diagrams shown in the main text in both the small and large $J$ limits. 
		\subsection{The small-$J$ phase diagram}
		\begin{figure}[!ht]
			\centering
			\begin{tabular}{c}
				\includegraphics[width=0.45\textwidth]{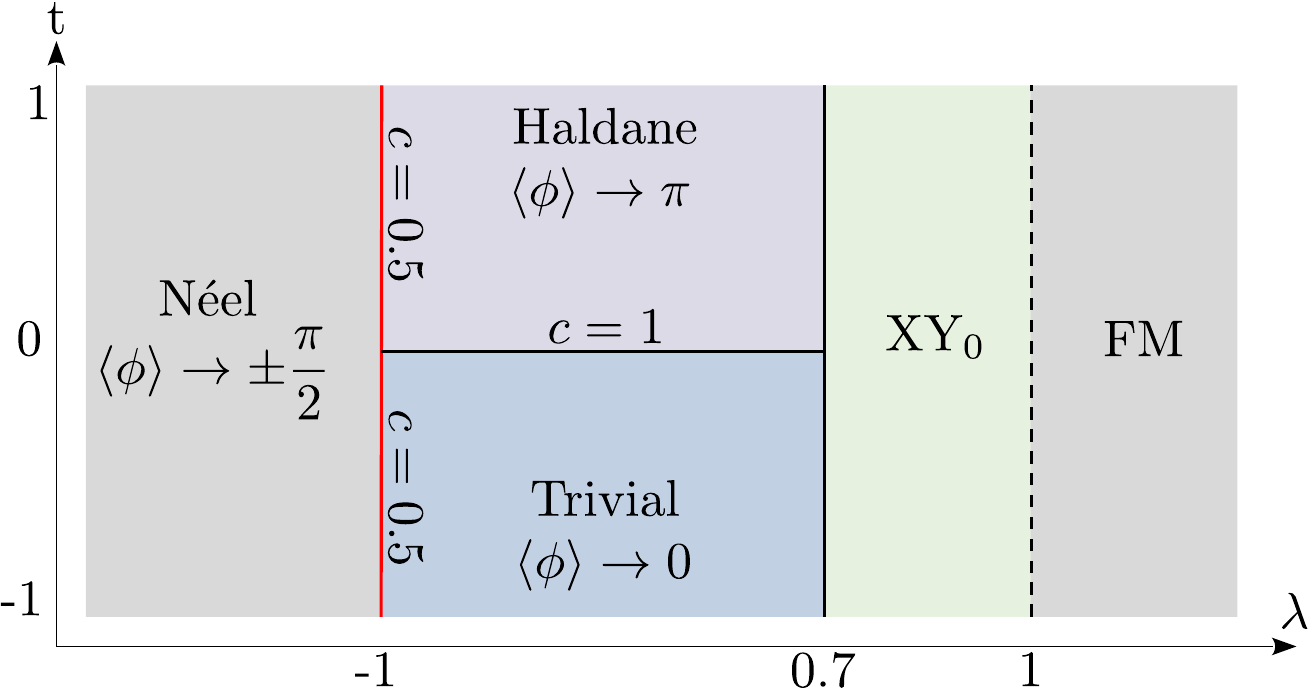} 
			\end{tabular}
			\caption{The small-$J$ phase diagram as determined from bosonization.}
			\label{fig:phasedig_smallJ_bosonization}
		\end{figure}
		Let us write down the form of the Hamiltonian at small $J$ shown in \cref{eq:Hamiltonian_smallJ}
		\begin{multline}
			H = \sum_j \left(1+ (-1)^j t\right) \left( S^x_{ j} S^x_{ j+1} +  S^y_{j} S^y_{j+1} - \lambda  S^z_{j} S^z_{ j+1} \right)\\ + J \sum_j \left( S^x_{ j} S^x_{ j+2} +  S^y_{j} S^y_{j+2} - \Delta  S^z_{j} S^z_{ j+2} \right), \label{eq:Hamiltonian_smallJ_appendix}
		\end{multline}
		and its bosonized version shown in \cref{eq:H_bosonize_smallJ},
		\begin{multline}
			H  \approx  \frac{v}{2 \pi} \int dx \left[\frac{1}{4K} \left(\partial_x \phi\right)^2 + K \left(\partial_x \theta\right)^2\right] \\+ 2 \mathcal{A} \mathcal{C} t  \int dx~    \cos \phi  - \frac{\mathcal{B}^2 \lambda}{2}   \int dx~    \cos 2\phi  + \ldots \label{eq:H_bosonize_smallJ_appendix}
		\end{multline}
		The Luttinger parameter $K$ and velocity $v$ depend on Hamiltonian parameters and can be determined from the Bethe ansatz solution of the XXZ spin chain~\cite{Haldane_BetheLL_1981153} 
		\begin{equation}
			K = \frac{\pi}{2 \arccos \lambda}, ~ v = \frac{K}{(2K-1)} \sin \left(\frac{\pi}{2K}\right). \label{eq:Bethe_Luttinger_single}
		\end{equation}
		Let us comment on a few limits of \cref{eq:Hamiltonian_smallJ_appendix}. If we switch off both the nnn coupling $J$ and dimerization $t$, we have the XXZ model, which can be solved by Bethe ansatz~\cite{XXZ_Bethe_Orbach_PhysRev.112.309,XXZ_Bethe_YangYang1PhysRev.150.321,XXZ_Bethe_YangYang2PhysRev.150.327} with the phases shown in the $t=0$ line of the figure in \cref{fig:phasedig}. The phase diagram with $t \neq 0$  and $J \neq 0$ can be easily understood as a perturbation of the XXZ spin chain~\cite{KohmotoNijsKadanoff_XXZDimerization_PhysRevB.24.5229} using the bosonized Hamiltonian shown in \cref{eq:H_bosonize_smallJ_appendix}. This is done by tracking the relevance (in the RG sence) of the two vertex operators $\cos \phi$ and $\cos 2 \phi$ which have scaling dimensions $K$ and $4K$, respectively, as follows:
		
		\emph{The XY$_0$ phase:} In the regime when $K>2$, which corresponds to $\frac{1}{\sqrt{2}} < \lambda <1$ from the formula in \cref{eq:Bethe_Luttinger_single}, both $\cos \phi$ and $\cos 2 \phi$ are irrelevant, and we get a gapless phase, XY$_0$.
		
		\emph{The Haldane and Trivial phases:} When $\frac{1}{2} < K < 2$ which corresponds to $-1 < \lambda < \frac{1}{\sqrt{2}}$, $\cos\phi$ is relevant while $\cos2\phi$ is irrelevant. Therefore, we get gapped phases for $t \neq 0$ where $\moy{\phi} \rightarrow \pi$ for $t >0$ corresponds to the Haldane phase and $\moy{\phi} \rightarrow 0$ for $t < 0$ corresponds to the trivial phase. 
		
		\emph{The N\'{e}el phase:} When $K<\frac{1}{2}$ which corresponds to $\lambda < -1$, both $\cos \phi$ and $\cos 2 \phi$ are relevant. When $\cos2\phi$ is dominant (eg: when $t=0$), we get a N\'{e}el phase with $\moy{\phi} \rightarrow \pm \frac{\pi}{2}$. The transition between the Haldane/ trivial phase and N\'{e}el phase is second-order and corresponds to the Ising universality class. See \cite{DelfinoMussardo_1998675} for an explanation of this. 
		
		\emph{The Ferromagnet:} As $\lambda \rightarrow 1$, we get $K \rightarrow \infty$ and $v \rightarrow 0$ and the Luttinger liquid description becomes invalid as the system transitions to a ferromagnet through a first-order transition.
		
		Putting these various pieces together, we reproduce the topology of the small-$J$ phase diagram seen for small $t$. This is shown in \cref{fig:phasedig_smallJ_bosonization}. 
		
		\subsection{The large-$J$ phase diagram}
		\begin{figure}[!ht]
			\centering
			\begin{tabular}{c}
				\includegraphics[width=0.49\textwidth]{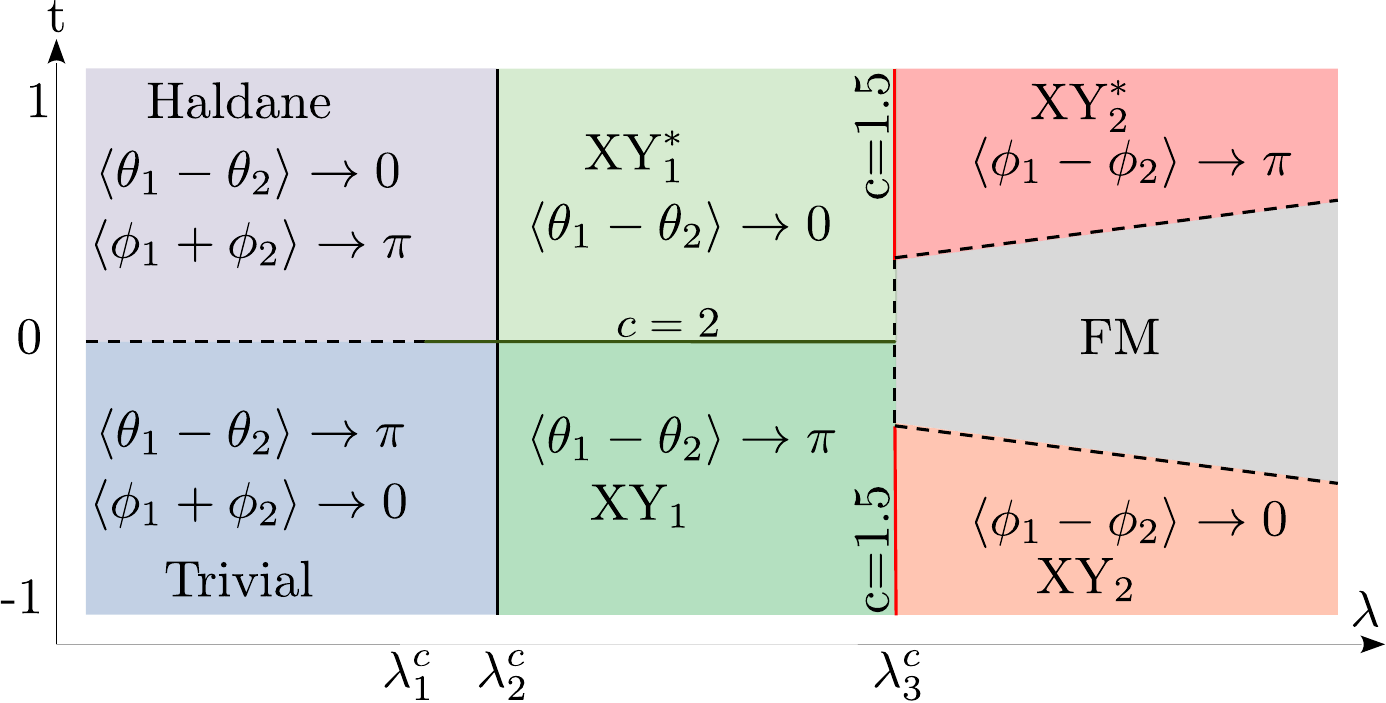} 
			\end{tabular}
			\caption{The large-$J$ phase diagram as determined from bosonization.}
			\label{fig:phasedig_largeJ_bosonization}
		\end{figure}
		Let us now write down the Hamiltonian form appropriate for large-$J$
		\begin{align}
			H &= H_1 + H_2 + H_\perp + H'_\perp,  \label{eq:Hamiltonian_largeJ_appendix} \text{ where, } \\
			H_{\alpha} &=  J \sum_j \left( S^x_{\alpha j} S^x_{\alpha j+1} +  S^y_{\alpha j} S^y_{\alpha j+1} - \Delta  S^z_{\alpha j} S^z_{\alpha j+1} \right), \nonumber\\ 
			H_\perp &=  (1-t) \sum_j \left(S^x_{1 j} S^x_{2 j}+S^y_{1 j} S^y_{2 j}-\lambda S^z_{1 j} S^z_{2 j} \right), \nonumber \\
			H'_\perp &=  (1+t) \sum_j \left( S^x_{2 j} S^x_{1 j+1}+ S^y_{2 j} S^y_{1 j+1}-\lambda  S^z_{2 j} S^z_{1 j+1} \right),  \nonumber
		\end{align}
		and its bosonized form
		\begin{multline}
			H \approx \frac{v}{2 \pi} \sum_{\alpha = 1,2} \int dx \left(\frac{1}{4K}( \partial_x \phi_\alpha)^2 + K ( \partial_x \theta_\alpha)^2\right) \\- \frac{\lambda}{2 \pi^2} \int dx~   \partial_x \phi_1 \partial_x \phi_2 -4\mathcal{A}^2t \int dx~   ~\cos \left(\theta_1-\theta_2\right) \\
			- \mathcal{B}^2  t \int dx~  \lambda~ \left(\cos \left(\phi_1+\phi_2\right) - \cos \left(\phi_1-\phi_2\right)\right) \\ +2 \mathcal{C}^2 \int dx \cos\left(\theta_1 - \theta_2\right) \cos\left(\phi_1 + \phi_2\right)+ \ldots \label{eq:H_bosonize_largeJ_appendix}
		\end{multline}	
		We now reproduce qualitative features of its diagram shown in \cref{fig:phasedig}. We will focus on the phases surrounding the $c=2$ line over which we have good analytical control. The leading term in \cref{eq:H_bosonize_largeJ_appendix} is a c=2 CFT of two identical compact bosons. The operator $\partial_x \phi_1 \partial_x \phi_2$ has scaling dimensions $2$ and is therefore exactly marginal. It generates motion in the space of $c=2$ CFTs where the compact bosons are no longer identical and have different compactification radii. We also have operators $\mathcal{V}_\pm \equiv \cos \left( \phi_1 \pm \phi_2\right)$, $\mathcal{W}_- \equiv \cos \left(\theta_1 - \theta_2\right)$ and $ \cW_- \cV_+\equiv \cos\left(\theta_1 - \theta_2\right) \left(\phi_1 + \phi_2\right)$ whose scaling dimensions can be obtained perturbatively to the leading order in $\lambda$  as ~\cite{Giamarchi}
		\begin{align}
			[\mathcal{V}_\pm] &=  K_\pm\approx 2K~ \left(1 \pm \frac{\lambda K}{ \pi v }\right)\nonumber ,\\~~[\mathcal{W}_-] & = \frac{1}{K_-}\approx \frac{1}{2K}~ \left(1 +\frac{\lambda K}{ \pi v }\right) \text{ and  } \nonumber \\  [ \cW_- \cV_+] &=  \frac{1}{K_-} + K_+ \approx \left(\frac{1}{2K} + 2K\right) \left(1 +\frac{\lambda K}{ \pi v }\right) \label{eq:Scaling_dim_perturbative_largeJ}
		\end{align}
		where, again, relationship of the Luttinger parameter $K$ and velocity $v$ with the parameters in the Hamiltonian is determined from the Bethe ansatz solution of the XXZ spin chain~\cite{Haldane_BetheLL_1981153} as
		\begin{equation}
			K = \frac{\pi}{2 \arccos \Delta}, ~ v = \frac{JK}{(2K-1)} \sin \left(\frac{\pi}{2K}\right). \label{eq:Bethe_Luttinger}
		\end{equation}
		
		Note that we have $[\cV_-] [\cW_-] = 1$. As a result, it is impossible for both $\cV_-$ and $\cW_-$ to be irrelevant at the same time. Consequently, for any $t \neq 0$, the $c=2$ theory is unstable and flows a gapless phase with $c<2$ or a gapped phase ~\cite{StrongMillis_Spinladder_PhysRevLett.69.2419,OrignacGiamarchi_SpinLadder_PhysRevB.57.5812,Giamarchi} as seen in \cref{fig:phasedig}.

		\subsubsection{The phases and transitions}
		Let us begin in the limit $t \rightarrow 0$ in \cref{eq:H_bosonize_largeJ_appendix} when $\cV_+ \cW_-$ is irrelevant, giving us a $c=2$ theory.  Recall that one of the two operators $\cW_- \equiv \cos\left(\theta_1 - \theta_2\right)$ or  $\cV_- \equiv \cos\left(\phi_1 - \phi_2\right)$ is always relevant and, therefore, for $t \neq 0$, the theory flows to either a gapless state with $c<2$ or gaps out completely. We are interested in the case where the system does not gap out completely which occurs when $\cV_+ \equiv \cos \left( \phi_1 + \phi_2\right)$ is irrelevant and the theory flows to effective single-component Luttinger liquid  gapless phases. The nature of the phase depends on (i) which among $\cW_-$ and $\cV_-$ dominates at large distances, pinning $\theta_1 - \theta_2$ or $\phi_1 - \phi_2$ and (ii) the sign of $t$ which determines the value to which the fields are pinned $\moy{\theta_1 - \theta_2} = 0/\pi$ or $\moy{\phi_1 - \phi_2} = 0/\pi$. We label these four cases XY$_{1,2}$ and XY$^*_{1,2}$ as shown in \cref{fig:phasedig_largeJ_bosonization}. All four are distinct phases. The universality class of a direct continuous transition between XY$_{1/2}$ and XY$^*_{1/2}$ is the parent $c=2$  theory obtained by tuning $t\rightarrow 0$. The transition between XY$_1$ and XY$_2$ or between XY$^*_1$ and XY$^*_2$  corresponds to a compact boson plus Ising CFT with central charge $c= \frac{3}{2}$\cite{LECHEMINANT_SelfDualSineGordon_2002502,Schulz_Higherspinbosinization_PhysRevB.34.6372,AP2022multiversality}. In the parameter regime we study the model numerically, a direct transition between XY$_2$ and XY$_2^*$ is not observed.  
		
		When we are in the XY$_1$ or XY$_1^*$ phases where $\cW_-$ pins the value of $\theta_1 - \theta_2$, a transition to a gapped phase can occur when $\cV_+$ also becomes relevant. The gapped phases resulting when $\theta_1 - \theta_2$ and $\phi_1 + \phi_2$ are pinned correspond to the Haldane or trivial phase~\cite{OrignacGiamarchi_SpinLadder_PhysRevB.57.5812} as shown in \cref{fig:phasedig_largeJ_bosonization}. A different transition can occur when we are in any of the four gapless phases, XY$_{1,2}$ and XY$^*_{1,2}$ and the Luttinger velocity vanishes, resulting in a first-order transition to a FM similar to the single-component small-$J$ case. 
		
		\subsubsection{The $t=0$ line and its proximate phases}
		We now analyze the $t=0$ line and its proximity in detail. First, let us analyse which gapless phase results when $t\neq 0$ is switched on. This is determined by which operator $\cW_-$ or $\cV_-$ has the smaller scaling dimension. In the parameter regime we studied numerically, we only find the former situation as shown in \cref{fig:phasedig_largeJ_bosonization}. When $\cV_+$ becomes relevant along with $\cW_-$, we see that $t \neq 0$ results in gapped phases. Let us denote $\lambda^c_2$ as the location along the $t=0$ line when $\cV_+$ is marginal, i.e. $[\cV_+] = 2$ where the XY$_1$ to the trivial phase boundary and the XY$_1^*$ -to- Haldane phase boundary meets the $c=2$ line at $t=0$. 
		
		Now, as seen in \cref{eq:H_bosonize_largeJ_appendix}, the $c=2$ theory is destroyed by either (i) the composite operator $\cW_- \cV_+$ becomes relevant leading to a gapped state with two degenerate vaccua  $\moy{\phi_1 + \phi_2} =  \pi - \moy{\theta_1 - \theta_2} = 0/ \pi$  or (ii) the Luttinger velocity for one of the sectors vanishes rendering the continuum description invalid and we get a first-order transition to a FM. Let us denote the critical values of $\lambda$ that result in each of these as $\lambda^c_1$ and $\lambda^c_3$ respectively. From the perturbative result shown in \cref{eq:Scaling_dim_perturbative_largeJ}, we can get rough estimates for $\lambda^c_{1} - \lambda^c_{3}$ although these estimates are not very reliable when they result in large values of $|\lambda^c_k|$ where the validity of perturbation theory no longer holds.
		
		The nature of the phase transition between the trivial and Haldane phases that occurs at $t=0$ depends on whether we are at $\lambda < \lambda^c_1$ or $ \lambda^c_1 < \lambda < \lambda^c_2$. As shown in \cref{fig:phasedig_largeJ_bosonization}, the latter results in a first-order phase transition in which the vacua of the Haldane and the trivial phase are degenerate whereas the former results in a second-order transition with $c=2$. Putting all this together, we get the form shown in \cref{fig:phasedig_largeJ_bosonization}.
		
		\subsubsection{Multiversality}
		A curious observation is that although the small-$J$ and large-$J$ gapped Haldane and trivial phases are adiabatically connected, the nature of the second-order transitions between them is different at small-$J$ and large-$J$. For small $J$, it is a $c=1$ critical theory whereas for large-$J$ it is $c=2$. Both are obtained by tuning a single parameter and are therefore generic. This phenomenon, called multiversality, has received attention in recent studies~\cite{BiSenthil_Multiversality_PhysRevX.9.021034,AP2022multiversality} although microscopic models that exhibit them are rare.

		\subsubsection{A nice possible proximate phase diagram}
		
		\begin{figure}[!ht]
			\centering
			\begin{tabular}{c}
				\includegraphics[width=0.49\textwidth]{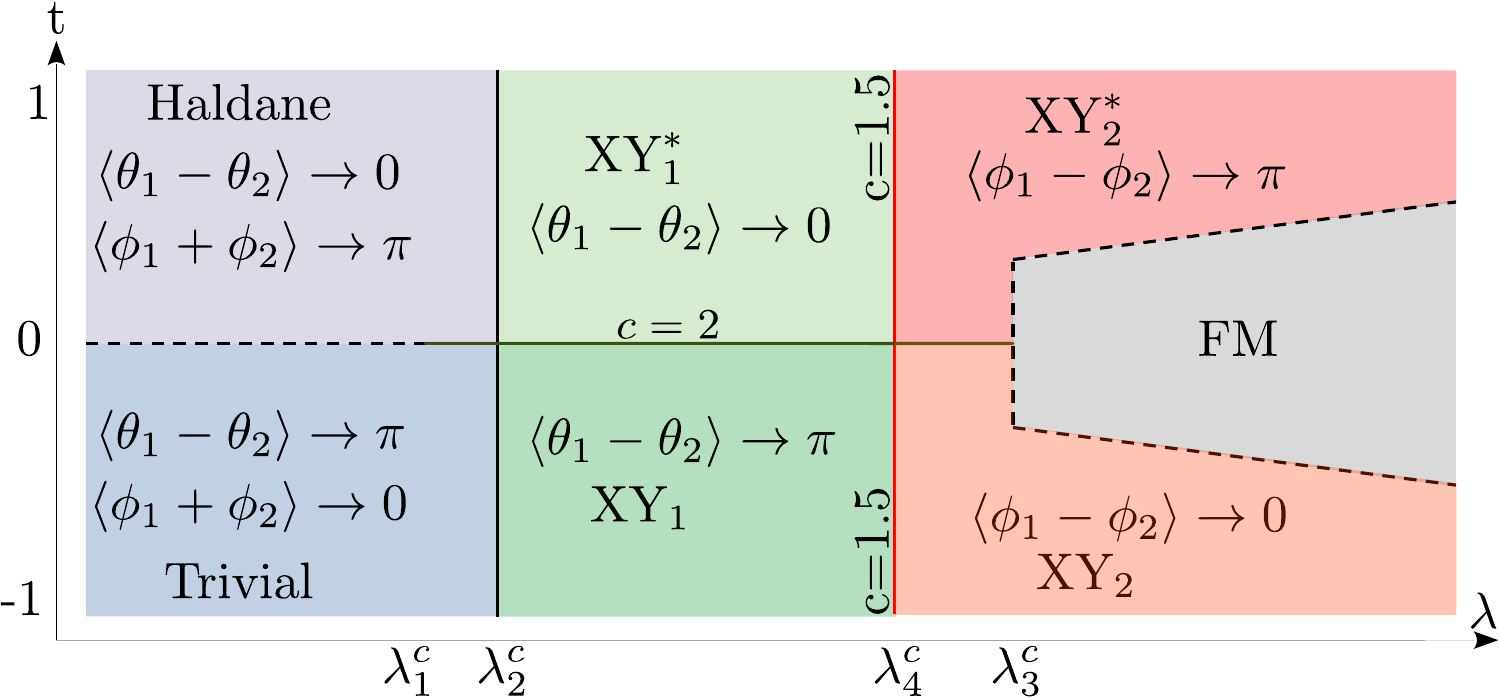} 
			\end{tabular}
			\caption{A nice proximate phase diagram at large $J$ suggested by bosonization.}
			\label{fig:phasedig_largeJ_proximate_bosonization}
		\end{figure}
		
		In the parameter regime when $\cV_+$ is irrelevant, we previously argued that close to the $t=0$ line $t \neq 0$ resulted in a gapless XY$_1$ or (t$<$0) XY$^*_1$ (t$>$0) phases if $[\cW_-] < [\cV_-]$ and XY$_2$ (t$<$0) or XY$^*_2$ (t$>$0) phases if $[\cW_-] > [\cV_-]$. If the c=2 theory survived as $[\cW_-] = [\cV_-]$ (at some putative value $\lambda^c_4$, say) then it would open a direct transition between the phases XY$_2$ and XY$^*_2$. The $c=\frac{3}{2}$ lines discussed previously that separated the phases XY$_1$ and XY$_2$ (t$<$0) and XY$_1^*$ and XY$_2^*$  (t$>$0) would meet the line $t=0$ at this point $\lambda^c_4$. Alternatively, the gapless theory becomes unstable before this can happen, giving us the situation shown in \cref{fig:phasedig_largeJ_bosonization} which we observe in our numerical investigation. We postulate that there is some proximate parameter regime of our microscopic Hamiltonian where $\lambda^c_3 > \lambda^c_4$ can be realized. In this case, we should see a phase diagram as shown in \cref{fig:phasedig_largeJ_proximate_bosonization} which contains all the same phases as in \cref{fig:phasedig_largeJ_bosonization} but also a direct transition between XY$_2$ and XY$^*_2$.

		\section{Bosonizing string operators}
		\label{app:String}
		\subsection{Bosonizing $C(x,y)$ for small $J$}
		Bosonizing string order parameters is known to be tricky and rife with ambiguities~\cite{Kim_StringBosonization_PhysRevB.62.14965,NAKAMURA_string}. Let us try to naively apply \cref{eq:Bosonization_single} to bosonize the string operator in \cref{eq:C} in the small -$J$ limit. 
		\begin{multline}
			C(x,y) \propto  e^{\pm  i \pi \sum_{l=x}^{y} S^z_{l} }  \sim e^{ \pm  \frac{i}{2} \int_x^y ds~ \partial_s\phi(s)} \\\sim  e^{ \pm \frac{i}{2} \left( \phi(x)- \phi(y)\right)}.  \label{eq:wrong_string_bosonize}
		\end{multline}
		\Cref{eq:wrong_string_bosonize} leads to the conclusion that $\moy{C(x,y)} \neq 0$ anytime $\moy{\phi} \neq 0$, in particular both in the Haldane and in the trivial phases. This is incorrect. We now use symmetries to identify the correct bosonized form of $C(x,y)$. We begin by postulating the following general bosonized form for $C(x,y)$
		\begin{align}
			C(x,y) &\sim C_L(x)~ C_R(y) \text{ where }, \label{eq:CLCR decomposition_smallJ} \\
			C_{L/R}(x) &\sim \sum_{m  \in \bZ} A^{L/R}_{m} ~e^{\frac{i}{2} m \phi(x) }.\label{eq:CLR expansion_smallJ}
		\end{align}
		While the form in \cref{eq:CLCR decomposition} appears as though the string operator $C(x,y)$ has been written in terms of local operators with support at $x$ and $y$, this is not so. The half-integer prefactor to the fields $\frac{\phi_\alpha}{2}$ ensures that the operators in $C_{L/R}$ are not part of the spectrum of local operators $\cX_{m,n}\equiv \exp\left(i \left(m\theta + n \phi\right)\right)$  and are therefore nonlocal. Furthermore, we have used the fact that $C(x,y)^2 = 1$ to restrict the coefficients to multiples of $\frac{1}{2}$. We now impose constraints on $A^{L/R}_{m}$ using symmetry. First, observe that the end-points of $C(x,y)$ defined in terms of spin operators as shown in \cref{eq:C} are charged under $\ztwo^R$ ($S^z_{j} \mapsto - S^z_{j}$).  Using the action of $\ztwo^R$ on the boson fields shown in \cref{tab:symmetries_bosonization}, we obtain a constraint on $A^{L/R}_{m}$ as
		\begin{equation}
			\ztwo^R: C_{L/R}\xrightarrow{\phi \mapsto -\phi} -C_{L/R} \implies A^{L/R}_{m} = -A^{L/R}_{-m}. \label{eq:Amn_Z2R_smallJ}
		\end{equation}
		We now impose the action of $\ztwo^P$ shown in \cref{tab:symmetries_orderparameters} on the bosonized form of $C(x,y)$    using \cref{tab:symmetries_bosonization} which gives a relationship between $A^L_{m}$ and $A^R_{m}$ as
		\begin{multline}
			\ztwo^P: C_L(x) \xrightarrow{\phi(x)\mapsto- \phi(-x)} C_R(-x) \\\implies A^R_{m} = ~A^L_{-m} = -A^L_{m}.\label{eq:Amn_Z2P_smallJ}
		\end{multline}
		Using \cref{eq:Amn_Z2R_smallJ,eq:Amn_Z2P_smallJ} in \cref{eq:CLR expansion_smallJ}, we get the final bosonized form for $C(x,y) \sim C_L(x) C_R(y)$ with $C_L(x) = - C_R(x)$ and 
		\begin{equation}
			C_{L} \sim \sum_{m \in \bZ^+} \alpha_m \sin \left(\frac{m\phi}{2}\right) \approx \alpha_1\sin \left(\frac{\phi}{2}\right).
		\end{equation}
		where the coefficients $\alpha_m$  are linear combinations of $A^{L/R}_{m}$. This correctly reproduces the numerically observed behaviour of $\moy{C(x,y)}$, which is nonzero when $\moy{\phi} = \pi$ such as in the Haldane phase but not when $\moy{\phi} = 0$ such as in the trivial phase.

		\subsection{Bosonizing $C(x,y)$ for large $J$}
		We now bosonize the string operator in the large-$J$ version. We follow the same line of reasoning as shown previously for the small $J$ version. Let us begin by attempting to bosonize $C(x,y)$ using the formulas shown in \cref{eq:Bosonization_ladder}:
		\begin{multline}
			C(x,y) \propto e^{i \pi \left(\sum_{l=x}^{y-1} S^z_{1,l}  +  \sum_{l=x+1}^{y} S^z_{1,l}\right)  }   \sim e^{ \frac{i}{2} \int_x^y ds~ \partial_s\left(\phi_1 + \phi_2\right)}\\\sim e^{ \frac{i}{2} \left( \phi_1(x)+ \phi_2(x)\right)} ~ e^{- \frac{i}{2} \left( \phi_1(y)+ \phi_2(y)\right)}. \label{eq:wrong_string_bosonize1}
		\end{multline}
		We may just as well have gone a different route to get
		\begin{multline}
			C(x,y) \propto e^{ i \pi \left(\sum_{l=x}^{y-1} S^z_{1,l}  -  \sum_{l=x+1}^{y} S^z_{1,l}\right) }  \sim e^{\frac{i}{2} \int_x^y ds~ \partial_s\left(\phi_1 - \phi_2\right)} \\\sim e^{ \frac{i}{2} \left( \phi_1(x)- \phi_2(x)\right)} ~ e^{- \frac{i}{2} \left( \phi_1(y)- \phi_2(y)\right)}.\label{eq:wrong_string_bosonize2}
		\end{multline}
		The bosonized expressions in \cref{eq:wrong_string_bosonize1,eq:wrong_string_bosonize2} lead to very different physics. We have $\moy{C(x,y)} \neq 0$  when $\moy{\phi_1 + \phi_2} \neq 0$ according to \cref{eq:wrong_string_bosonize1} and when $\moy{\phi_1 - \phi_2} \neq 0$ according to \cref{eq:wrong_string_bosonize2} which corresponds to very different phases as seen in \cref{fig:phasedig_largeJ_bosonization}. Now we use symmetries to write down the correct bosonized form of $C(x,y)$. We again begin by postulating the following form for $C(x,y)$
		\begin{align}
			C(x,y) &\sim C_L(x)~ C_R(y) \text{ where }, \label{eq:CLCR decomposition} \\
			C_{L/R}(x) &\sim \sum_{m , n \in \bZ} A^{L/R}_{m,n} ~e^{\frac{i}{2} \left(m \phi_1(x) + n \phi_2(x)\right) }.\label{eq:CLR expansion}
		\end{align}
		We now impose constraints on $A^{L/R}_{m,n}$ using symmetry. First, we use the fact that the end-points of $C(x,y)$ are charged under $\ztwo^R$ ($S^z_{\alpha j} \mapsto - S^z_{\alpha j}$).  Using the action of $\ztwo^R$ on the boson fields shown in \cref{tab:symmetries_bosonization}, we get 
		\begin{multline}
			\ztwo^R: C_{L/R}(x)\xrightarrow{\phi_\alpha \mapsto -\phi_\alpha} -C_{L/R}(x) \\\implies A^{L/R}_{m,n} = -A^{L/R}_{-m,-n}. \label{eq:Amn_Z2R}
		\end{multline}
		We now impose the action of $\ztwo^P$  shown in \cref{tab:symmetries_orderparameters}  on the bosonized form of $C(x,y)$ using \cref{tab:symmetries_bosonization} which gives a relationship between $A^L_{mn}$ and $A^R_{mn}$ as
		\begin{multline}
			\ztwo^P: C_L(x) \xrightarrow[\phi_2(x) \mapsto \pi - \phi_1(-x)]{\phi_1(x) \mapsto \pm \pi - \phi_2(-x)} C_R(-x)   \implies \\A^R_{m,n} = \pm (i)^{m+n} ~A^L_{-n,-m}= \mp (i)^{m+n} ~A^L_{n,m}.\label{eq:Amn_Z2P}
		\end{multline}
		\Cref{eq:Amn_Z2R,eq:Amn_Z2P} are mutually compatible for non-zero $A$ iff $(m+n)$ is even. Note that we have allowed a sign ambiguity in the action of $\ztwo^P$,  $\phi_1 \mapsto \pm \pi - \phi_2$ which results in a harmless overall multiplicative sign factor in the final answer.  Using these in \cref{eq:CLR expansion}, we obtain the final bosonized form of $C(x,y) \sim C_L(x) C_R(y)$ with $C_L(x) = \pm  C_R(y)$ and  
		\begin{multline}
			C_{L}  \approx \alpha \sin\left(\frac{\phi_1+\phi_2}{2}\right)   +  \beta \sin\left(\frac{\phi_1-\phi_2}{2}\right) .
		\end{multline}
		where we have only shown  operators with the smallest scaling dimensions and the coefficients $\alpha, \beta$  are linear combinations of $A^{L/R}_{m,n}$. This reproduces the observations in \cref{sec:Numerics} that $\moy{C(x,y)} \neq 0$ when $\moy{\phi_1 \pm \phi_2} =\pi$ i.e. in the Haldane and XY$_2^*$ phases.
		
		\subsection{Bosonizing $U(\pi)$}
		We can obtain the bosonized form of the symmetry operator $U(\pi)$ defined on a finite interval $x \in [0,L]$, used in the main text using arguments similar to the above by treating it as a string operator defined for any interval.  In the small-$J$ limit, we can postulate the following form 
		\begin{align}
			U(\pi) &\sim U_L~U_R, \label{eq:ULR expansion}\\
			U_{L/R} &\sim \sum_m B^{L/R}_m e^{\frac{i}{2} m \phi}.
		\end{align}
		Unlike $C(x,y)$ which has $\ztwo^R$ charged end-points, $U_{L/R}$ do not carry any charge. Thus, we have 
		\begin{equation}
			\ztwo^R: U_{L/R}\xrightarrow{\phi \mapsto -\phi} U_{L/R} \implies B^{L/R}_{m} = B^{L/R}_{-m}. \label{eq:Bmn_Z2R_smallJ}
		\end{equation}		
		Imposing the action under $\ztwo^P$, we get
		\begin{multline}
			\ztwo^P: U_L(x) \xrightarrow{\phi(x)\mapsto- \phi(-x)} U_R(-x) \\ \implies B^R_{m} = ~B^L_{-m} = B^L_{m}.\label{eq:Bmn_Z2P_smallJ}
		\end{multline}
		Using \cref{eq:Bmn_Z2R_smallJ,eq:Bmn_Z2P_smallJ}, we get
		\begin{equation}
			U_{L/R} \sim \beta \cos \frac{\phi}{2} + \ldots
		\end{equation}
		where we have shown only the operator with the smallest scaling dimensions, and $\beta$ is some combination of $B^{L/R}_m$. In the large $J$ limit, we can postulate the form 
		\begin{align}
			U(\pi) &\sim U_L~U_R, \\
			U_{L/R} &\sim \sum_{m,n} B^{L/R}_{m,n} ~e^{\frac{i}{2} \left(m \phi_1(x) + n \phi_2(x)\right) }.
		\end{align}
		Again, imposing $\ztwo^R$ invariance of the endpoints, we get 		
		\begin{multline}
			\ztwo^R: U_{L/R}(x)\xrightarrow{\phi_\alpha \mapsto -\phi_\alpha} U_{L/R}(x) \\\implies B^{L/R}_{m,n} = B^{L/R}_{-m,-n}. \label{eq:Bmn_Z2R}
		\end{multline}
		The action of $\ztwo^P$ further gives us
		\begin{multline}
			\ztwo^P: B_L(x) \xrightarrow[\phi_2(x) \mapsto \pi - \phi_1(-x)]{\phi_1(x) \mapsto \pm \pi - \phi_2(-x)} B_R(-x) \implies \\  B^R_{m,n} = \pm (i)^{m+n} ~B^L_{-n,-m}= \pm (i)^{m+n} ~B^L_{n,m}.\label{eq:Bmn_Z2P}
		\end{multline}
		Again, \cref{eq:Bmn_Z2R,eq:Bmn_Z2P} are mutually compatible for non-zero B iff $(m+n)$ is even and we have retained the sign ambiguity in the action of $\ztwo^P$ as before when we bosonized $C(x,y)$. Using these in \cref{eq:ULR expansion}, we get the final form $U(\pi) \sim U_L U_R$ with $U_L = \pm U_R$ and
		\begin{equation}
			U_{L}\approx \gamma ~\cos\left(\frac{\phi_1+\phi_2}{2}\right) + \delta ~\cos\left(\frac{\phi_1-\phi_2}{2}\right).
		\end{equation}
		where we have only shown operators with the smallest scaling dimensions, and the coefficients $\gamma, \delta$  are linear combinations of $B^{L/R}_{m,n}$.

	\end{appendix}

	\bibliography{references,qtasep}
	
	
\end{document}